\documentclass{jfm}
\usepackage{graphicx}
\usepackage{stackengine}
\usepackage{booktabs}
\usepackage{subcaption}
\usepackage{float}
\usepackage{amssymb}
\usepackage{amsmath}
\usepackage{color}

\graphicspath{{images/}}

\shorttitle{Instability of viscoelastic plane Poiseuille flow}

\title{The center-mode instability of viscoelastic plane Poiseuille flow}

\author{Mohammad Khalid\aff{1}, Indresh Chaudhary\aff{1}, Piyush Garg\aff{2},
V.~Shankar\aff{1}\corresp{\email{vshankar@iitk.ac.in}}
\and  Ganesh Subramanian\aff{2}\corresp{\email{sganesh@jncasr.ac.in}}
}
\affiliation{\aff{1}Department of Chemical Engineering, Indian Institute of Technology, Kanpur 208016, India
\aff{2}Engineering Mechanics Unit, Jawaharlal Nehru Centre for Advanced Scientific Research, Bangalore 560064, India}
\begin{document}
	
	\maketitle

\begin{abstract}
A modal stability analysis shows that plane Poiseuille flow of an Oldroyd-B fluid becomes unstable to a `center mode' with phase speed close to the maximum base-flow velocity, $U_{max}$. The governing dimensionless groups are the Reynolds number $Re = \rho U_{max} H/\eta$, the elasticity number $E = \lambda \rho/(H^2 \eta)$, and the ratio of solvent to solution viscosity $\beta = \eta_s/\eta$; here, $\lambda$ is the polymer relaxation time, $H$ is the channel half-width, and $\rho$ is the fluid density. 
For experimentally relevant values (e.g., $E \sim 0.1$ and $\beta \sim 0.9$), the predicted critical Reynolds number, $Re_c$, for the center-mode instability is around $200$, with the associated
eigenmodes being spread out across the channel. In the asymptotic limit of $E(1-\beta) \ll 1$,  with $E$ fixed,  corresponding to strongly elastic dilute polymer solutions, $Re_c \propto (E(1-\beta))^{-3/2}$ and the critical wavenumber $k_c \propto (E(1-\beta))^{-1/2}$. The unstable eigenmode in this limit is confined in a thin layer near the channel centerline. 

The above features are largely analogous to the center-mode instability in viscoelastic pipe flow (Garg \textit{et al.}, \textit{Phys. Rev. Lett.}, \textbf{121}, 024502 (2018)), 
and suggest a universal linear mechanism underlying the onset of 
turbulence in both channel and pipe flows of sufficiently elastic dilute polymer solutions.
However, while the center-mode  instability continues down to  $\beta \sim 10^{-2} $ for pipe flow, it ceases to exist for $\beta < 0.5$ in channels.  Thus, while inertia, elasticity and solvent viscous effects are simultaneously required for this instability, a higher viscous threshold is required for channel flow.  Further, in the  opposite limit of $\beta \rightarrow 1$, the center-mode instability in channel flow continues to exist at $Re \approx 5$, again in contrast to pipe flow where the instability
ceases to exist below $Re \approx 63$, regardless of $E$ or $\beta$.
The predictions from our linear stability analysis are in excellent agreement with experimental observations for the onset of turbulence in the flow of polymer solutions through microchannels. 
\end{abstract}

%
	
\section{Introduction}
\label{sec:Introduction}
The onset of turbulence in the flow of Newtonian fluids through pipes and channels is now known to be dominated by nonlinear processes \citep{eckhardt_etal_2007,white_mungal_2008}, with the actual transition being preceded by the emergence of three-dimensional solutions of Navier-Stokes equations, dubbed `exact coherent states' \citep[abbreviated ECS;][]{Fabian_PRL_1998,waleffe_2001,wedin_kerswell_2004}, and with a concomitant reduction in the basin of attraction of the laminar state.  Experimentally, transition 
typically occurs at a Reynolds number $\Rey \approx 2000$ for pipe flows \citep{avila2011} and $\approx 1100$ for channel flows \citep{patel_head_1969}. In contrast, linear stability theory predicts  channel (plane Poiseuille) flow of a Newtonian fluid to become unstable at $\Rey \approx 5772$ \citep{Schimid-Henningson}, and pipe flow to be stable at all $\Rey$ \citep{meseguer_trefethen_2003}, implying that the presence or absence of a linear instability has no relevance to the observed subcritical transition.
The mechanisms underlying transition in pipe and channel flows of viscoelastic polymer solutions has, however, received much less attention.
%
While the addition of polymers ($\sim 10$ppm onward) is well known to result in drag reduction in the fully turbulent regime \citep{Tom1977,Virk,white_mungal_2008,Graham2014,Xi2019DRreview}, the onset of turbulence in polymer solutions has attracted attention only recently. In their experiments on pipe flow of polymer solutions, \cite{Samanta2013} showed that, for concentrations greater than $300$ppm, transition  occurs at an $Re$ lower than $2000$, and the ensuing flow state was  referred to as `elasto-inertial turbulence' (abbreviated EIT). Recent experiments by  \cite{Choueiri2018} and \cite{Bidhan2018,chandra_shankar_das_2020} have corroborated these findings using micro-PIV and pressure-drop measurements.	
While most of the experiments on viscoelastic transition have been carried out in the pipe geometry, the study of \cite{Srinivas-Kumaran2017} showed, using PIV measurements, that transition in the 
flow of dilute polymer solutions (with concentrations in the range $30$--$50$ppm), through a rectangular channel with 
a gap width of $160 \mu$m and a cross-sectional aspect ratio of $1:10$, occurred at an $\Rey \sim 300$, again significantly lower than the Newtonian threshold. Importantly, \cite{Samanta2013} showed that, for concentrations greater than $300$ppm, turbulence onset in pipe flow occurred at the same $\Rey$ irrespective of whether the flow is perturbed or not, implying that the flow becomes unstable to infinitesimal disturbances.
This suggests a common linear mechanism underlying  transition in the flow of polymer solutions through both pipes and channels, particularly for sufficiently concentrated polymer solutions for which the transition occurs at $Re$'s much lower than those corresponding to the Newtonian transition.  The proposed linear scenario for viscoelastic pipe and channel flows is thus in direct contrast with the Newtonian transition in these geometries, wherein the common underlying mechanism  has a nonlinear subcritical character.

The notion of a linear mechanism underlying the viscoelastic transition was reinforced by our recent discovery \citep{Piyush_2018} of pipe flow of an Oldroyd-B fluid  being linearly unstable, in sharp contrast to the Newtonian scenario, with the critical $Re$ being as low as $100$ for strongly elastic dilute solutions; a more detailed account of this instability is provided in \cite{chaudharyetal_2020}. In \cite{Piyush_2018}, we had alluded
to the existence of a similar instability in pressure-driven channel flow.   
In the present study, we show that an analogous instability does indeed exist for channel flow, and for $\Rey$'s much lower than $1000$.  We provide a comprehensive picture on the origin of the instability and the domain of its existence in the parameter space consisting of $\Rey = \rho U_{max} H/\eta$, elasticity number $E = \lambda \eta/(\rho H^2)$, and the ratio of solvent to solution viscosity $\beta = \eta_s/\eta$. Here, $\lambda$ is the microstructural relaxation time, $U_{max}$ is the maximum base-flow velocity, $\rho$ is the fluid density, and $H$ is the channel half-width.  
In addition, we discuss the similarities and differences between the center-mode instabilities of pipe and channel flows, in the aforementioned $Re$--$E$--$\beta$ space, ending with a discussion of the possible transition scenarios for viscoelastic channel flow. We also show that our predictions are in good agreement with the observations of \cite{Srinivas-Kumaran2017}.

\subsection{Stability of rectilinear viscoelastic shearing flows}
\label{subsec:stabilityshearflow}
We first provide a brief overview of relevant previous work on stability of viscoelastic channel flow; a detailed survey of this subject can be found in \cite{chaudhary_etal_2019}. 
Most earlier studies have employed the upper-convected Maxwell (UCM)/Oldroyd-B class of models to analyze the modal stability of both plane Couette and Poiseuille flows. 
Recall that the  dimensionless parameters governing the stability of an Oldroyd-B fluid are $Re$, $E$ and $\beta$, with $\beta = 0$ and $1$ being the UCM and Newtonian limits, respectively (note that, in lieu of $E$,  one may also use the Weissenberg number $W = E Re$).   
To begin with, it is useful to recall the broad features of the Newtonian spectrum for plane Poiseuille flow.
At sufficiently high $\Rey$, the spectrum has a characteristic `Y-shaped' locus with three distinct branches: the `A branch' comprising `wall modes' with phase speeds $c_r \rightarrow 0$, the `P branch' comprising `center modes' with phase speeds $c_r \rightarrow 1$, and the `S branch' which forms a vertical line in the $c_r$--$c_i$ plane comprising modes with a phase speed equalling two-thirds of the maximum base-flow velocity. A wall mode belonging to the A-branch, referred to as the Tollmien-Schlichting (TS) mode, becomes unstable for $Re > 5772$ for plane Poiseuille flow \citep{Schimid-Henningson}. 
While viscoelastic plane Poiseuille flow was found to be stable at low Reynolds number ($< 1$) by  \cite{HO_DENN1977,lee_finlayson_1986},
Denn and co-workers \citep{Porteous_Denn1972,HO_DENN1977} used the UCM model and showed that, for sufficiently high $Re$ ($> 2000 $) and $ E $, two new unstable wall modes appear in the eigenspectrum in addition to the elastically-modified TS mode and one of these new modes is the most unstable mode at sufficiently high $E$. 	
\cite{Sureshkumar1995} used an Arnoldi algorithm to identify the most unstable eigenmodes in plane Poiseuille flow of a UCM fluid, and showed that the critical $Re$ ($Re_c$) for the elastically modified TS mode showed a non-monotonic behaviour with increasing $E$. Consistent with the findings of \cite{Porteous_Denn1972}, at sufficiently high $E$, \cite{Sureshkumar1995} identified an unstable mode which is absent in Newtonian channel flow.
However, the new unstable mode was found to be suppressed on account of a finite solvent viscosity (using the Oldroyd-B model) or finite extensibility \cite[using the FENE-CR model; see][]{ChilcottRallison}.  Subsequently, 
\cite{sadanandan_sureshkumar_2002} carried out a modal stability analysis  to explore the effect of fluid elasticity on the TS mode at different $\beta$ and showed
a non-monotonic dependence of $Re_c$  on $E$, similar to the UCM limit. A similar non-monotonic behaviour was also reported by \cite{Zaki2013}  using the FENE-P model which, like the FENE-CR model above,  accounts for the finite extensibility of polymer chains.
The recent effort of \cite{Brandi2019} also explored the role of elasticity on the TS (wall) mode using 
the Oldroyd-B model,  focusing on smaller range of $E$'s ($0  < E < 0.003$).  Both linear stability analysis (using a shooting procedure) and DNS were used to analyze the unstable flow structures corresponding to the wall mode, and good agreement was found between the two.  

As mentioned above, viscoelastic plane Poiseuille flow is stable in the limit of low $Re$, and Kumar and co-workers \citep{hoda_jovanovic_kumar_2008,hoda_jovanovic_kumar_2009,jovanovic_kumar_2010,jovanovic_kumar_2011} have therefore explored the possibility of non-modal (transient) growth in these flows, with the non-modal mechanism being purely elastic, and therefore operative in the inertia-less limit. \cite{Zaki2013}, in contrast, examined non-modal growth in inertially dominated channel flows of both Oldroyd-B and FENE-P fluids, and found that stream-wise elongated structures exhibit the largest transient growth in the subcritical regime.
There have also been many studies that used a weakly nonlinear approach \citep{bertola_saarloos2003,meulenbroek_sarloos2004,morozov_saarloos2005,Pan_2012_PRL} to identify a subcritical instability in the inertia-less limit. These studies were motivated by a hoop-stress driven mechanism operative at the nonlinear order, which is caused by a curvature in the streamlines due to infinitesimal perturbations. 
However, these nonlinear analyses
were predicated on the rather simplistic structure of the viscoelastic spectrum in the inertialess limit, and as pointed out by \cite{chaudharyetal_2020}, may not be applicable at higher $Re$.

In a recent effort, \cite{chaudhary_etal_2019} employed a numerical shooting procedure along with the spectral method (over a wide range of $\Rey$ and $E$) to provide a comprehensive picture of the stability of both plane Couette and Poiseuille flows in the UCM limit. 
In contrast to the earlier efforts mentioned above, \cite{chaudhary_etal_2019} also analyzed the structure of the elastoinertial spectrum in detail, in addition to examining the unstable discrete modes found in earlier studies. In doing so, at sufficiently high $Re$ and $E$, the authors demonstrated the existence of a possibly infinite hierarchy of elasto-inertial instabilities in Poiseuille flow which are absent in the Newtonian limit. Further, both sinuous and varicose modes were shown to be unstable, in contrast to the Newtonian case where only the sinuous mode is unstable. For $\Rey \gg 1$, the unstable modes found
by \cite{chaudhary_etal_2019} belong to the class of wall modes, and the minimum  Reynolds number at which
the flow is unstable (at any $E$) was found to be $O(1000)$ in the UCM limit. It has recently been found \citep{khalid_solvent} that the inclusion of a solvent (viscous) contribution, corresponding to a small but finite $\beta$, has a strong stabilizing effect on these unstable modes, an effect that may be attributed to the presence of fine-scaled structures in the higher-order elasto-inertial modes.
 Thus, the wall mode instabilities examined in earlier studies do not pertain to the transition observed in channel flow of dilute polymer solutions with $\beta \sim 0.9$ \citep{Srinivas-Kumaran2017}.

While the aforementioned efforts focused on wall modes, \cite{Piyush_2018} reported a hitherto unexplored linear instability in pipe Poiseuille flow of an Oldroyd-B fluid, with the unstable eigenmode belonging to a class of elasto-inertial (axisymmetric) center modes with phase speed approaching the maximum base-flow velocity. The instability exists only in the presence of solvent viscous effects, and is surprisingly absent in the UCM limit.  This implies a destabilizing role of solvent viscosity on the center mode, in direct contrast to its stabilizing role on the aforementioned wall-mode instabilities in channel flow. 
Further, the threshold $Re$ for transition is significantly lower than the Newtonian threshold even for relatively modest $E$'s; for instance, $Re_c \sim 500$ for $\beta = 0.8$ and $E \sim 0.1$.
As was briefly reported in \cite{Piyush_2018}, a similar center-mode instability exists in plane channel flow of an Oldroyd-B fluid. The central objective of the present work is to expand further on the
origin and nature of this center-mode instability in viscoelastic channel flow, and to identify its domain of existence in the $Re$-$E$-$\beta$ space.

\subsection{Computational bifurcation studies and direct numerical simulations}
\label{subsec:dnsstudies}
We may classify computational efforts towards understanding viscoelastic transition and drag reduction into two broad categories: (i) bifurcation studies that have explored the role of viscoelasticity on the 3D Newtonian ECS solutions that helped shed light on the Newtonian transition scenario, and (ii) direct numerical simulations (DNS). Both classes of investigations almost exclusively use the FENE-P equation to model the polymer dynamics.
In direct contrast to the experimental scenario which, as already seen, is dominated by a focus on pipe flows, almost all of the computational studies 
(except that of \cite{lopez_choueiri_hof_2019}; see below)
have been carried out for the channel geometry. Implicit in this  focus on the channel geometry is the assumption of an identical physical mechanism underlying the transition in both the pipe and channel geometries. This
 is justified in the Newtonian case owing to the structural similarities of the Newtonian ECS solutions in all of the canonical rectilinear shearing flows including, in particular, the channel \citep{waleffe_2001} and pipe \citep{wedin_kerswell_2004} geometries; the ECS solutions in all cases are characterized by a staggered arrangement of counter-rotating vortices and streamwise streaks. 
Thus, although Newtonian pipe and channel flows yield very different results with regard to linear modal stability \citep{Drazinreid}, they nevertheless exhibit similar sub-critical transitions to turbulence, with this transition in either case being understood now in terms of a turbulent trajectory wandering chaotically amongst a multitude of the aforementioned ECS solutions in an appropriate phase space.  A series of papers by Graham and co-workers \citep{stone_graham2002,stone_graham2003,stone_graham2004,li_etal_2006,Graham2007} have shown that elasticity has a stabilizing effect on the simplest of the 3D ECS solutions (travelling waves) in viscoelastic plane Couette and Poiseuille flows, in terms of delaying the bifurcation birthing these solutions to a higher $Re$; the results for sufficiently high $E$ are suggestive of the ECS's being fully suppressed by elasticity. This, in turn, is suggestive of a delay in transition 
due to elasticity, a prediction that has some experimental support wherein the onset of turbulence, in pipe flow of polymer solutions, was delayed at lower polymer concentrations \citep{Samanta2013,Bidhan2018,Choueiri2018}.

Starting from the pioneering work of \cite{sureshkumar_1997}, there have been many DNS investigations
\citep{sibilla_baron2002,deangelies_etal,dubief_white_terrapon_shaqfeh_moin_lele_2004,xigrahamPRL2010,Xi_Graham_2012,Xi2019DRreview} 
carried out to study the  mechanisms underlying turbulent drag reduction. These efforts were able to successfully 
capture the moderate drag reduction regime (at $Re$'s below the so-called maximum drag reduction regime), and showed that turbulence production in the buffer layer is modified by the addition of polymers, as originally predicted by \cite{virk1975}. 
All of these early studies incorporated an additional diffusive term in the constitutive equation in order to preserve the positive definiteness of the polymer conformation tensor. However, the diffusivity $D$ used is orders of magnitude larger than the Brownian diffusivity of a polymer molecule. The Schmidt number $Sc = \nu/D$ should be $O(10^6)$ (where $\nu$ is the kinematic viscosity of the fluid) for realistic values of the polymer diffusivity, but the aforementioned simulations used $Sc \sim 0.5$. Recently, Dubief and co-workers \citep{Dubief2013,Samanta2013,Sid_2018_PRF} have carried out DNS of viscoelastic channel flow in the absence of stress diffusion to show that the deviation of friction factor from the laminar value occurred at $Re \sim 700$ (while it does so for $Re \sim 5000$ for the Newtonian case in their computations), 
thereby demonstrating the early onset of elastoinertial turbulence, in direct contradiction to the conclusions of Graham and co-workers based on their investigation of the elastically modified ECS's.
Crucially, the structures that dominated the onset of EIT were two-dimensional (span-wise elongated and stream-wise varying), in direct contrast with the 3D ECS structures (stream-wise elongated and span-wise varying) that dominate the Newtonian (and weakly elastic) transition. 
The recent work of \cite{Sid_2018_PRF} has shown that the 2D EIT structures are suppressed for $Sc < 9$, thus demonstrating the spurious stabilizing role played by the large stress diffusivities used in the earlier DNS studies (It is pertinent here to add a caveat that the aforementioned results of Graham and co-workers on the stabilization of the simplest ECS's were also obtained using artificially large stress diffusivities, and it would therefore be prudent to revisit the original conclusions of the authors, at $Sc \sim O(1)$, in light of the recent findings for $Sc = \infty$).
 Another recent DNS study \citep{lopez_choueiri_hof_2019},  the only one that pertains to the pipe geometry, showed that the onset to EIT is dominated by axisymmetric vortices oriented along the azimuthal direction (the analog of the span-wise direction in the pipe geometry).
The qualitative similarity between the nature of elasto-inertial structures seen in the aforementioned DNS of viscoelastic channel and pipe flows is, in fact, consistent with our earlier report \citep{Piyush_2018} of an analogous linear instability in these flows, thereby suggestive of a generic linear mechanism for turbulence onset in viscoelastic channel and pipe flows.  
Note, however, that the analogy is a qualitative one since the pipe-center mode eigenfunctions, even when confined to the neighborhood of the centerline, as happens at large $Re$ \citep[see][and Sec.~4 of this work]{chaudharyetal_2020}, do not still lend themselves to a two-dimensional approximation. Thus, as will be demonstrated below, there remain some important differences in the behavior of the threshold parameters for the pipe and channel flow cases.

As mentioned above, the ECS-driven 3D transition mechanism is suppressed for quite modest $E$'s, and on the other hand, it is shown in the present work that the center-mode instability exists only for sufficiently high $E$'s. Thus, for intermediate $E$'s,
there must be new (subcritical) nonlinear mechanisms that underlie the viscoelastic transition. In this regard, two very different mechanisms have been advanced in the recent literature. The first one by 
\cite{Shekar2019,Graham2019_Archive} proposes a 2D nonlinear mechanism  that entails strongly localized polymer stretch fluctuations near the
`critical layer' (the transverse location where the phase speed of disturbances equals the local laminar velocity) corresponding to the (least stable) elastically modified, TS (wall) mode. The second one by \cite{page2020exact} \citep[also see][]{dubief2020coherent} is rooted 
 in a novel nonlinear elastoinertial coherent structure that originates (subcritically) from the critical point corresponding to the center-mode instability. We argue below, in Sec.~\ref{ssec:Relative stability of center and wall modes}, that while the center mode is invariably the least stable mode for high $E$'s, even in the Newtonian or weakly elastic limit, there exist parameter regimes (based on the perturbation wavenumber and the elasticity number) where the center mode is less stable than all the wall modes, including the aforementioned TS mode. Thus, the 2D nonlinear mechanism rooted in the TS mode  \citep{Shekar2019} is likely be valid in restricted parts of the $Re$-$E$ parameter space, even for smaller $E$'s for which the center mode is linearly stable. Nevertheless, given the relevance of the least stable eigenmode(s) in the elasto-inertial spectrum to both of the aforementioned nonlinear mechanisms, in Sec.~\ref{ssec:ECS_vs_EIT},
we demarcate regions in the $Re$--$E$ plane where the center and wall modes are least stable. 
In light of the rather high-dimensional parameter space required even for a minimal description of viscoelastic shearing flow, such a demarcation should serve as a useful guide in the search for nonlinear mechanisms in the $Re$--$E$ plane, where the flow is linearly stable.
%
%

\begin{figure}
\centering {\includegraphics[width=0.65\textwidth]{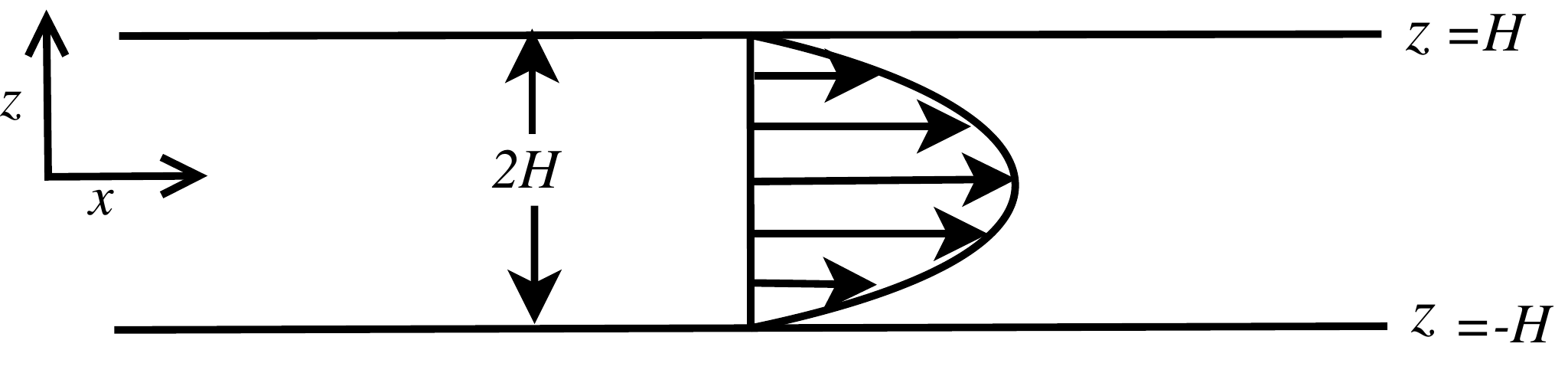}}
\caption{\small Schematic representation of the configuration consisting of pressure-driven flow in a
channel of half-width $H$.}
\label{fig:Prob_form}
\end{figure}

The rest of this paper is structured as follows. Section \ref{sec:problem formulation} provides the
linearized governing equations for viscoelastic channel flow, along with a discussion and validation
of the numerical methods used in this study. In Sec.~\ref{ssec:Newtonian and UCM limit}, we discuss the general features of the Oldroyd-B eigenspectrum and contrast it with its Newtonian counterpart. Section~\ref{sssec:Effect of elasticity} 
shows how the Oldroyd-B spectrum deviates from the Newtonian one as $E$ is increased at fixed $\beta$, and demonstrates the origin of the unstable center mode with increasing $E$.  Section~\ref{sssec:Effect of solvent viscosity} examines the deviation from the Newtonian limit at a fixed $E$, but with $\beta$ decreasing from unity, the focus again being on the emergence of the center mode below a threshold $\beta$.
 The relative importance of center modes, wall modes and modes belonging to the continuous spectrum in viscoelastic channel flow is discussed in Sec.~\ref{ssec:Relative stability of center and wall modes}, where it is argued that at sufficiently high $E$'s, it is either the continuous spectrum or the center mode which is  least stable (or even unstable, in  case of the center mode), and therefore, the recently proposed TS-mode-based transition scenario \citep{Shekar2019,Graham2019_Archive} might only have a restricted range of applicability. Neutral stability curves in the $Re$-$k$ plane are presented in Section~\ref{sec:Neutral stability curves for unstable elasto-inertial center mode}. Section~\ref{ssec:Data collapse} shows  the collapse of the neutral stability curves  in the limit $E \ll 1$ for a given $\beta$, and in the limit $E(1-\beta) \ll 1$ for fixed $E$. The variation of the critical parameters ($Re_c$, $k_c$) with $E (1-\beta)$ is discussed in Sec.~\ref{ssec:Critical parameters and Scalings}, while the absence of this instability
at lower $\beta$ is demonstrated in Sec.~\ref{ssec:Role of beta}. In Sec.~\ref{ssec:diffusion},
 the threshold $Re$ for the center-mode instability is shown to remain virtually unaltered for realistic polymer diffusivities, although the artificially large stress diffusivities used in many DNS simulations has a stabilizing effect.
Our theoretical predictions are shown to agree well with the observations of \cite{Srinivas-Kumaran2017} in Sec.~\ref{ssec: Comparison}. In Sec.~\ref{ssec:ECS_vs_EIT}, we discuss the  possible transition scenarios in viscoelastic channel flows by showing our results for the onset of transition via linear instability, alongside the results of \cite{Graham2007} for the  ECS-mediated nonlinear transition, in the $Re$-$E$ plane. The salient conclusions of the present study are provided in Sec.~\ref{sec:conclusions}.

%
 
\section{Problem formulation} 
\label{sec:problem formulation}
\subsection{Governing equations}
We consider pressure-driven flow of an    incompressible viscoelastic fluid in a  channel with walls separated by a distance $2H$ (Fig.~\ref{fig:Prob_form}). The viscoelastic fluid is modelled using the Oldroyd-B constitutive equation \citep{Larson}, which is applicable  to dilute polymer solutions wherein the polymer chains are assumed to be non-interacting, and each chain is modelled as an elastic dumbbell with beads connected by a linear infinitely extensible entropic spring. This model predicts a shear-rate independent viscosity and first normal stress difference in viscometric shearing flows. Many authors have used this model  in the past to analyze instabilities in the flow of dilute polymer solutions in
rectilinear \citep{Sureshkumar1995,wilson1999,Zaki2013,MOROZOV_2007,Piyush_2018},
 curvilinear \citep{Shaqfeh_1996}, and cross-slot \citep{Poole2007} geometries with considerable success. 
To render the governing equations dimensionless, we use the centerline maximum velocity of the laminar base state, $U_{max}$, as the velocity scale,  channel half-width $H$ as the length scale, $H/U_{max}$ as the time scale, and $\eta U_{max}/H$ as the scale for the stresses and pressure. Here, $\eta = \eta_p + \eta_s$ is the solution viscosity which is a sum of the polymer  $\eta_p$  and solvent $\eta_s$ contributions. The dimensionless continuity and momentum equations are given by
\begin{eqnarray}
\label{eq:fluid-continuity}
\nabla \cdot \mathbf{u}=0,\\
\label{eq:fluid-momentum}
Re \, \Big(\frac{\partial \mathbf{u}}{\partial t}+(\mathbf{u}\cdot\nabla) \, \mathbf{u} \Big)=-\nabla {p}+ {\beta} \,\nabla^2 \, \mathbf{u} +\nabla \cdot \boldsymbol{\tau} .
\end{eqnarray}
Here, $Re = \rho U_{max} H/\eta$ is the Reynolds number based on the  solution viscosity and $\beta = \eta_s/\eta$. The Oldroyd-B constitutive relation for the polymeric stress tensor, $\boldsymbol{\tau}$, in dimensionless form is given by
\begin{eqnarray}
\boldsymbol{\tau}+{W} \,\Big(\frac{\partial \boldsymbol{\tau}}{\partial t}+(\mathbf{u}\cdot\nabla )\boldsymbol{\tau}-(\nabla \mathbf{u})^\intercal \cdot \boldsymbol{\tau}- \,\boldsymbol{\tau} \cdot(\nabla \mathbf{u})\Big)=\Big( 1-\beta \Big)\Big(\nabla \mathbf{u}+\nabla \mathbf{u}^\intercal \Big).
\label{eq:stress constitutive eq}
\end{eqnarray}
Here, $W = \lambda U_{max}/H$ is the Weissenberg number and $\lambda$ is the microstructural relaxation time. The upper-convected Maxwell (UCM) model, which ignores the solvent contribution to the stress, is obtained from the Oldroyd-B model by setting $\beta = 0$, while the limit of a Newtonian fluid is obtained by setting $\beta = 1$.

\subsection{Base flow}
The laminar base state whose stability is of interest here is the steady fully-developed pressure-driven  channel flow of an Oldroyd-B fluid, with the base-state velocity profile  $U(z) = 1-z^2$ being identical to that of plane Poiseuille flow of a Newtonian fluid. However, unlike its Newtonian counterpart, the Oldroyd-B fluid exhibits a nonzero first normal stress difference ($T_{xx} - T_{zz}) = 8 (1-\beta) W z^2$. Here, and in what follows, the velocity and stress fields corresponding to the base flow are denoted by upper case alphabets.
%
\subsection{Linearized governing equations}
A temporal linear stability analysis of the aforementioned base flow  is carried out by imposing  infinitesimal perturbations (denoted by primes) to the base flow:
$\mathbf{u}=\mathbf{U}+\mathbf{u'}, \,\, p=P+p', \,\, \boldsymbol{\tau}=\mathbf{T}+\boldsymbol{\tau'}$. Since Squire's  theorem is valid for plane Poiseuille flow of an  Oldroyd-B fluid \citep{BISTAGNINO2007},  we restrict our analysis to two-dimensional perturbations, which are
considered as elementary  Fourier modes  of  the form $f'(x,z,t)=\tilde{f}(z)\exp[ik(x-ct)]$, where $f'$  is the  relevant disturbance field, $\tilde{f}(z)$ is the eigenfunction, $k$ is the dimensionless  wavenumber, and the eigenvalue $c
=c_r+i c_i$ is the complex wavespeed of perturbations. If $c_i > 0$, the perturbations grow exponentially with time leading to an instability. Substituting the Fourier mode representation for the perturbations in the linearized governing equations yields 

\begin{eqnarray}
\label{eq:fluid-continuity-NM}
d_z \tilde {v}(z)+i k\tilde {u}(z)=0,
\\
\label{eq:x-fluid-momentum-NM}
Re \, \Big[i k \,(U-c)\tilde{u}(z)+\tilde{v}(z)U'\Big]=-ik\tilde{p}(z)+\beta(d_z^2-k^2)\tilde{u}(z)+ik\tilde{\tau}_{xx}(z)+d_z\tilde{\tau}_{xz}(z), \\
\label{eq:z-fluid-momentum-NM}
Re \, i k \,(U-c)\tilde{v}(z)=-d_z\tilde{p}(z)+\beta (d_z^2-k^2) \tilde{v}(z) 
+ik\tilde{\tau}_{xz}(z)+d_z\tilde{\tau}_{zz}(z),
\\
\label{eq:Txx NM}
\Big[1+ i k W \,(U-c)\Big]\tilde{\tau}_{xx}(z)=(1-\beta)\,\Big[2 i k \tilde{u}(z)+ 4 i k W^2 (U')^2 \tilde{u}(z)  \nonumber \\
+2 W U' d_z\tilde{u}(z)-4W^2 U' U'' \tilde{v}(z)\Big]+2W U' \tilde{\tau}_{xz}(z),\\
\label{eq:Tzz NM}
\Big[1+i k W \,(U-c)\Big]\tilde{\tau}_{zz}(z)=2(1-\beta)\,\Big[
d_z \tilde{v}(z) + i k W U' \tilde{v}(z)\Big],
\\
\label{eq:Txz NM}
\Big[1+i k W \, (U-c)\Big]\tilde{\tau}_{xz}(z)=
(1-\beta)\,\Big[d_z \tilde {u}(z)+i k \tilde{v}(z)\nonumber \\
+2 i k W^2  (U')^2\tilde{v}(z) -W U'' \tilde{v}(z) \Big]+W U'\tilde{\tau}_{zz}(z)+2 i k W^2  (U')^2\tilde{v}(z).
\end{eqnarray}

  \begin{table*}
  \begin{center}
  \def\arraystretch{0.85} 
  \setlength\tabcolsep{10 pt}
  \def~{\hphantom{2}}
  \begin{tabular}{lccc}
  \hline
     $N$ & $Re, k, E, \beta$ & Sureshkumar and Beris & Present\\   
     \hline
     $257$ &$ 1990,1.2,0.003, 0$ & $0.34580+1.01\times10^{-4}$i & $0.34580 +  9.81\times10^{-6}$i \\
     $129$ &$ 1990,1.2,0.003,0$ & $0.34580 +  9.81\times10^{-5}$i & $0.34580 +  9.81\times10^{-6}$i \\ 
      $129$ &$ 3960,1.15,0.001, 0.5$ & $0.29643+1.71 \times 10^{-7}$i & $0.29643+1.73 \times 10^{-7}$i \\
    \hline       
  \end{tabular}
 \caption{ Validation of UCM ($\beta=0$) and Oldroyd-B $ (\beta=0.5) $ results with those of   \cite{Sureshkumar1995} for viscoelastic channel flow.}
 \label{tab:Validation}
  \end{center}
  \end{table*}

\subsection{Numerical procedure}
 In order to determine the complex eigenvalue ($c$),  we use a spectral collocation method \citep{Boyd1999,Weideman:2000:MDM:365723.365727}, where the dynamical variables (velocity, pressure and stress perturbations) are expanded as a finite sum of Chebyshev polynomials and substituted in the above linearized differential equations. 
 In our spectral formulation, we discretize all of the six Eqs.~\ref{eq:fluid-continuity-NM}-\ref{eq:Txz NM}, and the resulting generalized eigenvalue problem is of the form 
%
 \begin{equation}
\mathbf{Ax}=c\,\mathbf{Bx},
\label{eq:GEVP}
\end{equation} 
where $ \mathbf{A} $ and $ \mathbf{B} $ are coefficient matrices, and $\mathbf{x} = (\tilde{u},\tilde{v},\tilde{p},\tilde{\tau}_{xx},\tilde{\tau}_{xz},\tilde{\tau}_{zz})^\intercal$ is the vector comprising of the coefficients of the spectral expansion at the collocation points.   The size of the $ \mathbf{A} $ matrix is $6N \times 6N$, where $N$ is the number of Gauss-Lobatto collocation points. The generalized eigenvalue problem is solved using the `polyeig' eigenvalue solver of Matlab. To filter out the spurious eigenvalues associated with the spectral method, we run our spectral code for two different values of $N$, say, $400$ and $500$, and eliminate those eigenvalues that do not satisfy a prescribed tolerance criterion. 
In addition, a numerical shooting procedure \citep{HO_DENN1977,Schimid-Henningson,Finlayson} is used for further validation by providing the results from the spectral method as initial guesses. The numerical shooting procedure involves an adaptive Runge-Kutta integrator coupled with a Newton-Raphson iterative scheme to solve for the eigenvalues. Only physically genuine modes from the spectral method converge with the shooting code. 
To benchmark the implementation of our numerical methodology, we compare (Table~\ref{tab:Validation}) results from our procedure with those of \cite{Sureshkumar1995}  for both UCM and Oldroyd-B fluids.  The unstable eigenvalues are in good agreement for $N=129$ and $N=257$. In addition, we have benchmarked our results with those of \cite{chaudhary_etal_2019} for the UCM case.


\begin{figure*}
	\centering
	\begin{subfigure}[b]{0.5\textwidth}
		\centering
		\includegraphics[width=\textwidth]{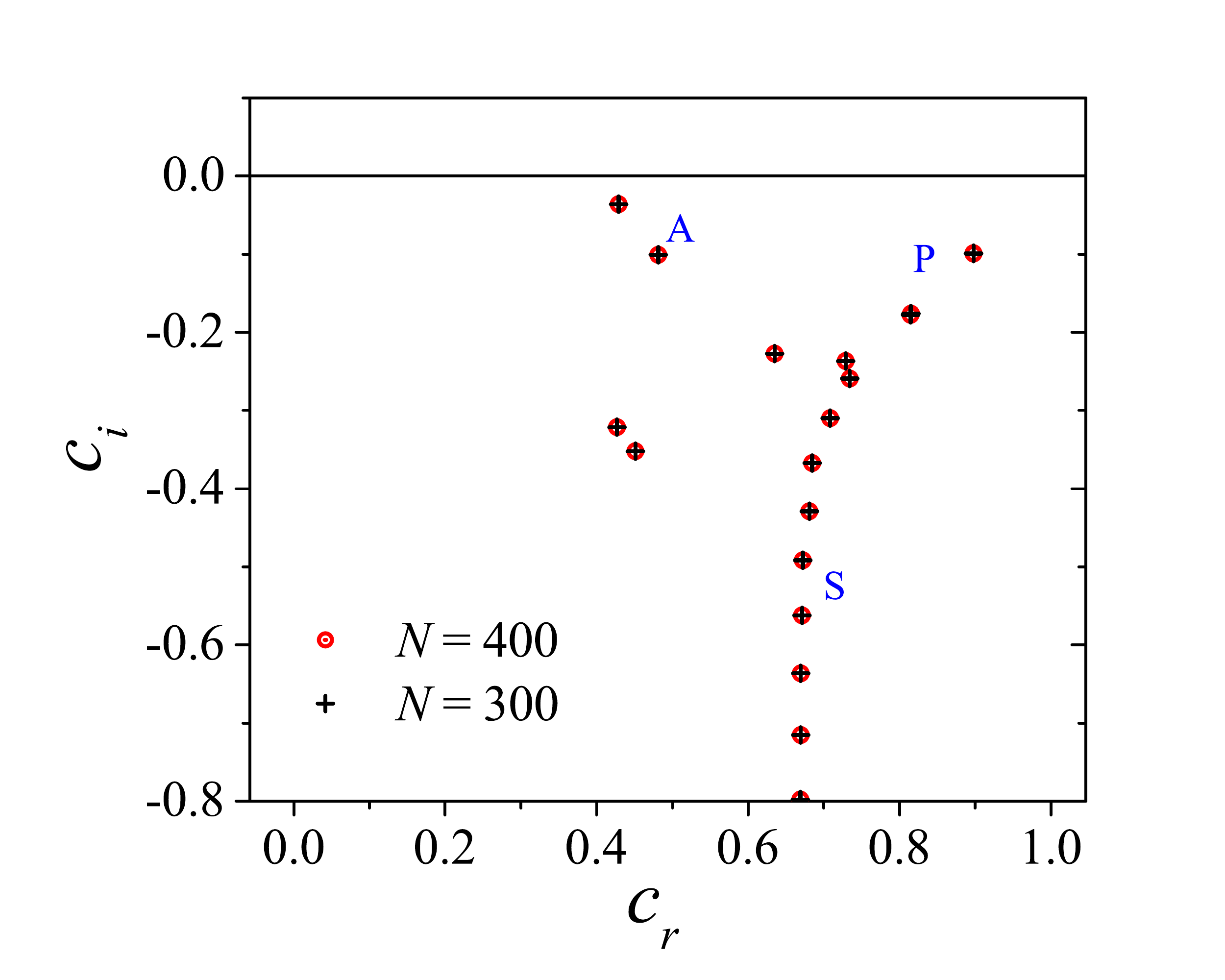}
		\caption{Newtonian }
		\label{fig:Re800_k1p5_E0p1_B1}
	\end{subfigure}%
	~ 
	\begin{subfigure}[b]{0.5\textwidth}
		\centering
		\includegraphics[width=\textwidth]{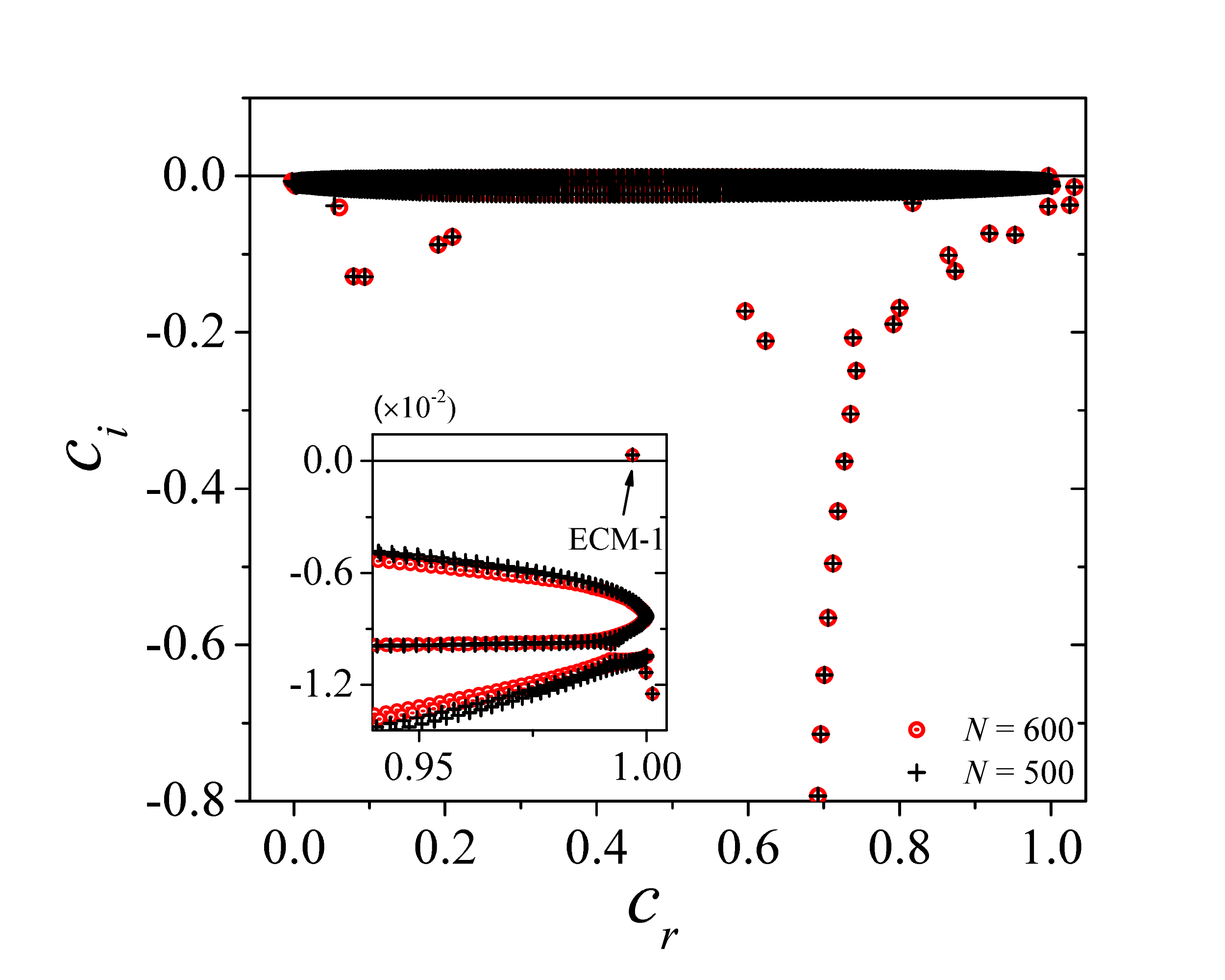}
		\caption{Oldroyd-B }
		\label{fig:Re800_E0p1_k1p5_B0p8}
	\end{subfigure}
	
	\caption{\small  Eigenspectra for plane Poiseuille flow of (a) Newtonian ($E= 0$), and (b) Oldroyd-B ($E = 0.1$) fluids at $Re = 800, k = 1.5$, and $\beta = 0.8$. The A, P, and S branches of the Newtonian spectrum are indicated in panel~(a). The inset in panel (b) zooms over the region near the unstable eigenvalue. }
	\label{fig:Eigenspectrum_Newtonian-Oldroyd-B}
\end{figure*}

\section{The elasto-inertial spectrum of an Oldroyd-B fluid}   
\label{sec:Results and discussion}
\subsection{Newtonian and Oldroyd-B spectra}
\label{ssec:Newtonian and UCM limit}

\begin{figure*}
	\centering
	\begin{subfigure}[b]{0.5\textwidth}
		\centering
		\includegraphics[width=\textwidth]{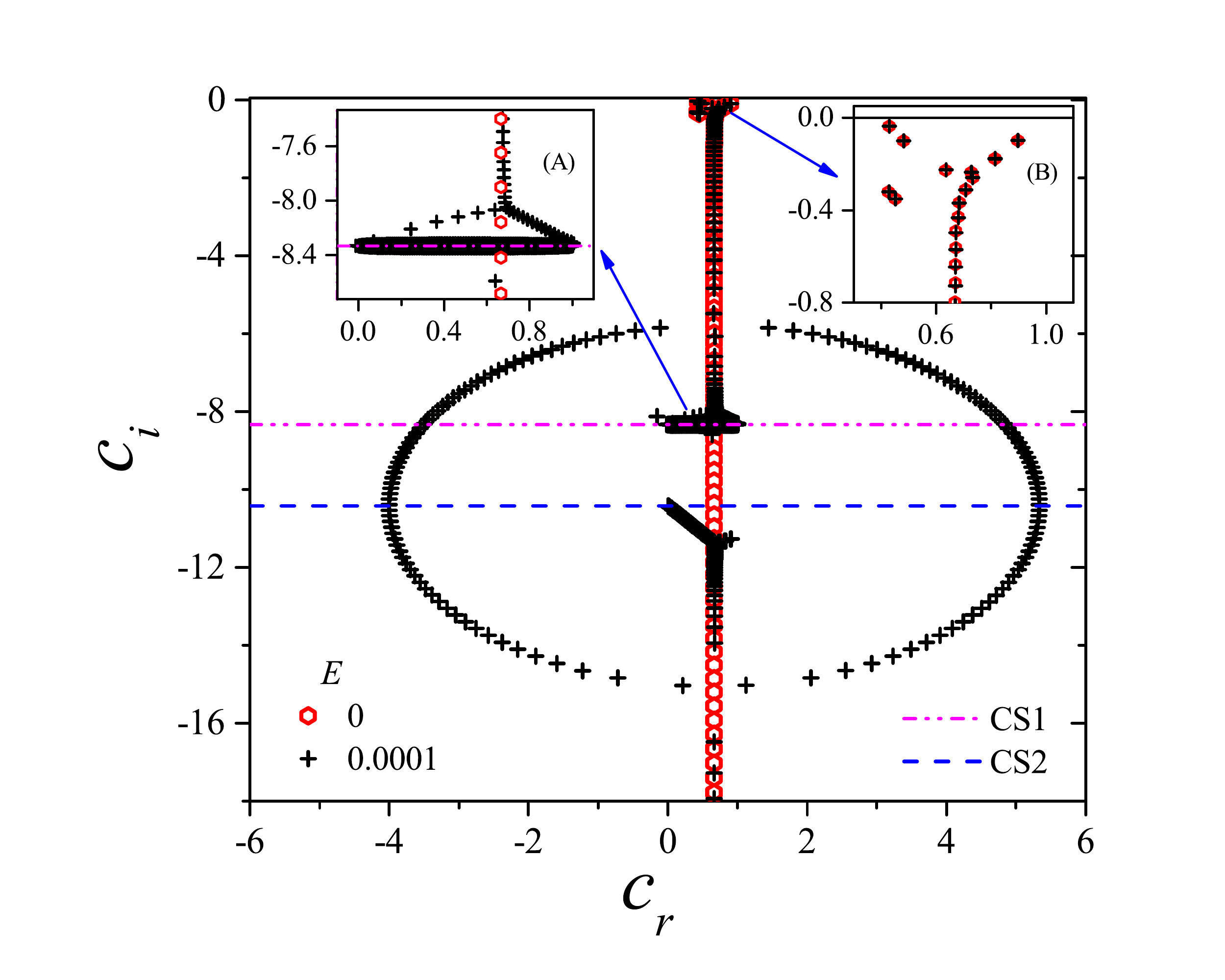}
		\caption{$  E = 10^{-4} $}
		\label{fig:E_1e-4}
	\end{subfigure}%
	~ 
	\begin{subfigure}[b]{0.5\textwidth}
		\centering
		\includegraphics[width=\textwidth]{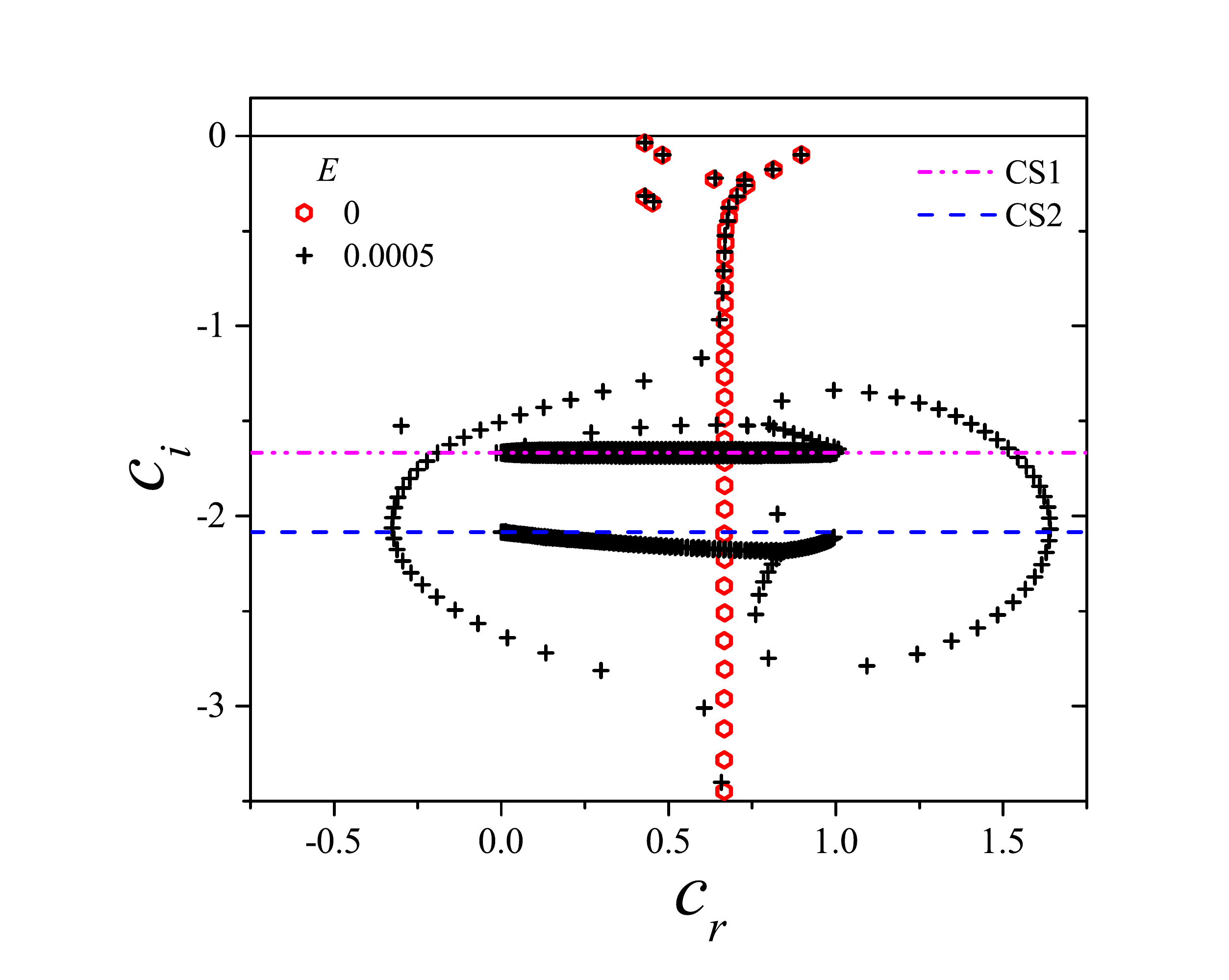}
		\caption{$ E = 5\times 10^{-4} $}
		\label{fig:E_5e-4}
	\end{subfigure}
	~
	\begin{subfigure}[b]{0.5\textwidth}
		\centering
		\includegraphics[width=\textwidth]{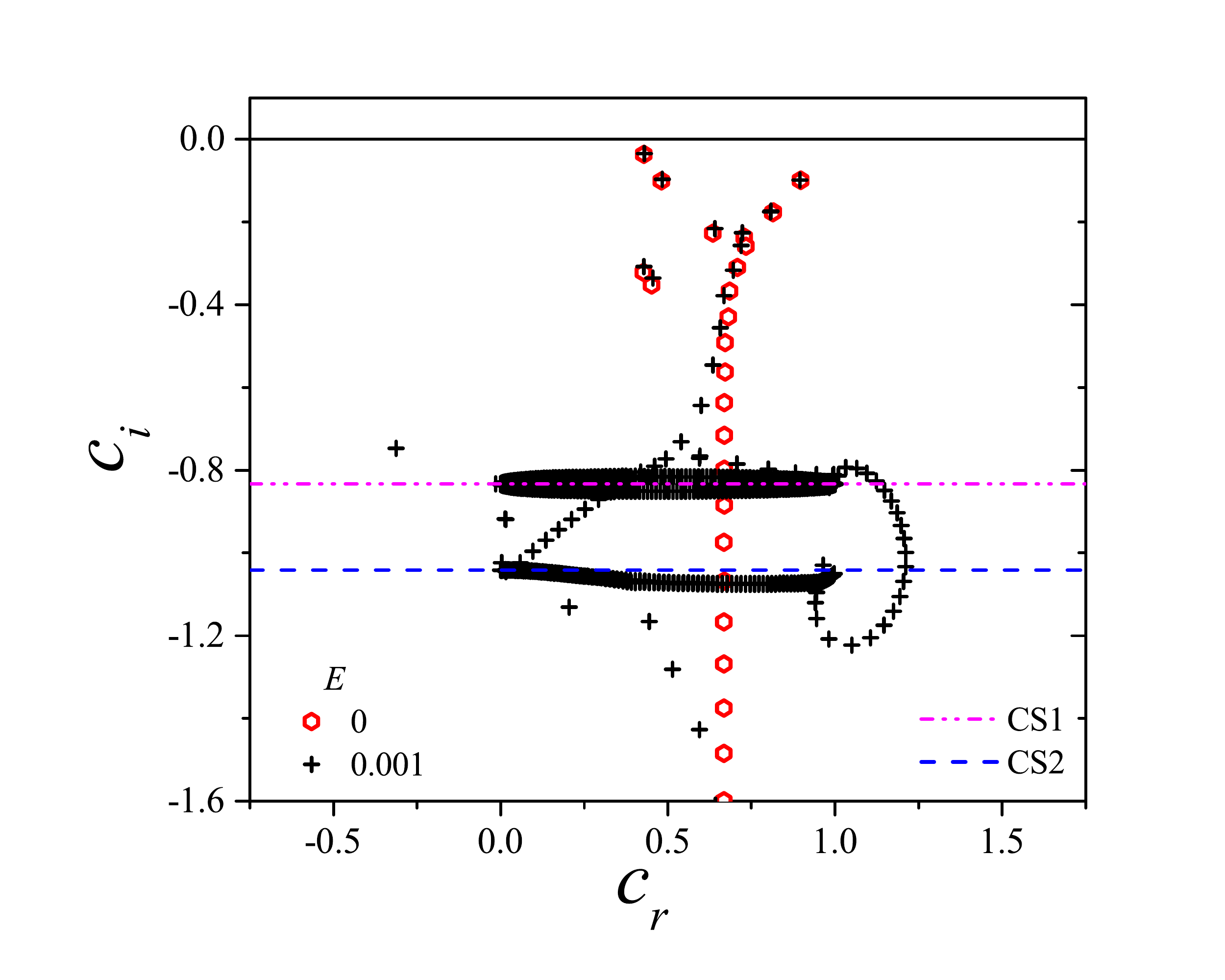}
		\caption{$  E = 10^{-3} $}
		\label{fig:E_0p001}
	\end{subfigure}%
	~
	\begin{subfigure}[b]{0.5\textwidth}
		\centering
		\includegraphics[width=\textwidth]{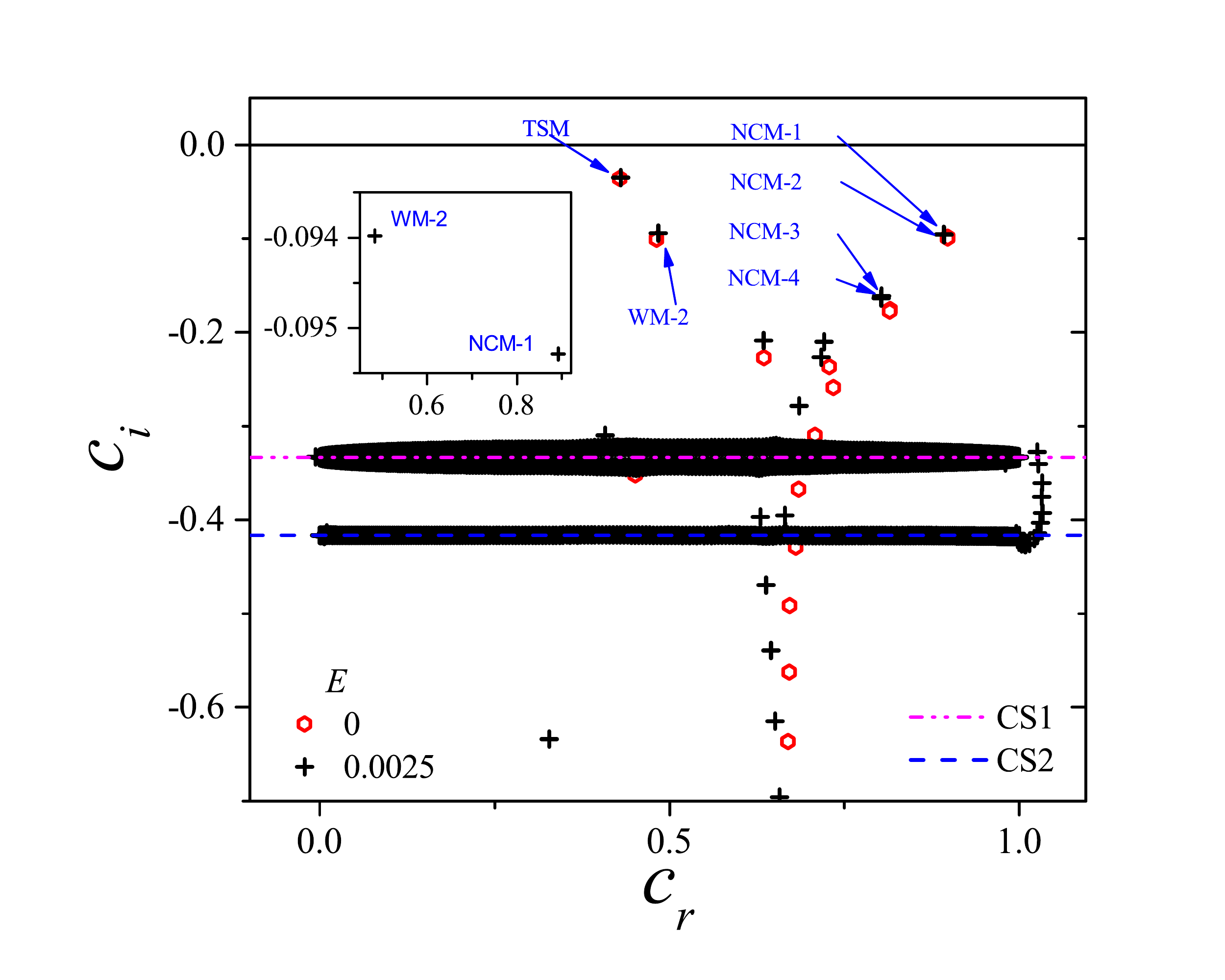}
		\caption{$ E = 2.5\times 10^{-3} $}
		\label{fig:E_0p0025}
	\end{subfigure} 
	~
	\begin{subfigure}[b]{0.5\textwidth}
		\centering
		\includegraphics[width=\textwidth]{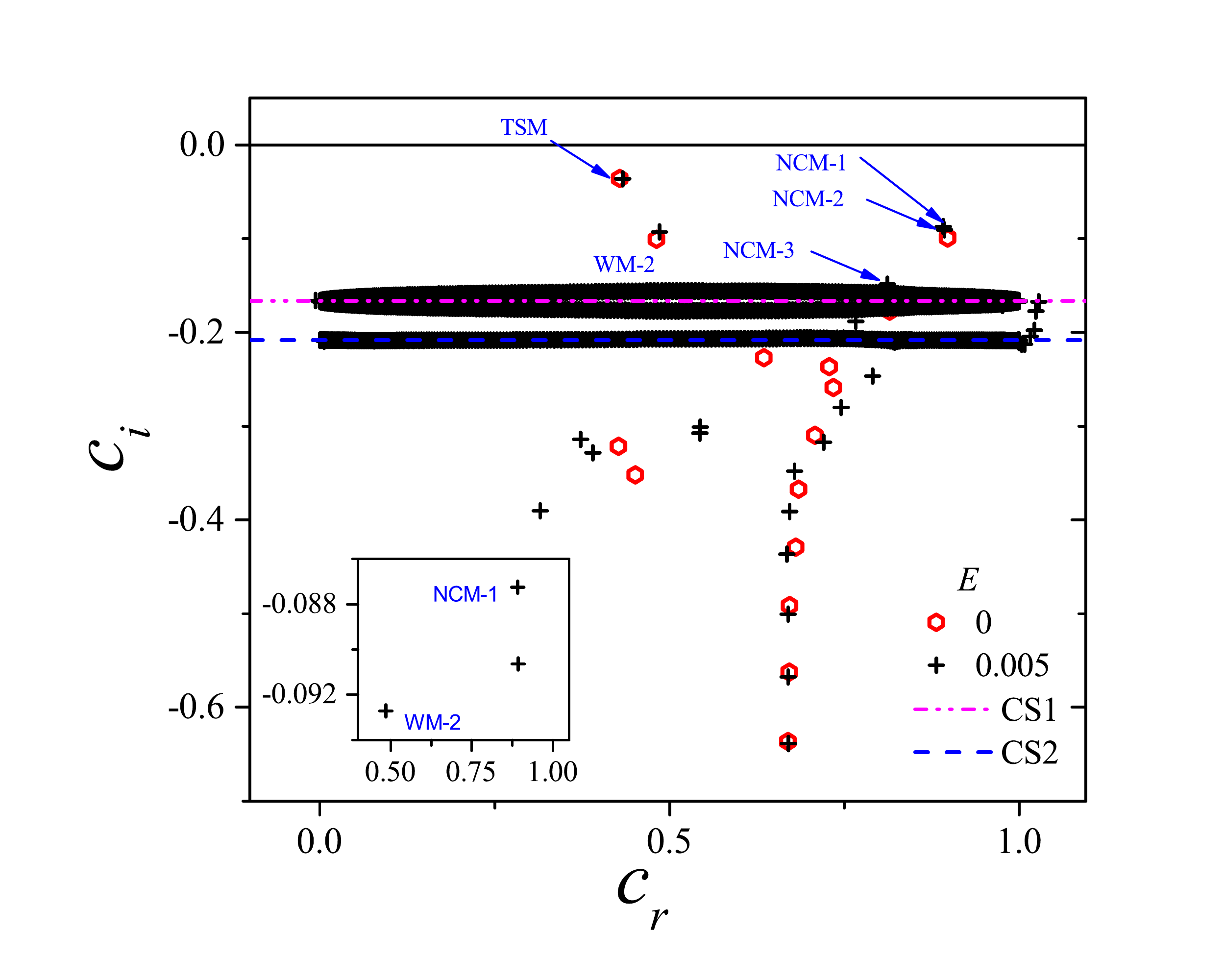}
		\caption{$ E = 5\times 10^{-3} $}
		\label{fig:E_0p005}
	\end{subfigure}%
	~
	\begin{subfigure}[b]{0.5\textwidth}
		\centering
		\includegraphics[width=\textwidth]{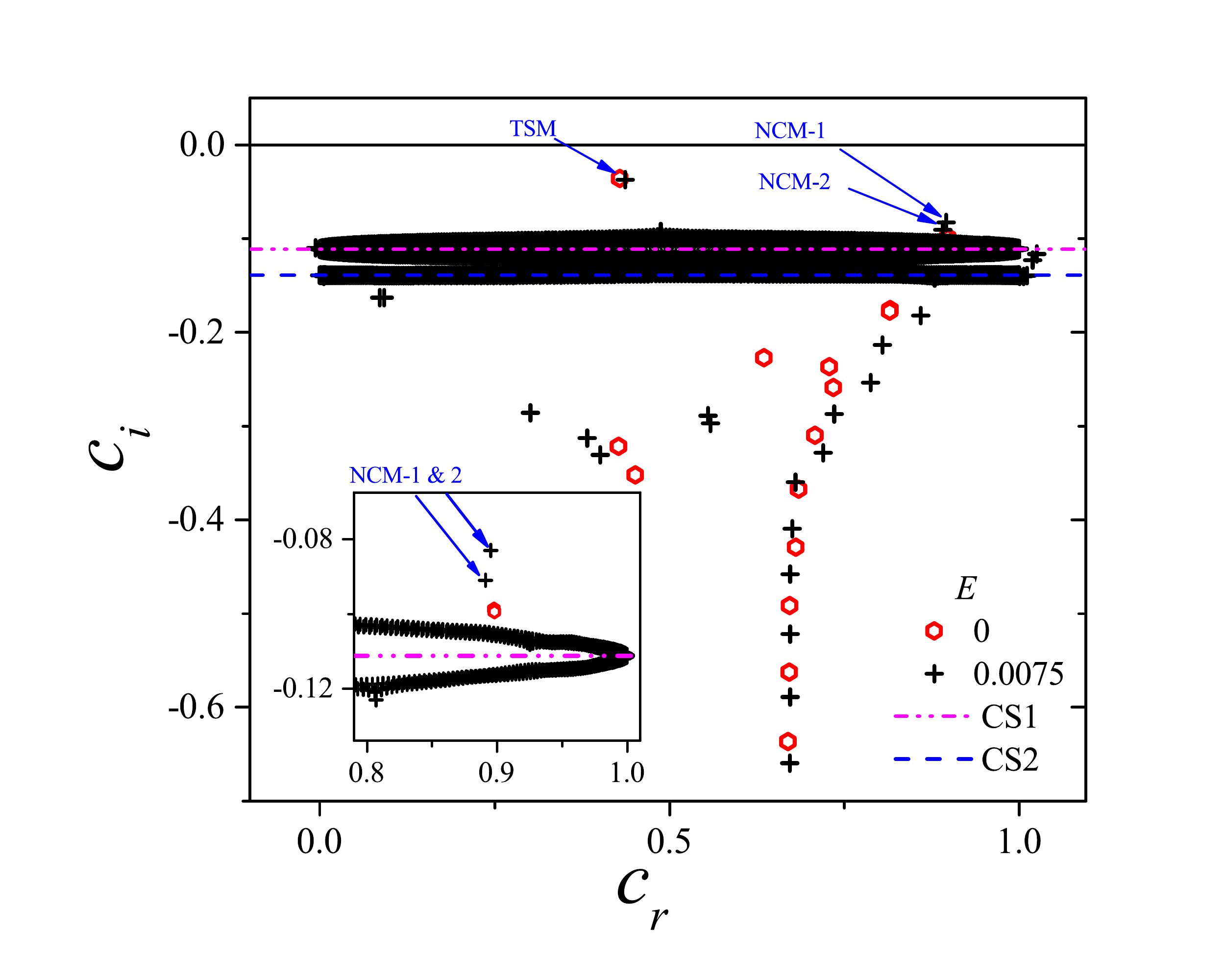}
		\caption{$ E = 7.5\times 10^{-3} $}
		\label{fig:E_0p0075}
	\end{subfigure}%
	
	\caption{\small Elasto-inertial spectra for plane channel flow as $E$ is increased from zero; $Re=800$, $ k=1.5, \beta=0.8 $. In panels (a) and (b), the HFGL line bends back as an elliptical ring so as to merge with the S branch below the CS; in panel (c) further increment in $ E $ leads to collapsing of HFGL line with discrete modes wrapping near $ c_r \approx 1 $ of the CS; in panels (d), (e), and (f), as $E$ is increased, both the CS move up and the elastically-modified NCM's disappear into the CS. The insets (A) and (B) in (a) show the zoomed-in region near CS1 and the modified Y-shaped structure respectively. The insets in (d) and (e) show the zoomed-in regions showing the second least-stable wall mode (WM-2) and NCM-1. The inset in (f) shows the two least stable NCMs. Here, CS1 and CS2 denote the two continuous spectra.}
	\label{fig:Re800_k1.5_B0.8}
\end{figure*}

We first discuss the key differences between the Oldroyd-B eigenspectrum and the Newtonian one. Note that the Oldroyd-B eigenspectrum reduces to the Newtonian one  when either $E  = 0$ (for any $\beta$) or $\beta = 1$ (for any $E$). As mentioned in Sec.~\ref{sec:Introduction}, the Newtonian eigenspectrum for plane Poiseuille flow (see Fig.~\ref{fig:Re800_k1p5_E0p1_B1}), at sufficiently high $Re$, has a characteristic `Y-shaped' structure. For $Re > 5772$, a wall mode belonging to the A branch becomes unstable \citep {Schimid-Henningson}, this being the `Tollmien-Schlichting' (TS) instability. The eigenspectrum at $Re = 800$, $E = 0.1$,  $\beta = 0.8$ and $k = 1.5$ (Fig.~\ref{fig:Re800_E0p1_k1p5_B0p8}) shows that in addition to the elastic modification of the discrete modes of the Newtonian spectrum, 
the spectrum for the Oldroyd-B fluid  has a pair of continuous spectrum `balloons' \citep{Graham_1998,wilson1999,chaudhary_etal_2019}.
The vertical location of the two continuous spectra is obtained by setting the coefficient of the highest order derivative (\textit{viz.,} $ 1+i k W [U-c] $ and $ 1+i \beta k W [U-c] $) in the governing differential equation to zero.
The continuous spectrum with $c_i = -1/(kW)$ is present even in the absence of solvent (i.e. the UCM limit), and henceforth will be referred to as `CS1'.
The second continuous spectrum (abbreviated as CS2),  characterized by modes with $c_i = -1/(\beta k W)$, is present only for non zero $\beta$. Theoretically,  both the CS are `lines' in the $c_r$-$c_i$ plane with the aforementioned $c_i$, and with $c_r$ taking any value in the base state range  of velocities $[-1,1]$. Since the eigenfunctions corresponding to the eigenvalues belonging to CS's are singular, these are resolved 
only approximately by the finite number of collocation points used in the spectral method. Thus, both the CS's appear as balloons whose spread only decreases slowly with increasing $N$.
In addition to the elastically modified Newtonian discrete modes and the CS balloons, 
new discrete modes (absent in the Newtonian spectrum) also appear, of which one of the  center modes 
is unstable at $E = 0.1$ (see inset of Fig.~\ref{fig:Re800_E0p1_k1p5_B0p8}); all other discrete modes, including the continuation of the TS (wall) mode, remain stable for $Re = 800$. An analogous center-mode instability for 
viscoelastic pipe flow  (over a similar range of parameters) was
first  reported by \cite{Piyush_2018}, and has since been examined in more detail by \cite{chaudharyetal_2020}. 
The presence of analogous center-mode instabilities  for both channel and pipe flows of an Oldroyd-B fluid is in direct contrast to the Newtonian scenario, where pipe flow is stable at any $Re$. 


\subsection{Evolution of the unstable elasto-inertial center mode }
\label{ssec:origination_ECM}

\begin{figure*}
	\centering
	\begin{subfigure}[b]{0.5\textwidth}
		\centering
		\includegraphics[width=\textwidth]{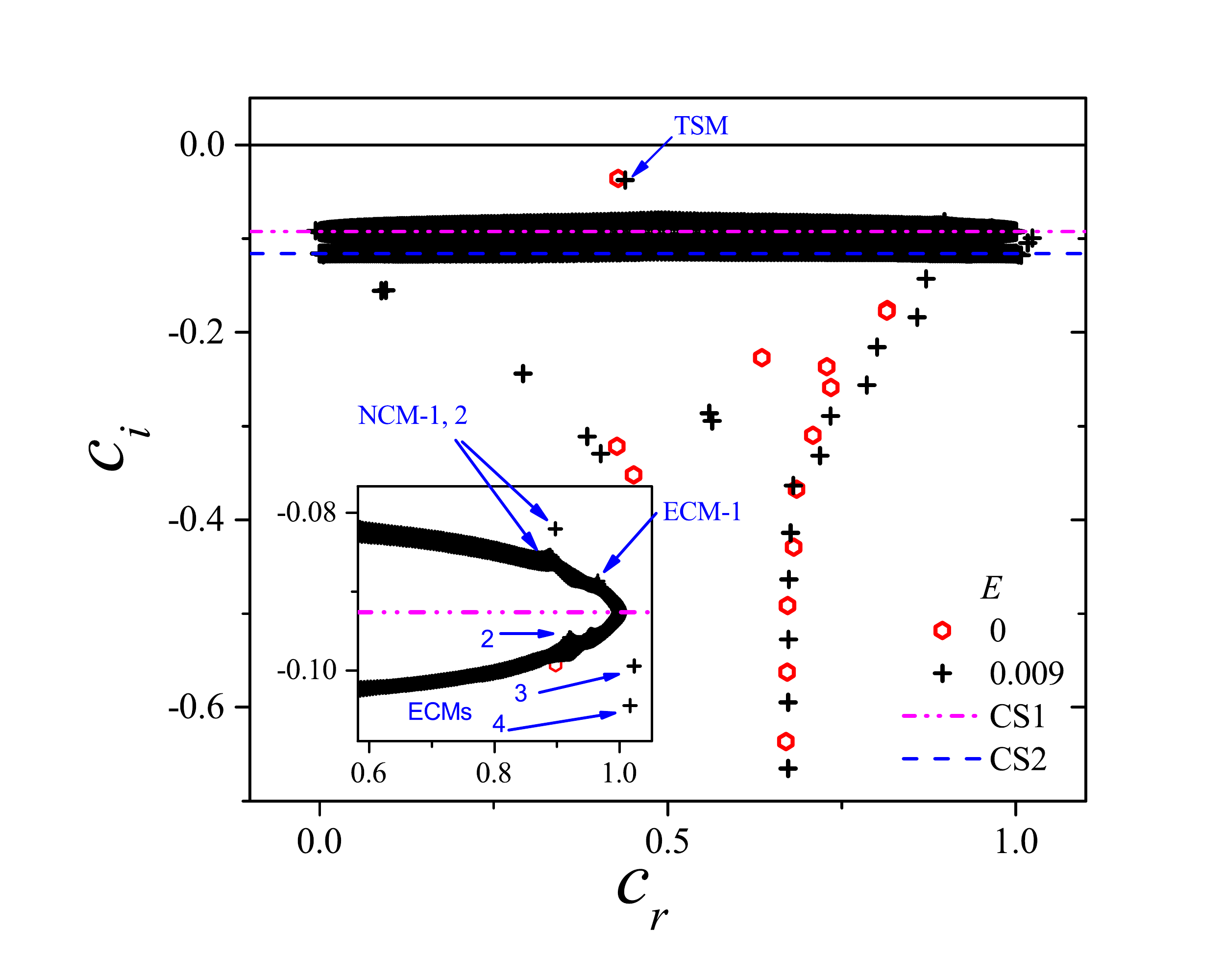}
		\caption{$E = 0.009 $}
		\label{fig:E_0p009}
	\end{subfigure}%
	~ 
	\begin{subfigure}[b]{0.5\textwidth}
		\centering
		\includegraphics[width=\textwidth]{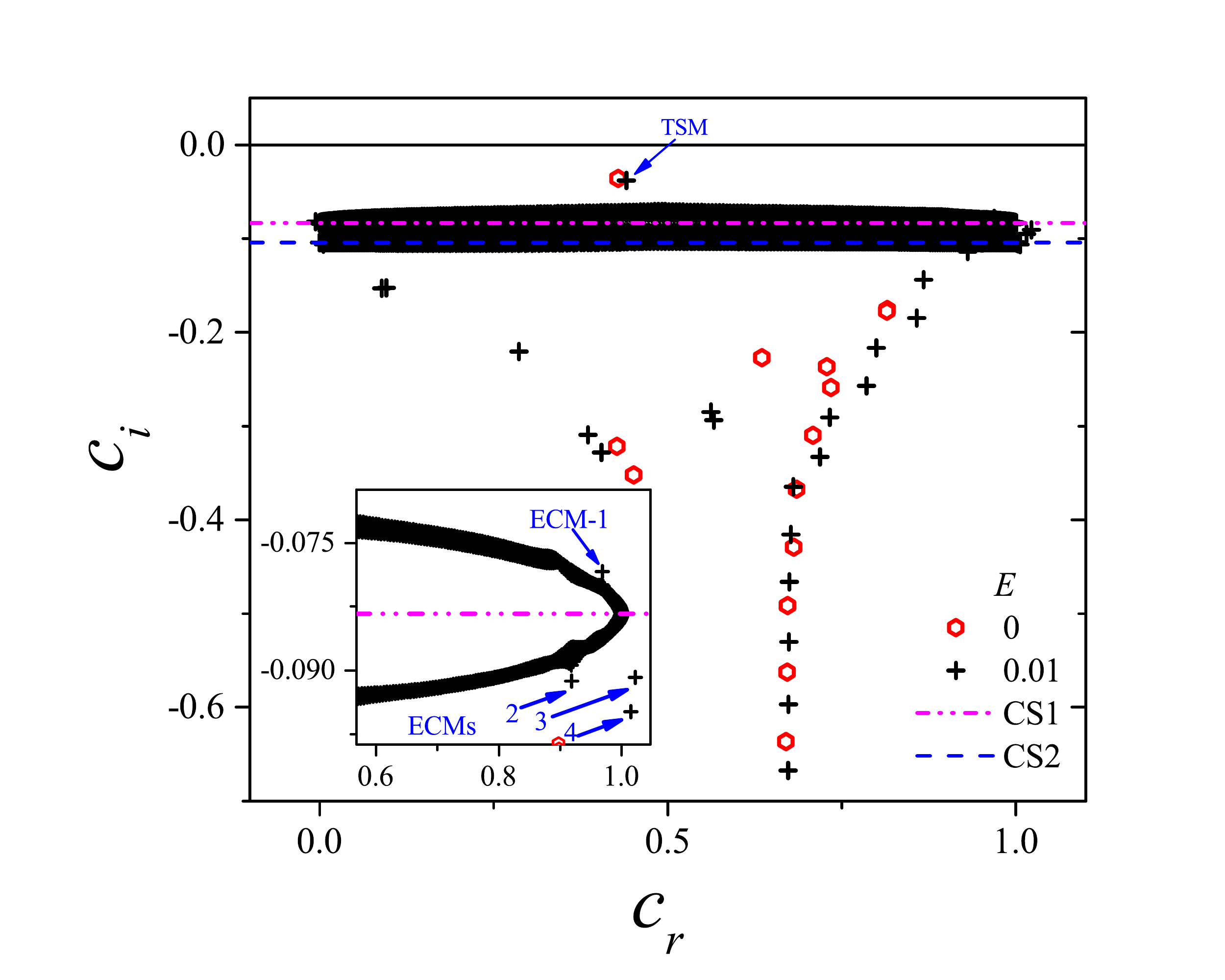}
		\caption{$E = 0.01 $}
		\label{fig:E_0p01}
	\end{subfigure}
	~
	\begin{subfigure}[b]{0.5\textwidth}
		\centering
		\includegraphics[width=\textwidth]{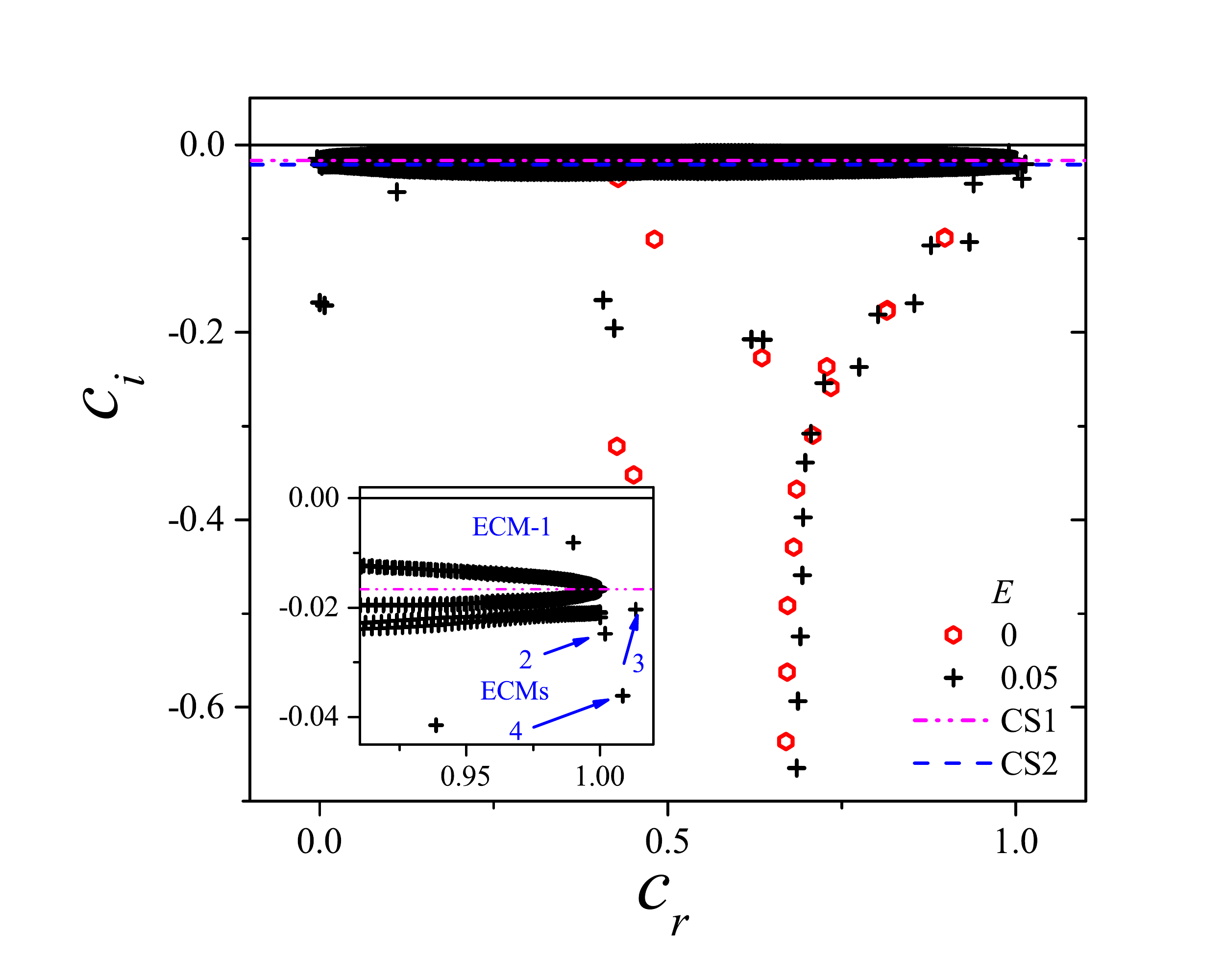}
		\caption{$ E = 0.05 $}
		\label{fig:E_0p05}
	\end{subfigure}%
	~
	\begin{subfigure}[b]{0.5\textwidth}
		\centering
		\includegraphics[width=\textwidth]{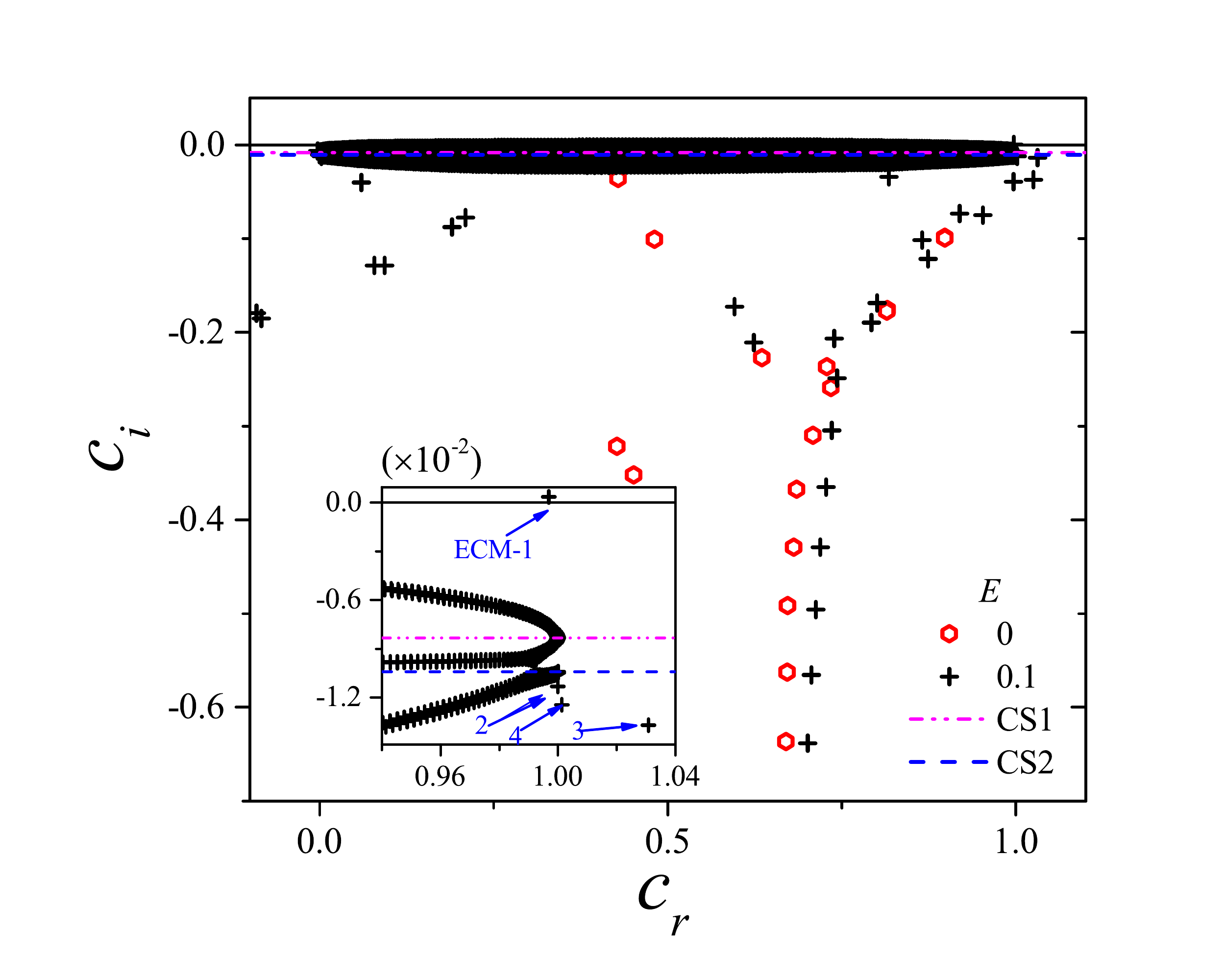}
		\caption{$ E = 0.1 $}
		\label{fig:E_0p1}
	\end{subfigure} 
	
	\caption{\small Elasto-inertial spectra at $Re = 800$, $k = 1.5$, and $\beta = 0.8$ as $E$ is varied in the range $0.009$-$0.1$.  For higher $  E $, both NCM-1 and 2 merge with CS1 (panels (a) and (b)), and a new elasto-inertial center mode (ECM-1) emerges above CS1 (panels (b) and (c)); (d) ECM-1 becomes unstable at $ E=0.1 $. }
	\label{fig:Re800_k1.5_B0.8_high-E}
\end{figure*}

In this section, we discuss the emergence and trajectory of the  elasto-inertial center mode that eventually becomes unstable (henceforth labelled as ECM-1), and other discrete stable modes, by examining two different paths in the parameter space, both starting from the Newtonian limit: (i)  increasing $E$ (from zero) at  fixed $\beta$, and (ii) decreasing $\beta$ (from unity) at fixed $E$. 

\subsubsection{Effect of varying $E$ at fixed $\beta$}
\label{sssec:Effect of elasticity}

Figures~\ref{fig:Re800_k1.5_B0.8} and \ref{fig:Re800_k1.5_B0.8_high-E} show the unfiltered eigenspectra for $Re = 800$, $k = 1.5$, and $\beta = 0.8$ for $E$ ranging from $ 10^{-4}$ to $10^{-1}$.  The Newtonian eigenspectrum ($E = 0$) is shown in each figure as a reference.  The original Y-shape of the Newtonian spectrum is modified only slightly for very low values of $E$ (inset (B) in Fig.~\ref{fig:E_1e-4}),  although there is the appearance of an additional inverted Y-shape just above CS1. In addition to this modified Newtonian locus, the two CS balloons are encircled by a set of discrete modes which form an approximate ring-like structure (Figs.~\ref{fig:E_1e-4} and \ref{fig:E_5e-4}). We have verified (illustrated further below in Fig.~\ref{fig:HFGL_Bending}) that these modes are the continuation, to finite-$\beta$, of a class of damped shear waves in the UCM limit, termed the `high-frequency Gorodtsov-Leonov' (HFGL)  modes (after Gorodtsov and Leonov, 1967). 
The locus of these modes corresponds to $c_i = -1/(2kW)$  for $\beta = 0$ \citep{Kumar_2005,chaudhary_etal_2019}, but this line bends downwards upon increase
in $\beta$, leading to the ring-like structure seen in Fig. {\ref{fig:E_1e-4}.
%
For $ E>0.001 $, the bent locus collapses onto the two CS's, except for a small portion near the $c_r \approx 1$ (Fig.~\ref{fig:E_0p001}). 
Further, the discrete center modes belonging to the Newtonian P-branch are also modified with an increase in  $E$. Figures~\ref{fig:E_0p0025}-\ref{fig:E_0p0075} show that the elastically modified Newtonian center modes (referred to as `NCM's, with an index that labels them in order of increasing $|c_i|$) only change a little with increasing $E$, but both CS1 and CS2 move up and in this process, all the NCM's disappear into CS1 beyond a threshold $E$  ($\sim 7.5 \times 10^{-3}$) for $Re = 800$ in Fig.~\ref{fig:E_0p0075}. 
It is well known that the continuous spectrum (CS1) is a branch cut 
for any $Re$, allowing modes to collapse into it (crossing onto a different Riemann sheet in the process), and likewise, new modes to appear from it, with increasing $E$ \citep{wilson1999}.
 This behavior mimics that found earlier in viscoelastic pipe flow \citep{chaudharyetal_2020}.

Figure~\ref{fig:Re800_k1.5_B0.8_high-E} shows the spectra for a higher range of $E$, wherein all of the NCM's have collapsed into  CS1. 
For $E = 0.009$ and $0.01$ (Figs.~\ref{fig:E_0p009} and \ref{fig:E_0p01}), the lone discrete mode
that remains above the CS is the elastically modified TS mode.
This feature differs from that of the elasto-inertial spectrum for pipe flow, wherein there is no analogue of the TS mode, and the center modes remain the least stable, even for smallest $E$'s. However, even in the channel case, the elastically modified TS mode merges with  CS1 for higher $E$ (the absence of the TS mode is illustrated, for example, in Fig.~\ref{fig:E_0p05} for $E = 0.05$). Importantly, 
for $E \sim 0.01$, a new elasto-inertial center mode (labelled ECM-1) with phase speed close to the maximum base-state velocity, having no Newtonian counterpart, emerges above  CS1 (Fig.~\ref{fig:E_0p01}). 
This center mode (ECM-1) becomes unstable as $E$ is increased beyond $0.1$ (Fig.~\ref{fig:E_0p1}). 
New elasto-inertial center modes  (labelled  ECM-2, -3, and -4) also appear below CS1, but they remain stable as $E$ is increased.

 \begin{figure*}
	\centering
	\begin{subfigure}[b]{0.5\textwidth}
		\centering
		\includegraphics[width=\textwidth]{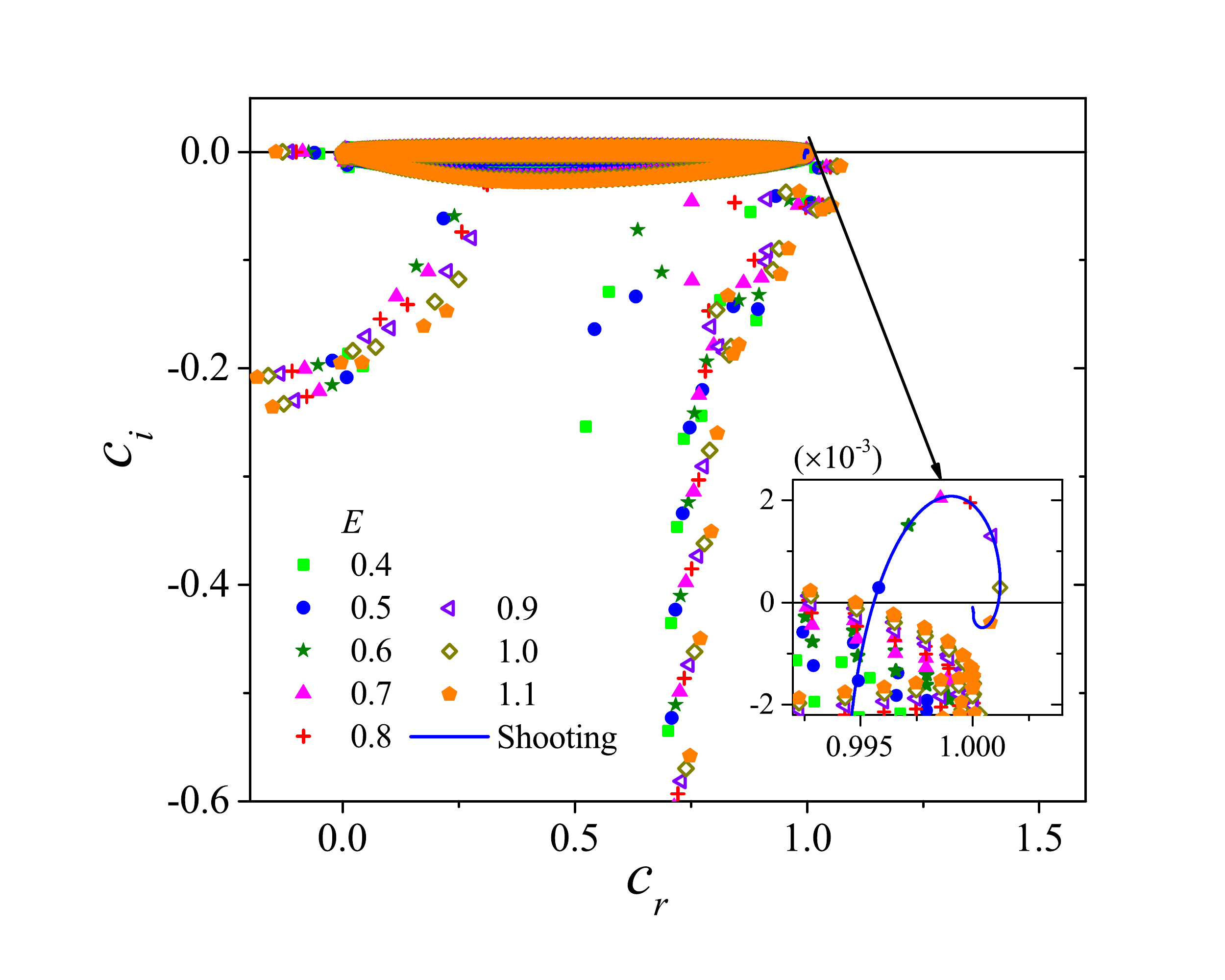}
		\caption{}
		\label{fig:E_0p4-1p1}
	\end{subfigure}%
	~ 
	\begin{subfigure}[b]{0.5\textwidth}
		\centering
		\includegraphics[width=\textwidth]{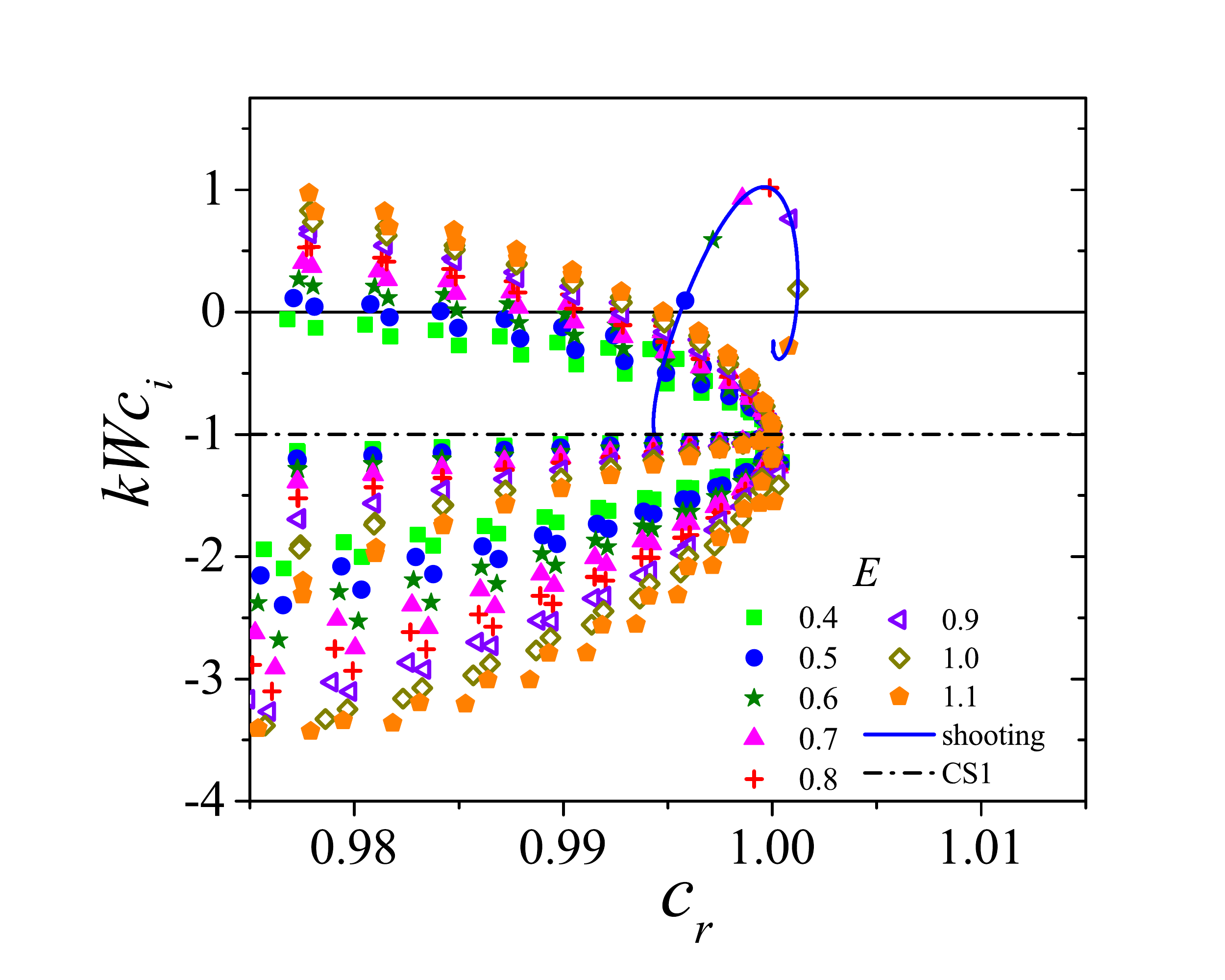}
		\caption{}
		\label{fig:origin_kwci}
	\end{subfigure}
		\caption{\small Eigenspectra for $ Re=650, k=1, \beta =0.96$ at different $ E $. (a) The full spectrum; (b) Enlarged view of panel (a) near the unstable eigenvalue expressed using the scaled growth rate $ kWc_i $. The continuous (blue) line showing the trajectory of ECM-1 is obtained using shooting method, while symbols show results from the spectral method.}
	\label{fig:Evolution_ECM}
\end{figure*}


\begin{figure*}
	\centering
\begin{subfigure}[b]{0.5\textwidth}
	\centering
	\includegraphics[width=\textwidth]{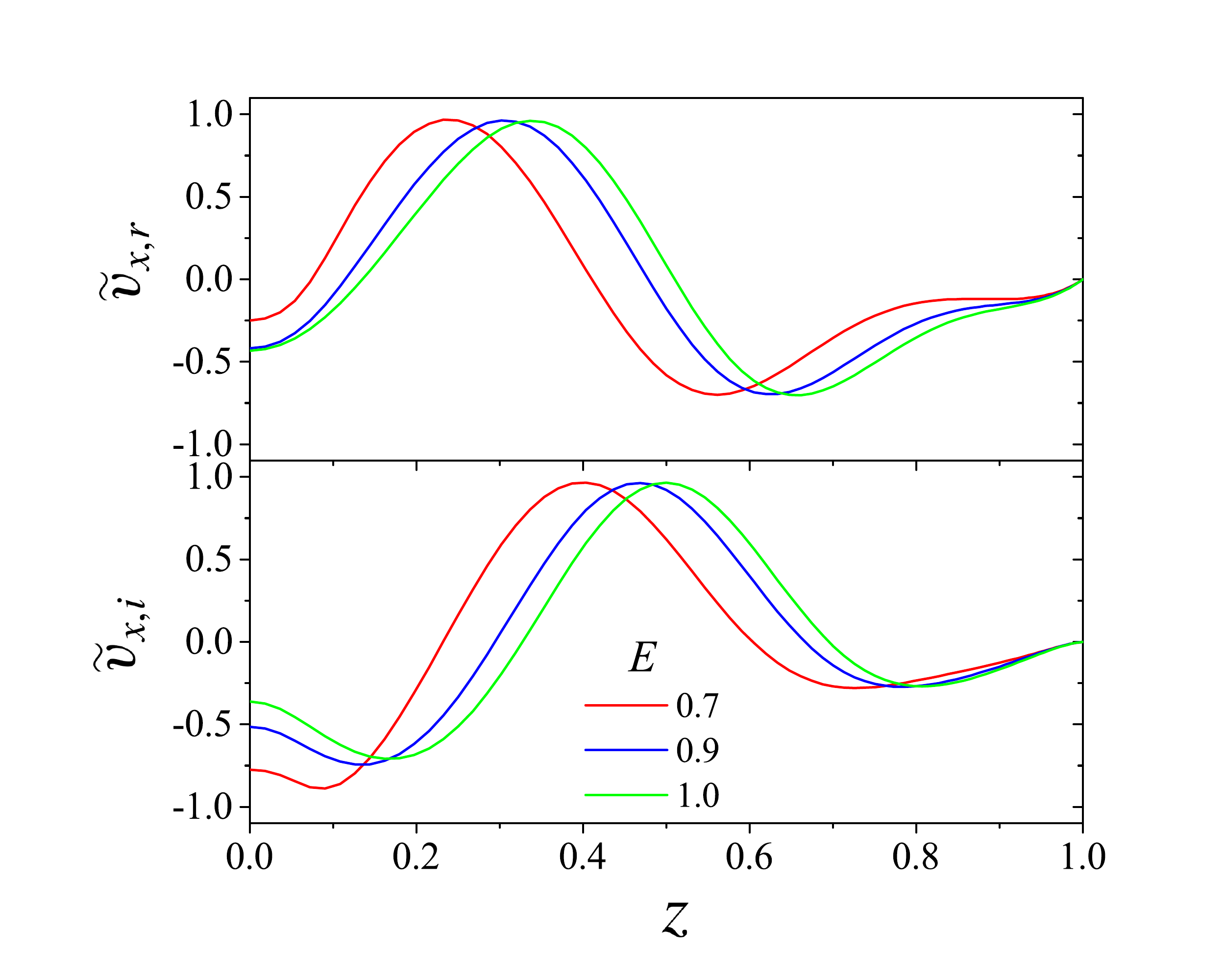}
	\caption{}
	\label{fig:vx}
\end{subfigure}%
~ 
\begin{subfigure}[b]{0.5\textwidth}
	\centering
	\includegraphics[width=\textwidth]{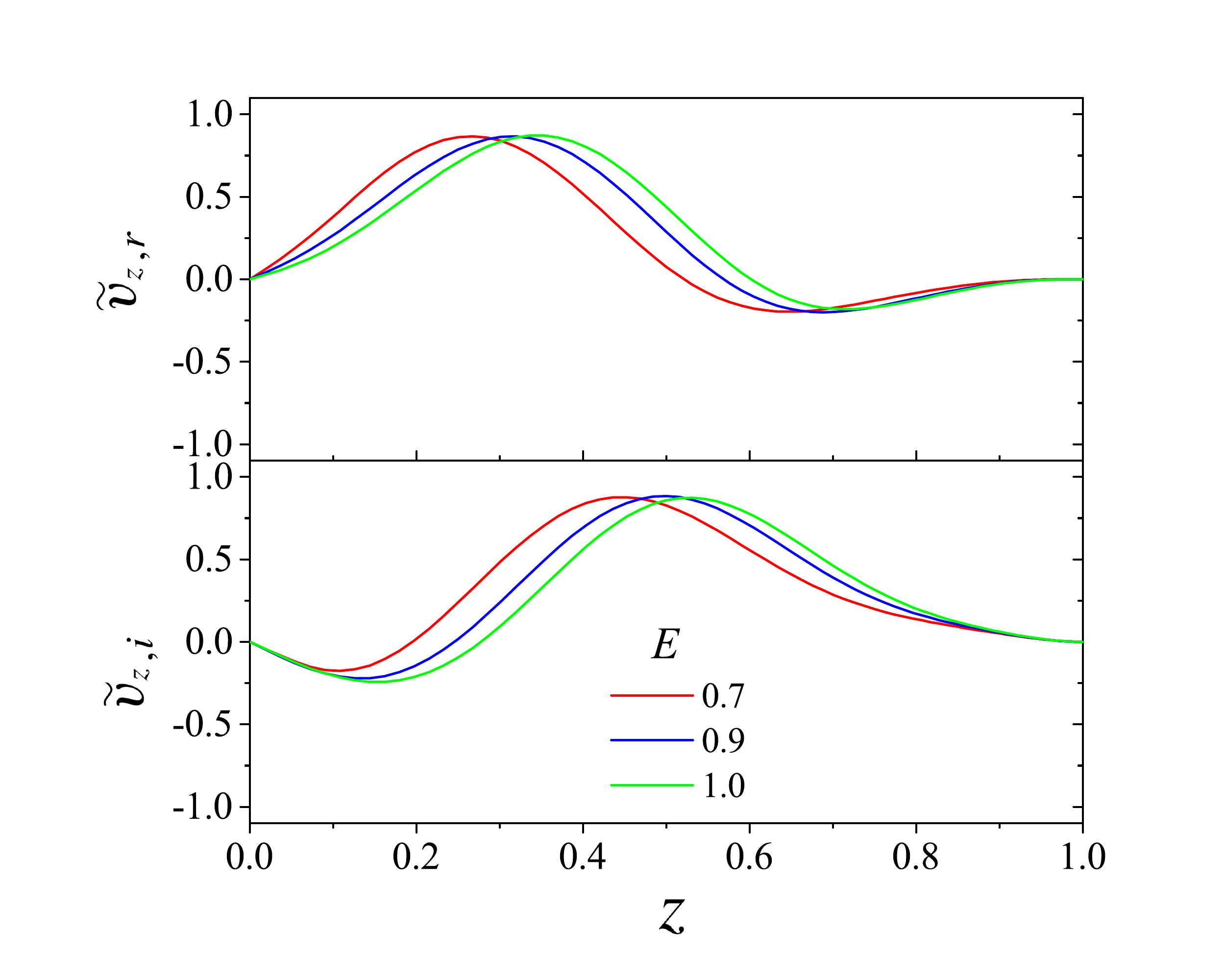}
	\caption{}
	\label{fig:vz}
\end{subfigure}
\caption{\small Velocity eigenfunctions corresponding to unstable eigenvalues in  Fig.~\ref{fig:Evolution_ECM} for $Re = 650$, $k = 1$, $\beta = 0.96$ and at different $E$. (a) Axial velocity, $ \tilde{v}_x $; (b) wall-normal velocity,  $ \tilde{v}_z $. The $ \tilde{v}_x $ eigenfunctions are symmetric about the channel center, and are shown over the half-domain $0 \leq z \leq 1$. The eigenvalues for which the eigenfunctions are shown here are 
$E = 0.7$, 
$c = 0.99856712	+	0.00204187i$; $E =0.9$, $c =  1.00087623	+   0.00130115i$; 
$E =1.0$,  $c = 1.00121782  +	2.88573410 \times 10^{-4}i$.}
\label{fig:Eigenfunction_vx_vz}
\end{figure*}

Figures~\ref{fig:E_0p4-1p1} and \ref{fig:origin_kwci} present the eigenspectra for different $E$ varying over the interval (0.4, 1.1) at a much higher value of $\beta = 0.96$, with Fig.~\ref{fig:origin_kwci} being plotted in terms of the scaled growth rate $k W c_i$, which ensures that the location of the two CS are fixed as  $E$ is changed (for a given $\beta$). Figure~\ref{fig:E_0p4-1p1} tracks the paths taken (with increasing $E$) by all discrete modes shown, while the continuous line in Fig.~\ref{fig:origin_kwci} represents the trajectory of the unstable elasto-inertial  center mode (ECM-1) alone obtained from the shooting method (the superposed symbols correspond to results obtained using the spectral method).
 The new elasto-inertial center mode, which emerges from above the CS1 at $ E \approx 0.4 $,   becomes unstable  for $0.48 < E < 1.04$, but becomes stable again for $ E>1.04 $, with $|c_i|$ eventually scaling as $1/E$ for large $E$, quite similar to pipe flow \citep{chaudharyetal_2020}. However, unlike pipe flow, $c_r$'s for the unstable mode exceed unity over some ranges of $E$.
 
\begin{figure*}
	\centering
	\begin{subfigure}[b]{0.5\textwidth}
		\centering
		\includegraphics[width=\textwidth]{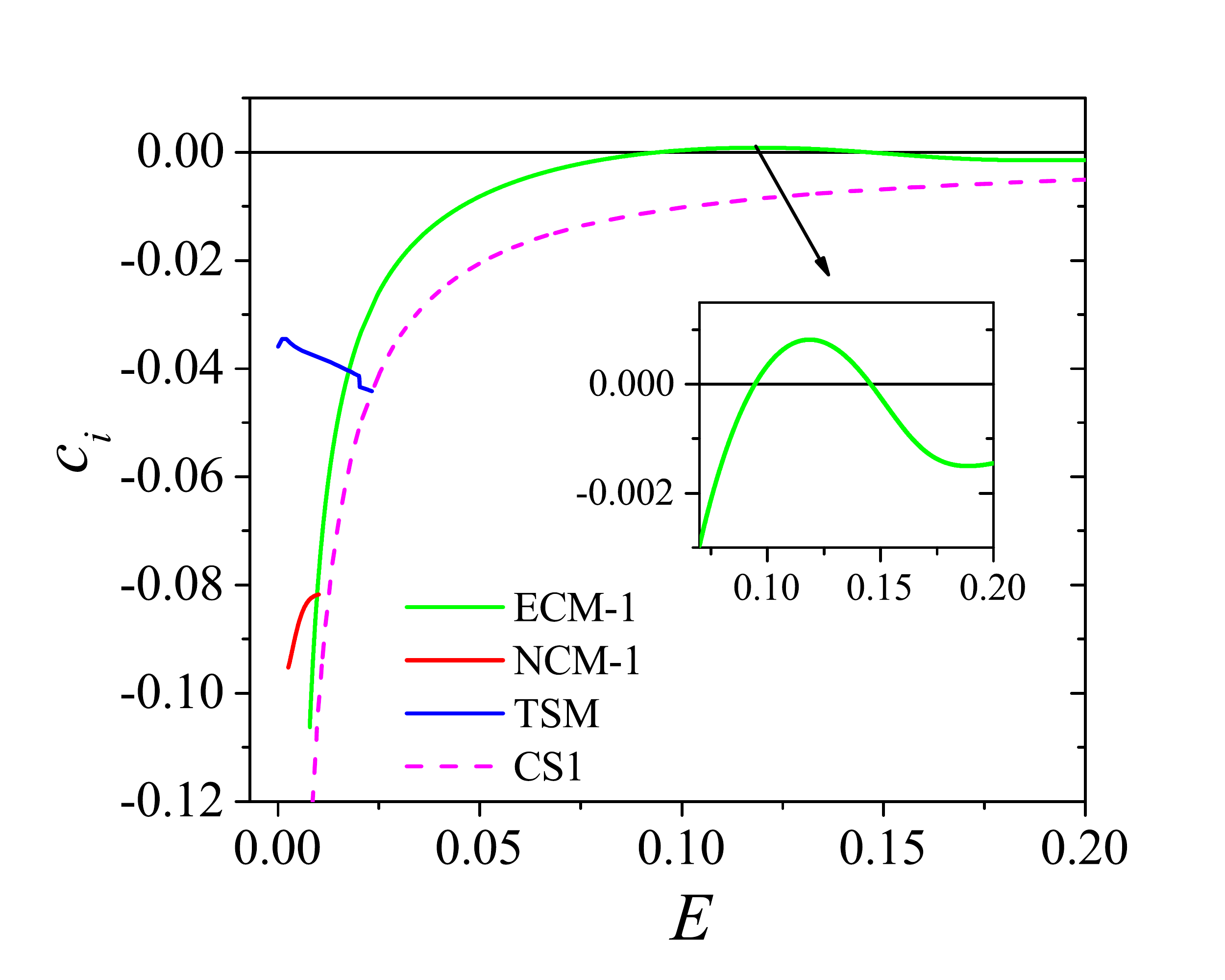}
		\caption{}
		\label{fig:Ci_E_for-all-modes_Beta0p8}
	\end{subfigure}%
	~ 
	\begin{subfigure}[b]{0.5\textwidth}
		\centering
		\includegraphics[width=\textwidth]{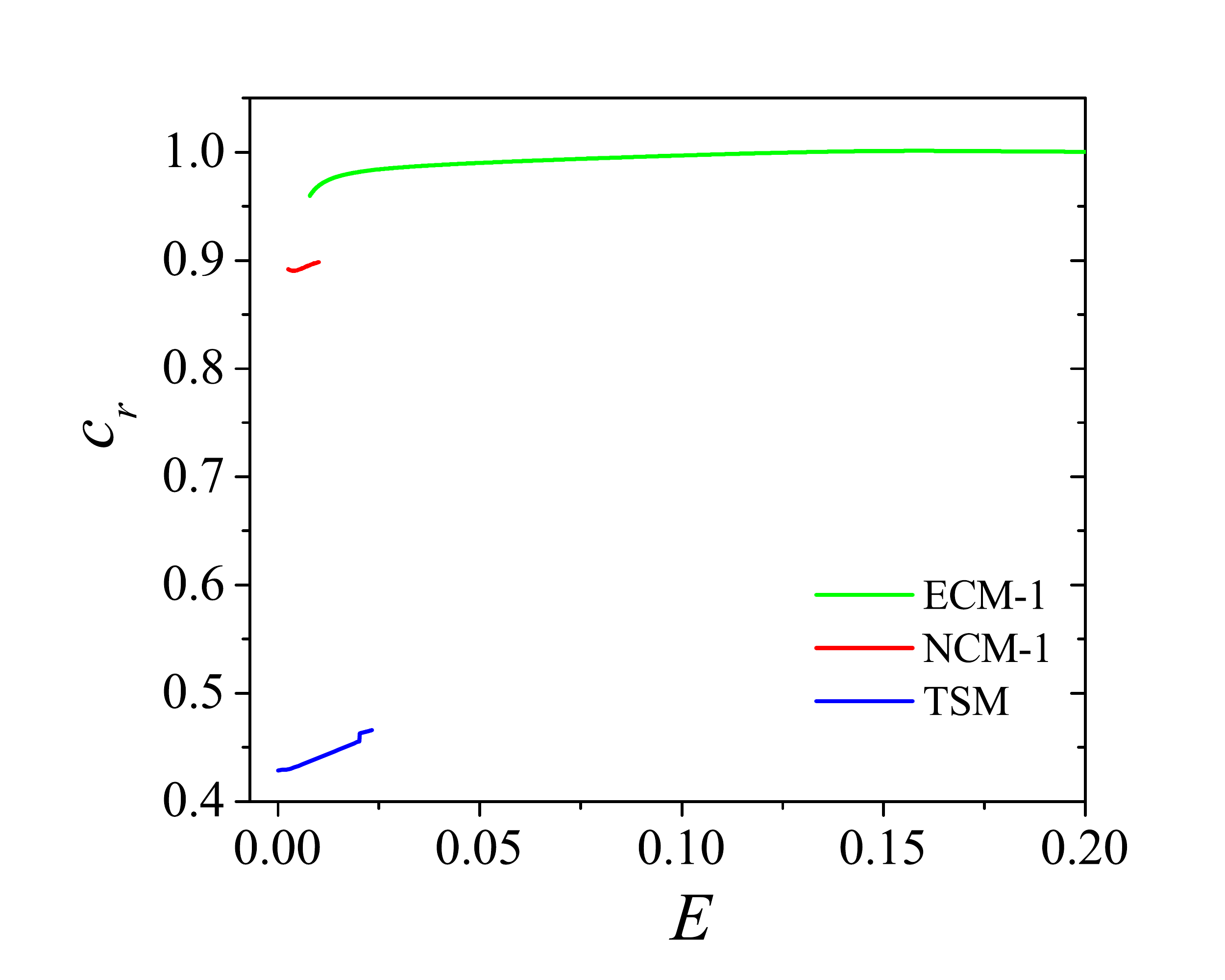}
		\caption{}
		\label{fig:Cr_E_for-all-modes_Beta0p8}
	\end{subfigure}
	\caption{\small Relative stability of the first three least stable eigenmodes \textit{viz}., Tollmien--Schlichting mode (TSM), elastically modified Newtonian center mode (NCM-1) and the new elasto-inertial center mode (ECM-1) at $ Re = 800, k = 1.5$, and $\beta =0.8$. (a) Variation of  $ c_i $ with $ E $ (inset shows the range of  $ E $ for which ECM-1 is unstable); (b) Phase speed $ (c_r) $ corresponding to the modes shown in panel (a). In the Newtonian limit $ (E\rightarrow 0) $, TSM is the least stable mode that governs the stability of the flow, while ECM-1 emerges from CS1 at $E \sim 0.01$. However, as $ E $ increases, both TSM and NCM-1 disappear into CS1 leaving behind ECM-1 as the least stable mode for $ E > 0.02$, which eventually becomes unstable at  $ E  \approx 0.1 $.}
	\label{fig:C_E_for-all-modes_Beta0p8}. 
\end{figure*}
 
%
Figure~\ref{fig:Eigenfunction_vx_vz} shows the velocity eigenfunctions ($ \tilde{v}_x , \tilde{v}_z$) for different $E$, corresponding to some of the unstable center modes shown in Fig.~\ref{fig:Evolution_ECM}. The $\tilde{v}_x$ eigenfunctions are symmetric about the channel center line (and are therefore shown only over one half of the channel), in marked contrast with the TS (wall) and NCM-1 modes, which are anti-symmetric about the channel centerline.  The eigenfunctions have their peak amplitudes closer to the channel centerline, but are nevertheless spread across the entire channel for the moderate $ Re $ considered here, similar to the center-mode instability in pipe flow \citep{chaudharyetal_2020}.
This latter fact, that the unstable eigenfunctions for moderate $Re$ and $E$ are not localized near the channel centerline despite the phase speed being close to the maximum velocity of the base flow, needs to be emphasized since this contradicts earlier interpretations of our original report on the center-mode instability \citep{Shekar2019}.

In the limit $ E\rightarrow 0 $, as demonstrated by the spectra
in Figs.~\ref{fig:Re800_k1.5_B0.8} and \ref{fig:Re800_k1.5_B0.8_high-E},
the first few least stable modes in the viscoelastic channel spectrum are the elastically modified Tollmien--Schlichting (TS) wall mode and Newtonian center mode (NCM-1) with former being the least stable one (the second wall mode becomes more stable than NCM-1 (Fig.~\ref{fig:E_0p0025}) as $E$ is increased, and is not considered in this discussion).
However, 
 this picture of relative stability does not hold as  $ E $  is increased. Figure~\ref{fig:Ci_E_for-all-modes_Beta0p8} shows the variation of $ c_i $ for the TSM, NCM-1 and ECM-1 modes with $ E $. In the near-Newtonian limit ($ E \rightarrow 0 $), TSM is the least stable mode followed by  NCM-1, while ECM-1 just emerges from the CS1 for $E \approx 0.01$.  For  $E \sim 0.01$, the decay rates of TSM and ECM-1 cross each other, and for all higher values of $E$, ECM-1 is the least stable/unstable mode. For
  $E > 0.02$, both TSM and NCM-1 collapse  into CS1 (Figs.~\ref{fig:Re800_k1.5_B0.8_high-E} and \ref{fig:Ci_E_for-all-modes_Beta0p8}) (we discuss this feature in more detail in Sec.~\ref{ssec:Relative stability of center and wall modes} where we compare the relative stability of these two modes for different values of $ Re, k  $ and $ \beta $). 
The mode ECM-1 is the least stable discrete mode for $E > 0.01$, and, in fact, is the only discrete mode that lies above the CS for $E > 0.02$; for $ E > 0.1$, ECM-1 becomes unstable (inset of Fig.~\ref{fig:Ci_E_for-all-modes_Beta0p8}). The corresponding behaviour of the phase speeds of the three modes is shown in Fig.~\ref{fig:Cr_E_for-all-modes_Beta0p8}, where the phase speeds for TSM and NCM-1  increases with $ E $, before eventually merging into CS1, while the phase speed of ECM-1 remains almost constant (close to unity) over the range of $ E $ spanned.

\begin{figure*}
	\centering
	\begin{subfigure}[b]{0.5\textwidth}
		\centering
		\includegraphics[width=\textwidth]{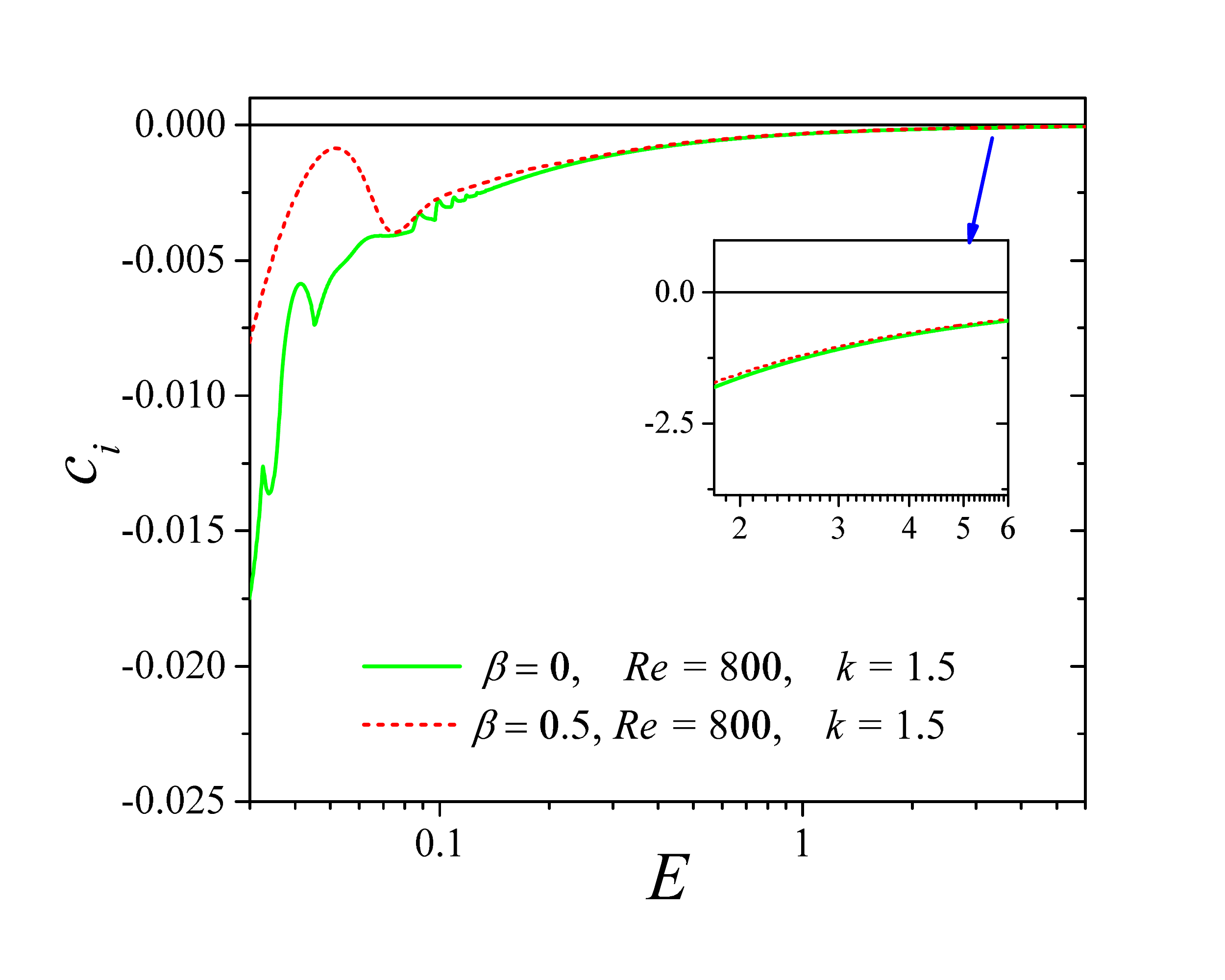}
		\caption{$ \beta =0, 0.5 $}
		\label{fig:ci_E_Beta-0-and-0p5}
	\end{subfigure}%
	~ 
	\begin{subfigure}[b]{0.5\textwidth}
		\centering
		\includegraphics[width=\textwidth]{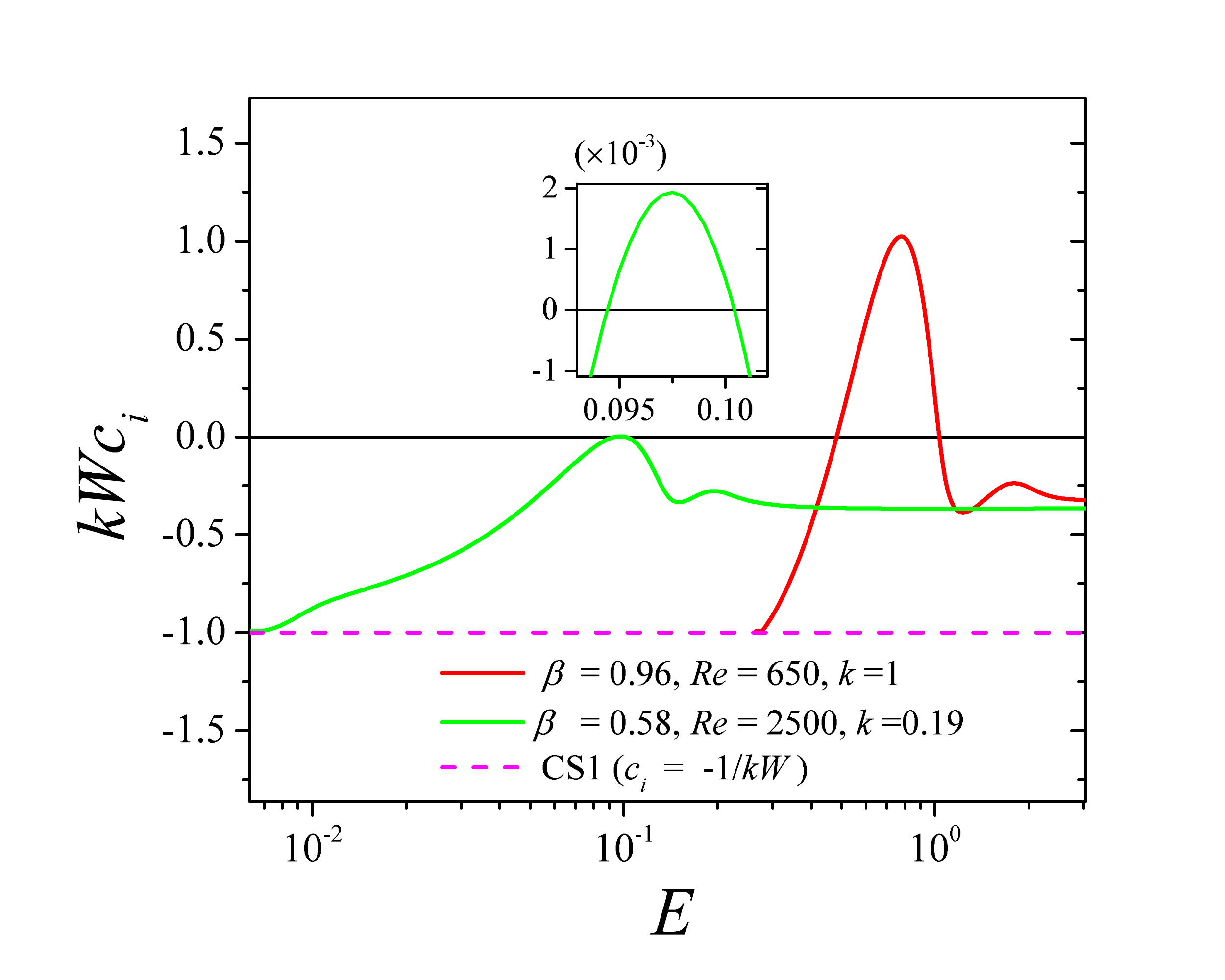}
		\caption{$ \beta =0.58, 0.96 $}
		\label{fig:ci_E_Beta0p58_and_0p96}
	\end{subfigure}
	
	\caption{Effect of increasing $ E $ on  the elasto-inertial center mode (ECM-1) for UCM and Oldroyd-B fluids. (a) $ c_i  $ for  $Re = 800, k = 1.5$ and $ \beta = 0 $ and $ 0.5 $. The center mode remains stable for $ \beta <0.5 $  even at very large values of $ E $, in stark contrast to pipe flow which remains unstable at much lower $ \beta$. (b) Scaled growth rate of ECM-1 for $ Re = 2500, k = 0.19, \beta = 0.58 $ and $ Re = 650, k = 1, \beta = 0.96$. Regardless of the value of $ \beta $, ECM-1 in channel always emerges from CS1 ($ c_i =-1/(kW) $) in the limit $ E \rightarrow 0 $.}
	\label{fig:mode_tracking_fixed-beta}
    \end{figure*}

Unlike elasto-inertial wall modes \citep{chaudhary_etal_2019}, the elasto-inertial center mode remains stable in the UCM limit ($ \beta =  0$) for channel flow, and remains so for $\beta$ below a finite threshold. Figure~\ref{fig:ci_E_Beta-0-and-0p5}  explores the effect of varying $ E $ on ECM-1 for $ \beta = 0$ and $0.5 $. In the UCM limit $ (\beta = 0) $,  as $ E $ is increased  from the Newtonian limit ($E \rightarrow 0$), $ |c_i| $ eventually decreases to very small values (Fig.~\ref{fig:ci_E_Beta-0-and-0p5}). However, 
$c_i$ remains negative even for very large $E$, and therefore, no center-mode instability is found in 
the UCM limit for channel flow. An analogous behaviour is found for $\beta = 0.5$.
%
%
%
%
%
%

\begin{figure*}
	\centering
	\begin{subfigure}[b]{0.5\textwidth}
		\centering
		\includegraphics[width=\textwidth]{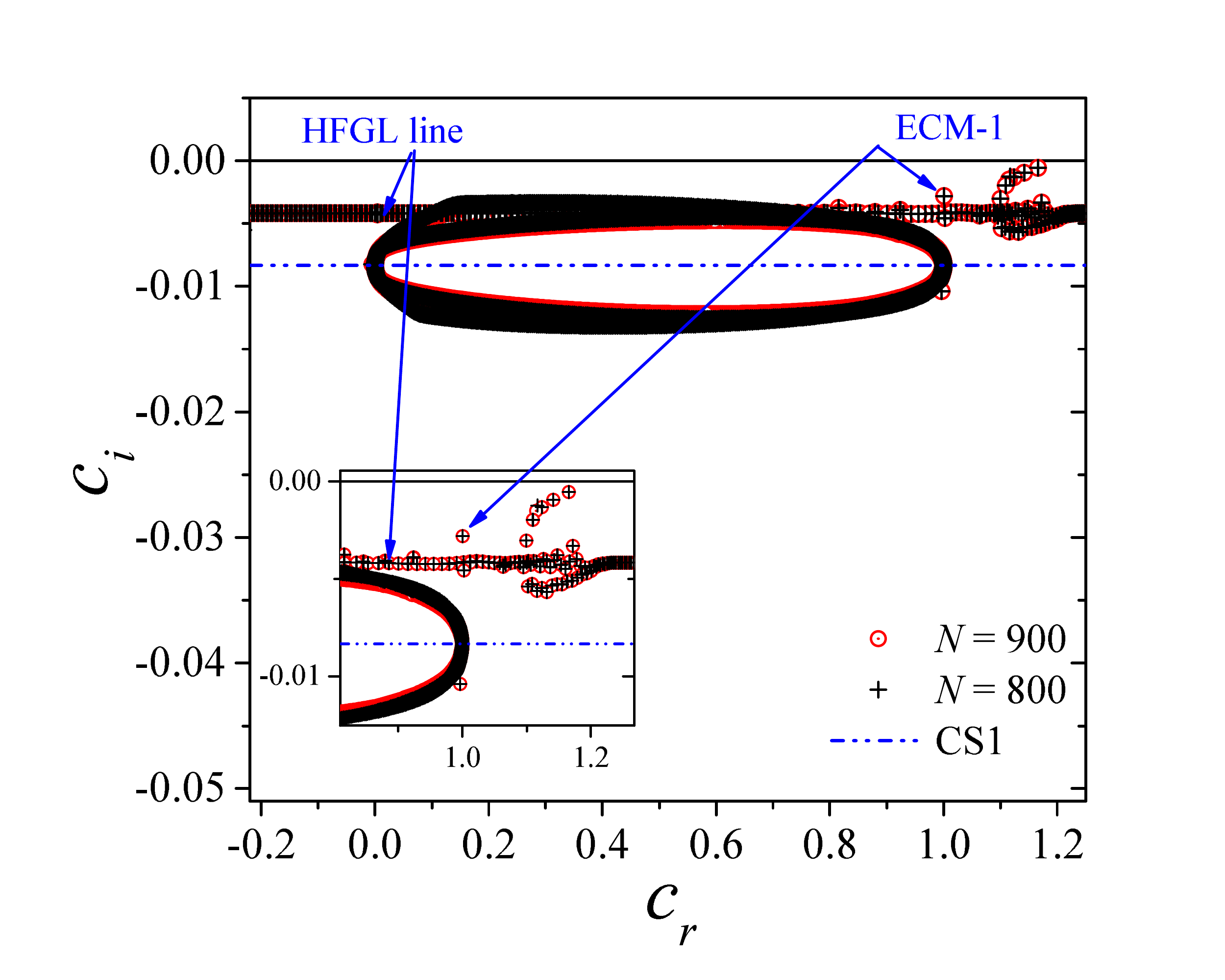}
		\caption{$\beta = 0, E = 0.1$}
		\label{fig:UCM}
	\end{subfigure}%
	~ 
	\begin{subfigure}[b]{0.5\textwidth}
		\centering
		\includegraphics[width=\textwidth]{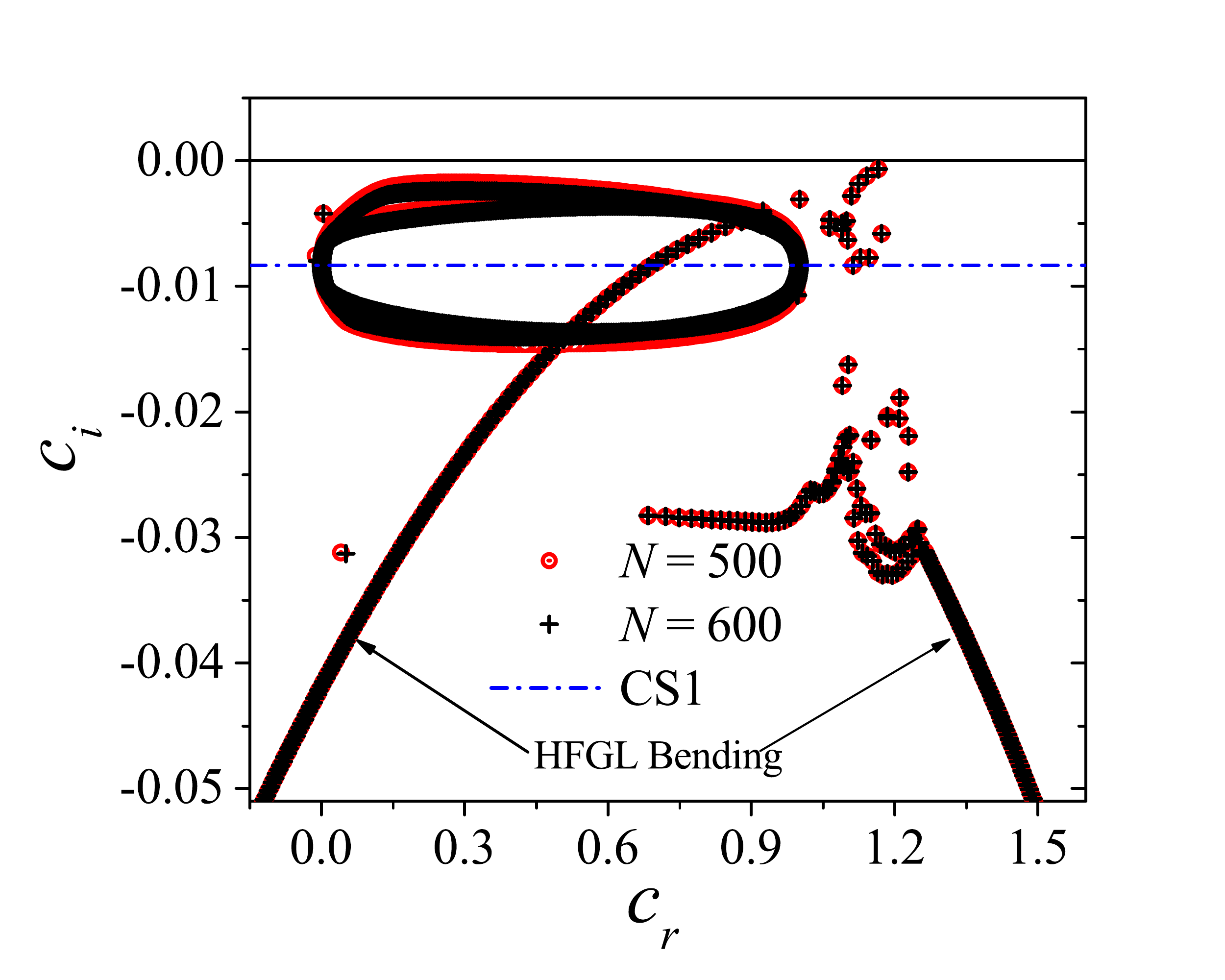}
		\caption{$\beta = 0.001, E = 0.1$}
		\label{fig:UCM_limit}
	\end{subfigure}
	~
	\begin{subfigure}[b]{0.5\textwidth}
		\centering
		\includegraphics[width=\textwidth]{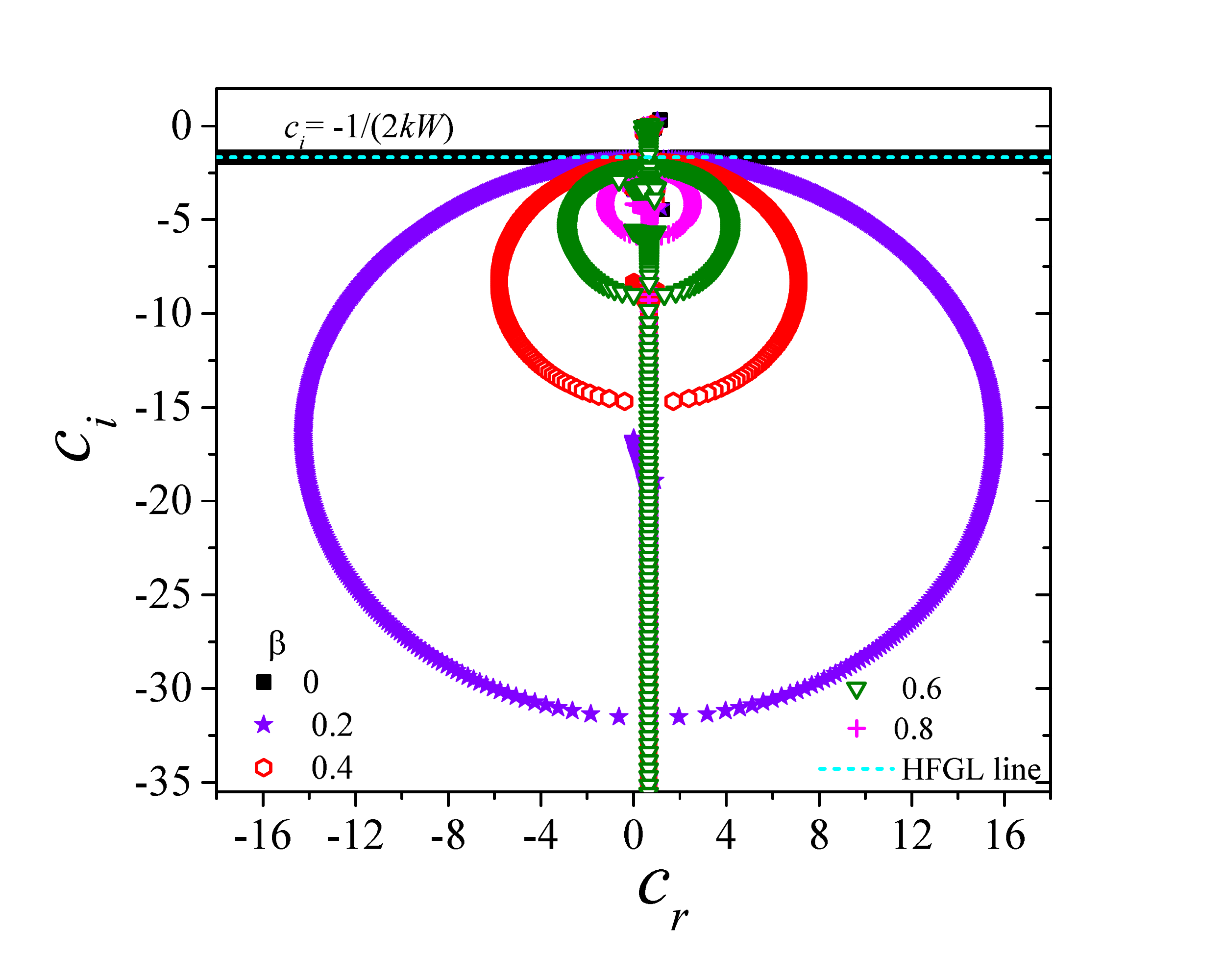}
		\caption{$\beta \in (0 , 0.8), E = 2.5 \times 10^{-4}$}
		\label{fig:HFGL_Bending}
	\end{subfigure}%
	~
	\begin{subfigure}[b]{0.5\textwidth}
		\centering
		\includegraphics[width=\textwidth]{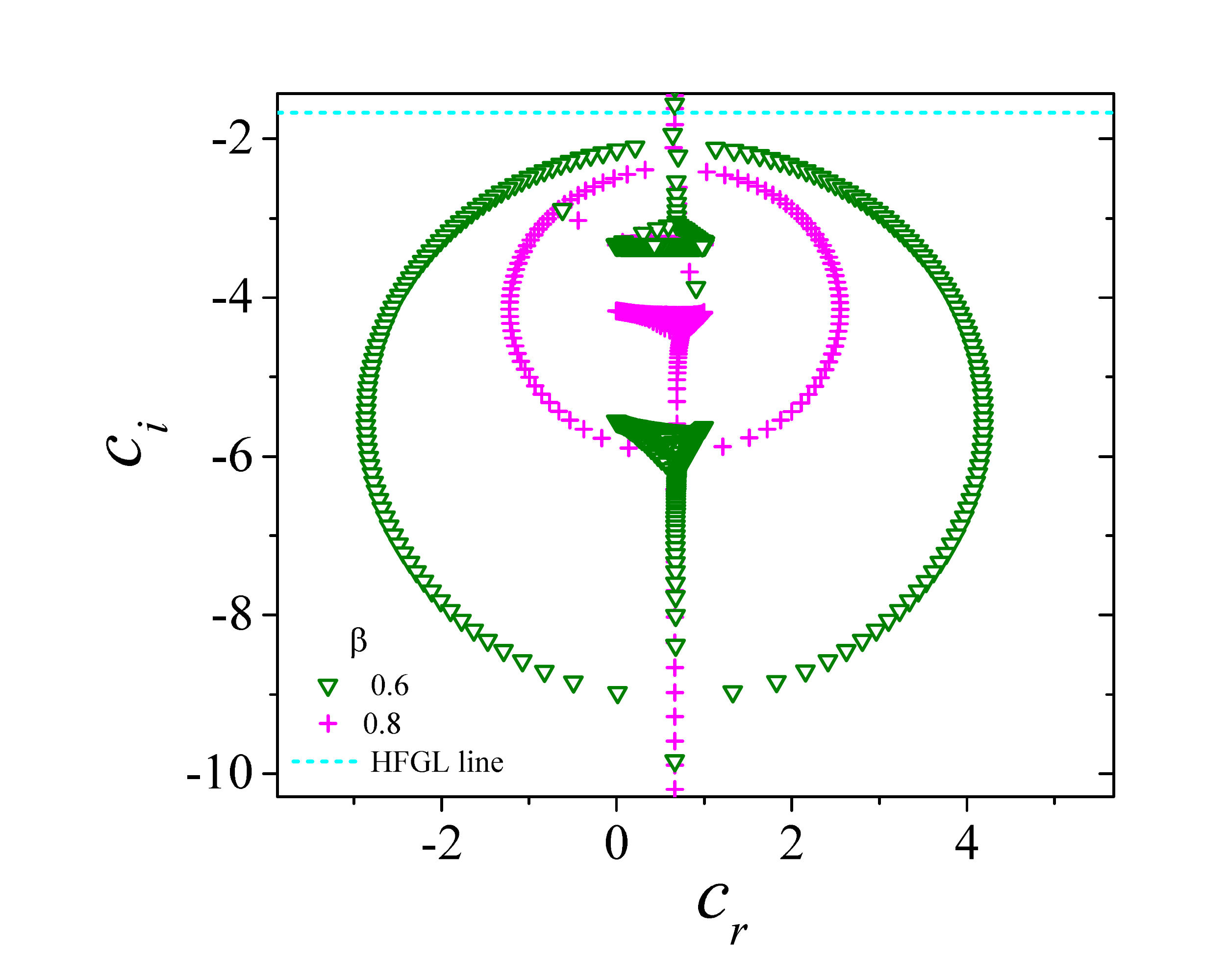}
		\caption{$\beta = 0.6, 0.8; E = 2.5 \times 10^{-4}$}
		\label{fig:HFGL_Bending_zoom}
	\end{subfigure}%
	\caption{Eigenspectrum of plane Poiseuille flow for  $Re=800, k=1.5$: (a) UCM model for $E=0.1$, (b) Oldroyd-B model for $\beta = 0.001$, $E =0.1$ showing the bending of HFGL, (c) Bending of HFGL with increasing $\beta $ illustrated for a very low value of $ E = 0.00025$, (d) zoomed-in version of panel~(c) showing the spectra at the higher $\beta$'s.}
	\label{fig:Eigenspectrum_UCM}
\end{figure*}

While discussing the evolution of the elasto-inertial center mode (ECM-1) in pipe flow at fixed $ \beta$, and  for different $ E $, \cite{chaudharyetal_2020} identified two qualitatively different trajectories of ECM-1 depending upon the value of $ \beta  $: For $\beta\geq0.85 $, ECM-1 collapses into CS1 in the limit $ E\rightarrow 0 $, and does not seem to have any connection with the 
Newtonian spectrum (and with the least stable Newtonian center mode NCM-1, in particular). However,  for $ \beta <0.85 $, the unstable center mode smoothly continues to the least stable center mode of the Newtonian eigenspectrum (labelled NCM-1 in this study).
For channel flow, in marked contrast, the unstable elasto-inertial center mode never smoothly continues to its Newtonian counterpart with decreasing $E$,  within the parameter regimes explored. 
 This is because ECM-1 and NCM-1 are modes with opposite symmetry (as will be seen later in Fig.~\ref{fig:contours}, the tangential velocity eigenfunction for NCM-1 is antisymmetric about the channel centerline, while it is symmetric for ECM-1 as already seen in Fig.~\ref{fig:Eigenfunction_vx_vz}), with the former emerging out of CS1 at a (non-zero) threshold $E$, and the latter collapsing into CS1 at a smaller $E$, for any fixed $\beta$.
It is worth contrasting this feature with  that in the pipe-flow elasto-inertial spectrum, where the least stable Newtonian and elastic center modes remain smoothly connected for $\beta < 0.85$, the connection made possible by the axisymmetry of both modes.
Figure~\ref{fig:ci_E_Beta0p58_and_0p96} reinforces this trend by showing the  
scaled growth rate of the least stable elasto-inertial center mode for two different $\beta$ (\textit{viz.,} 0.58 and 0.96).  The range of $ E $ for which elasto-inertial center mode remains unstable increases with $ \beta $. 
For both $ \beta $, ECM-1 follows a  trajectory similar to the one  shown in Figs.~\ref{fig:Evolution_ECM} and \ref{fig:Ci_E_for-all-modes_Beta0p8}. Thus, the
elasto-inertial center mode, whether unstable or otherwise, is not the continuation/elastic modification of least stable Newtonian center mode (NCM-1) for any $ \beta $.
The behavior in Fig.~\ref{fig:ci_E_Beta0p58_and_0p96}  holds true even if one were to choose a $\beta$ where the flow remains stable (regardless of $Re$ or $E$).


\subsubsection{Effect of varying $\beta$ at fixed $ E $}
\label{sssec:Effect of solvent viscosity}

In Fig.~\ref{fig:Eigenspectrum_UCM}, we explore the effect of increasing $\beta$ from 0 (the UCM limit)  on the elasto-inertial spectrum, at a fixed $E$.  The structure of the elasto-inertial spectrum in the UCM limit (Fig.~\ref{fig:UCM}) is now well understood \citep{chaudhary_etal_2019}, comprising of the HFGL class of modes and the ballooned-up continuous spectrum CS1. In addition, at sufficiently high $Re$ and $E$, \cite{chaudhary_etal_2019} also showed the existence of a hour-glass like structure which, however, is not prominent for the moderate $Re$ and $E$ considered in Fig.~\ref{fig:Eigenspectrum_UCM}. 
The center mode (ECM-1) remains stable for $\beta = 0$ in Fig.~\ref{fig:UCM}.  As $\beta$ is increased to $0.001$ in Fig.~\ref{fig:UCM_limit}, the HFGL  modes are seen to be heavily damped even at this small $\beta$. Thus, for $E = 0.1$, the continuation of the HFGL modes are not important in determining the stability of the flow in the (experimentally relevant) dilute limit ($\beta \sim 0.8$ and higher).  
As pointed out earlier in Sec.~\ref{sssec:Effect of elasticity}, for nonzero $\beta$, the HFGL line in the UCM limit bends leading to the formation of an ellipse.
The formation of the ellipse-like structure is best illustrated at a lower $E = 2.5 \times 10^{-4}$ (Fig.~\ref{fig:HFGL_Bending}). The extent of the ellipse shrinks as $\beta$ is increased  to $0.4$, leading to an enhanced stability of the HFGL modes. Thus, regardless of $E$,  in the limit of dilute polymer solutions, the continuation of the HFGL modes are not relevant in determining the stability, and we do not discuss them hereafter.
In our earlier study on viscoelastic channel flow \citep{chaudhary_etal_2019}, we showed that an increasing number of wall modes belonging to the upper bulb of the hour-glass structure become unstable in the UCM spectrum with increasing $Re$ and $E$. The effect of nonzero $\beta$ on these elasto-inertial wall modes, however, was found to be strongly stabilizing  \citep{khalid_solvent}, akin to its stabilizing effect on the continuation of the TS mode found in earlier studies  \citep{Sureshkumar1995,sadanandan_sureshkumar_2002,Zaki2013}. This stabilizing role of $\beta$ on wall modes is in direct contrast to its destabilizing effect on the elasto-inertial center mode examined in the present study.


\begin{figure*}
	\centering
	\begin{subfigure}[b]{0.5\textwidth}
		\centering
		\includegraphics[width=\textwidth]{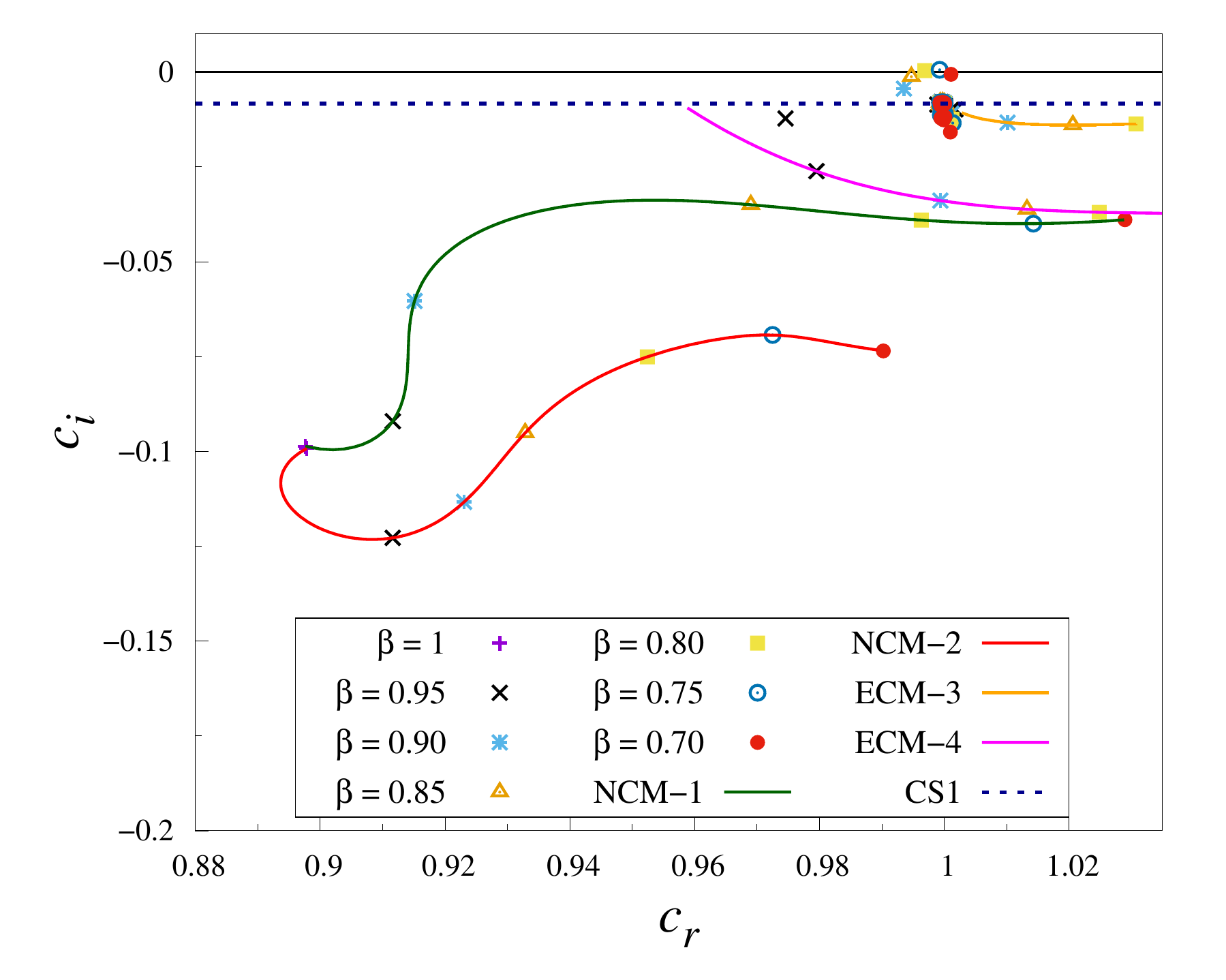}
		\caption{}
		\label{fig:ECM3_ECM4}
	\end{subfigure}%
	~ 
	\begin{subfigure}[b]{0.5\textwidth}
		\centering
		\includegraphics[width=\textwidth]{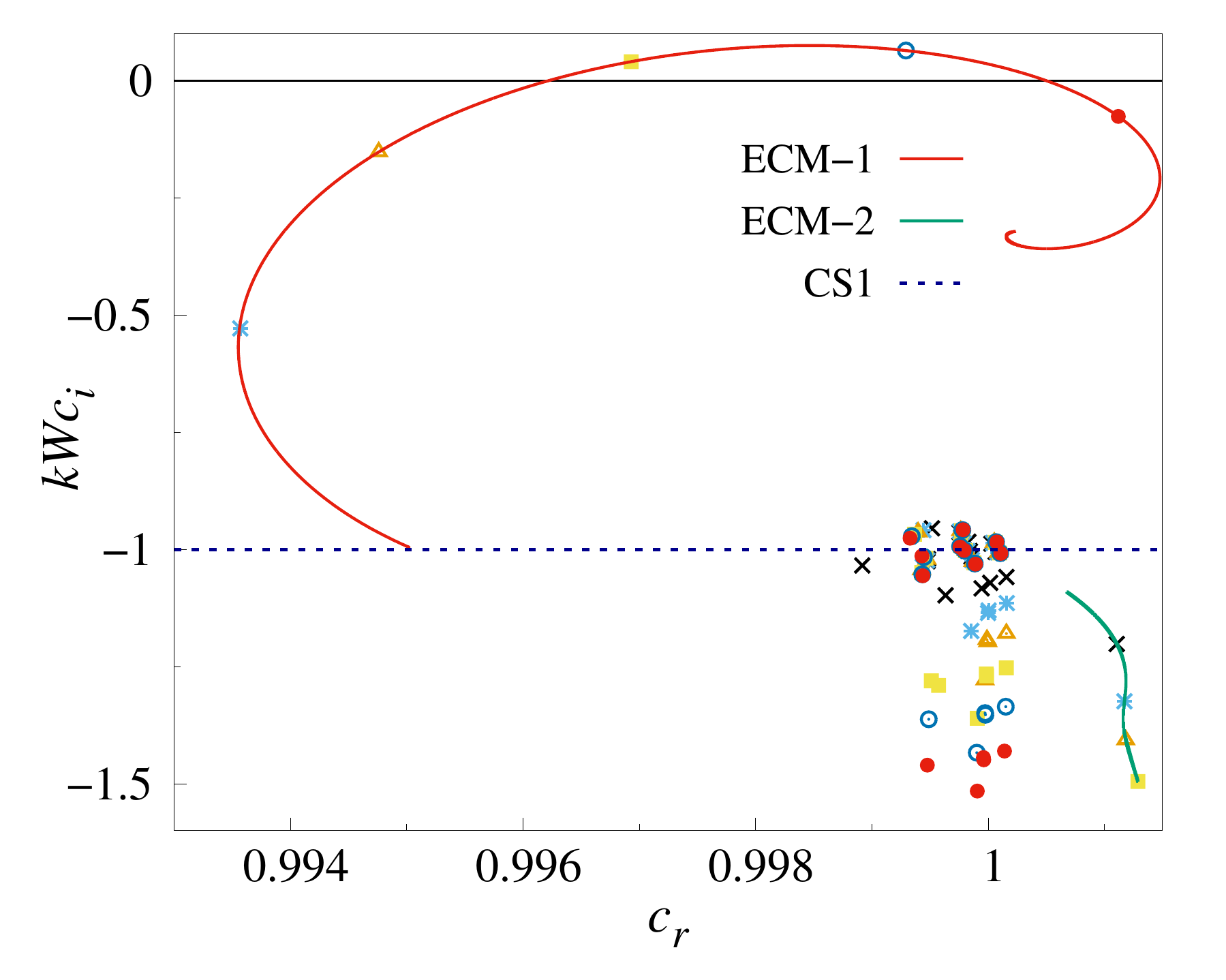}
		\caption{}
		\label{fig:ECM-1_ECM-2}
	\end{subfigure}
	
	\caption{\small Modification of Newtonian center modes (NCM-1,-2) in the viscoelastic spectrum and the emergence of new elasto-inertial center modes (ECM-1,-2,-3,-4) as $ \beta$ is decreased from unity at $Re=800,  k=1.5 ,E=0.1$. (a) NCM-1 and -2 and  ECM-3, ECM-4; (b) ECM-1 and ECM-2. All the new elasto-inertial center modes emerge from CS1 as $ \beta  $ is reduced from unity. In panel~(a), the modes NCM-1 and 2 are distinct, but closely placed,  in the Newtonian limit. The continuous lines represent results from the shooting method while symbols denote results from the spectral method. For clarity, only the filtered eigenspectrum is shown (with the CS balloons being absent). The theoretical location of CS1 is shown using dotted lines.}
	\label{fig:Elastic_modes}
\end{figure*}

\begin{figure*}
	\centering
	\begin{subfigure}[b]{0.5\textwidth}
		\centering
		\includegraphics[width=\textwidth]{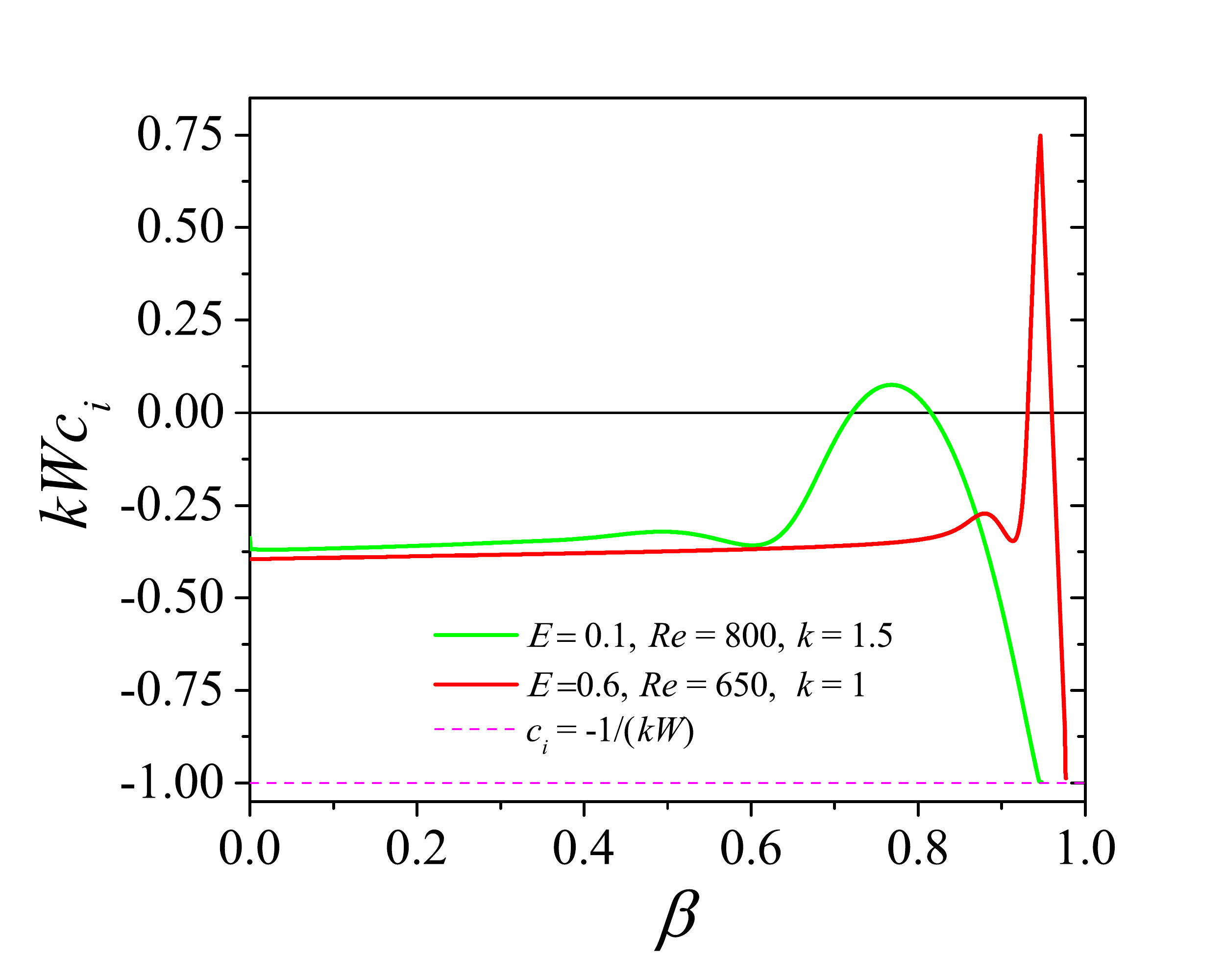}
		\caption{}
		\label{fig:kWci_Beta_E0p1_and_E0p6}
	\end{subfigure}%
	~ 
	\begin{subfigure}[b]{0.5\textwidth}
		\centering
		\includegraphics[width=\textwidth]{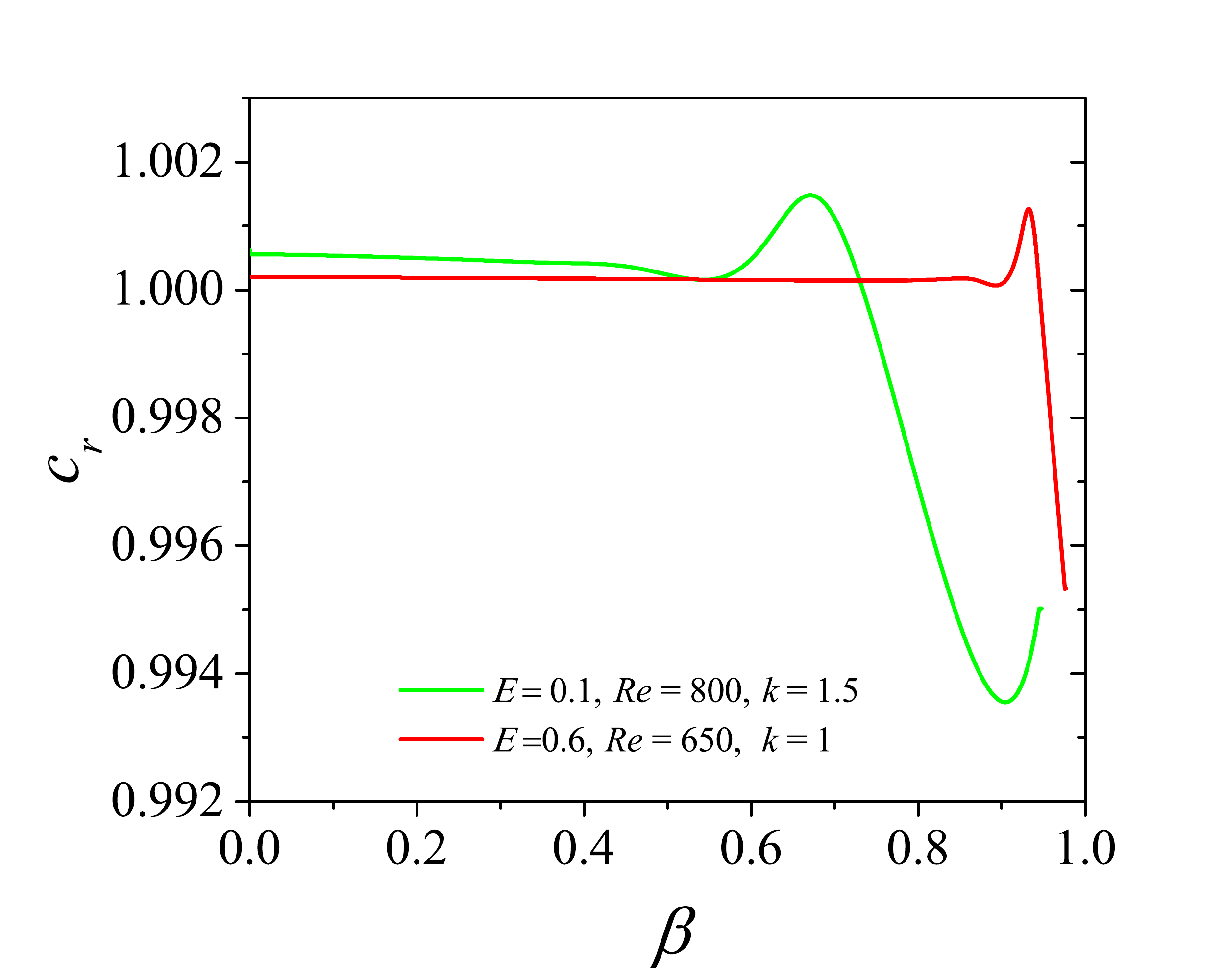}
		\caption{}
		\label{fig:cr_Beta_E0p1_and_E0p6}
	\end{subfigure}
\caption{Effect of variation in $ \beta $ on the scaled growth rate ($k W c_i $) of unstable elasto-inertial center mode (ECM-1) for  $Re=800, k=1.5$. (a) In the UCM limit ($ \beta \rightarrow 0 $), the center mode remains stable even at very large values of $ E $, illustrating the role of solvent viscosity in the center-mode instability in channel flow. For a fixed $ E = 0.1$, as $\beta$ is decreased from unity, ECM-1 emerges from CS1 and becomes unstable over a small range of  $ \beta $ ($0.8 $--$ 0.7$). The unstable range of $\beta$ shifts towards $\beta \rightarrow 1$ for $E = 0.6$. Panel~(b) shows the corresponding variation of $c_r$ with $\beta$.}
\label{fig:mode_tracking_fixed-E}
\end{figure*}

\begin{figure*}
	\centering
	\begin{subfigure}[b]{0.5\textwidth}
		\centering
		\includegraphics[width=\textwidth]{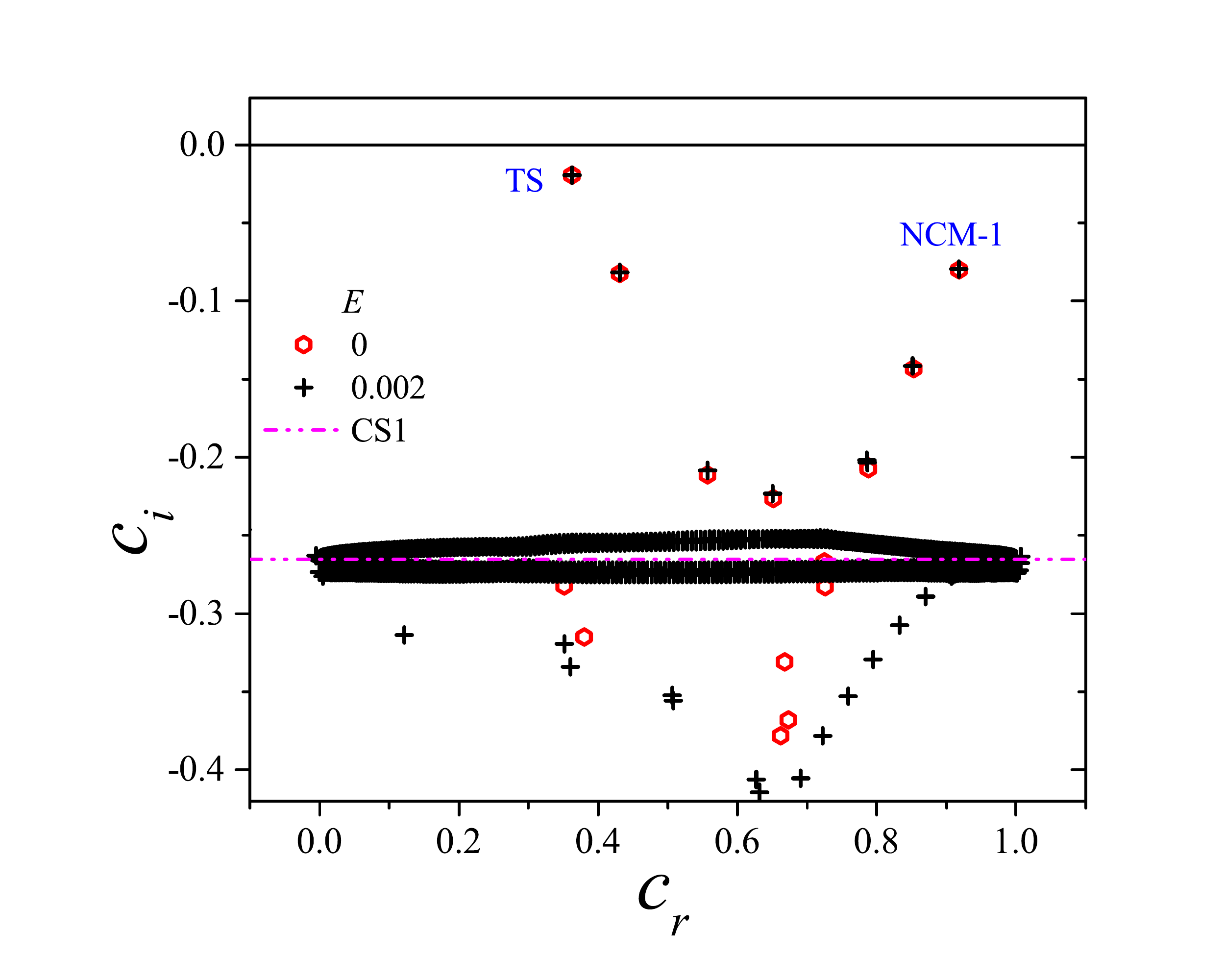}
		\caption{$ E = 0.002 $}
		\label{fig:E_0p002_Re1500}
	\end{subfigure}%
	~ 
	\begin{subfigure}[b]{0.5\textwidth}
		\centering
		\includegraphics[width=\textwidth]{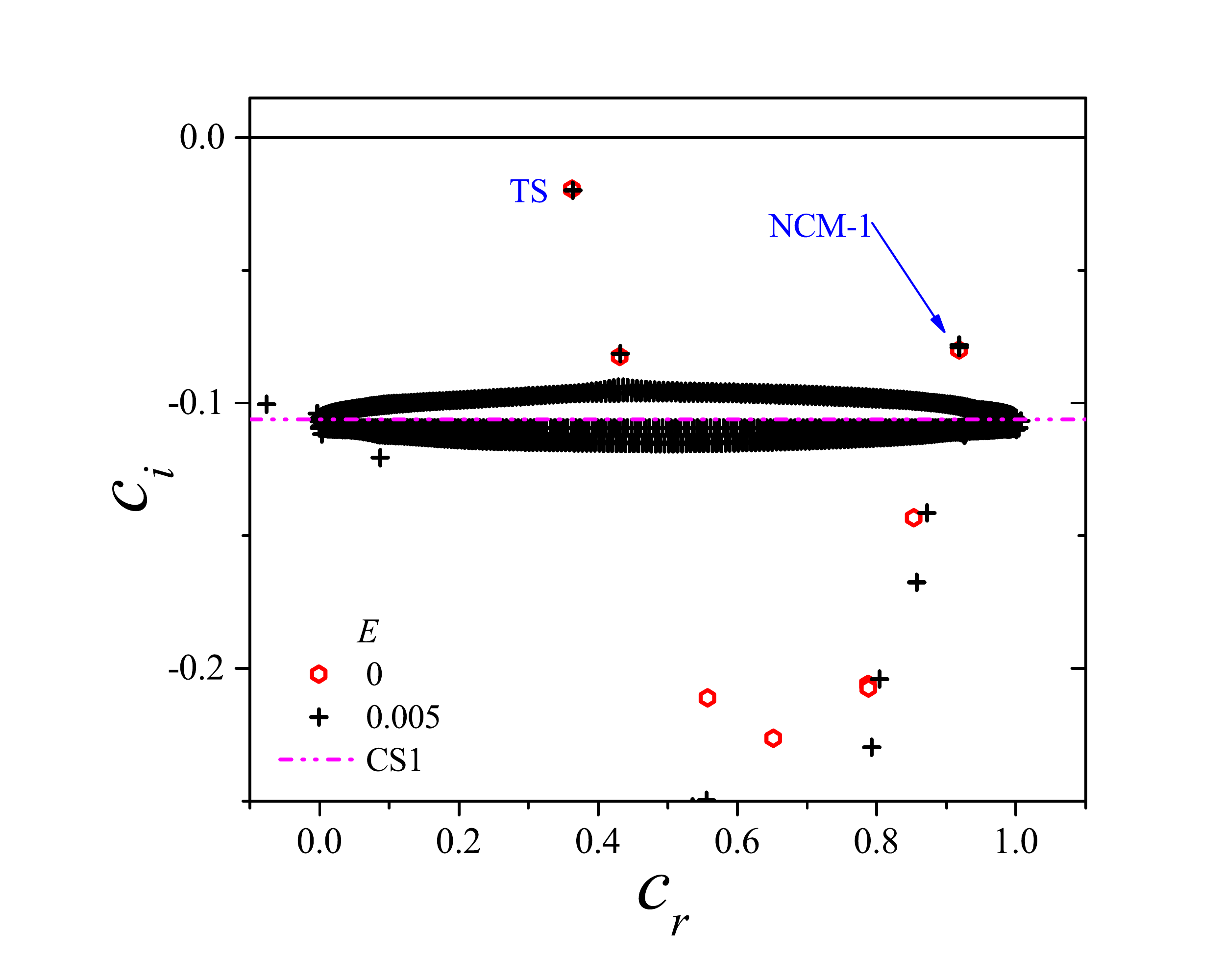}
		\caption{$ E = 0.005$}
		\label{fig:E_0p005_Re1500}
	\end{subfigure}
	~
	\begin{subfigure}[b]{0.5\textwidth}
		\centering
		\includegraphics[width=\textwidth]{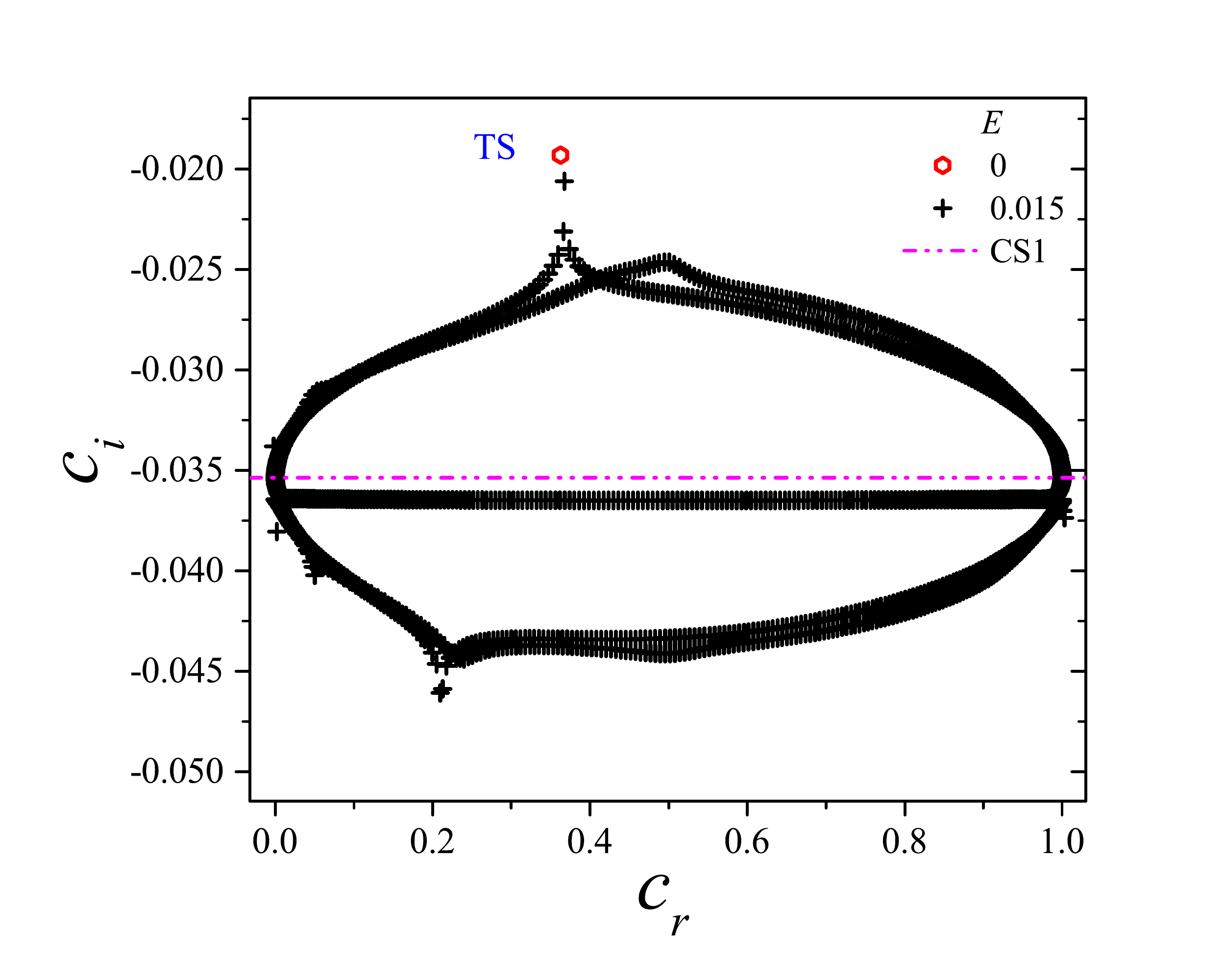}
		\caption{$ E = 0.015 $}
		\label{fig:E_0p015_Re1500}
	\end{subfigure}%
	~
	\begin{subfigure}[b]{0.5\textwidth}
		\centering
		\includegraphics[width=\textwidth]{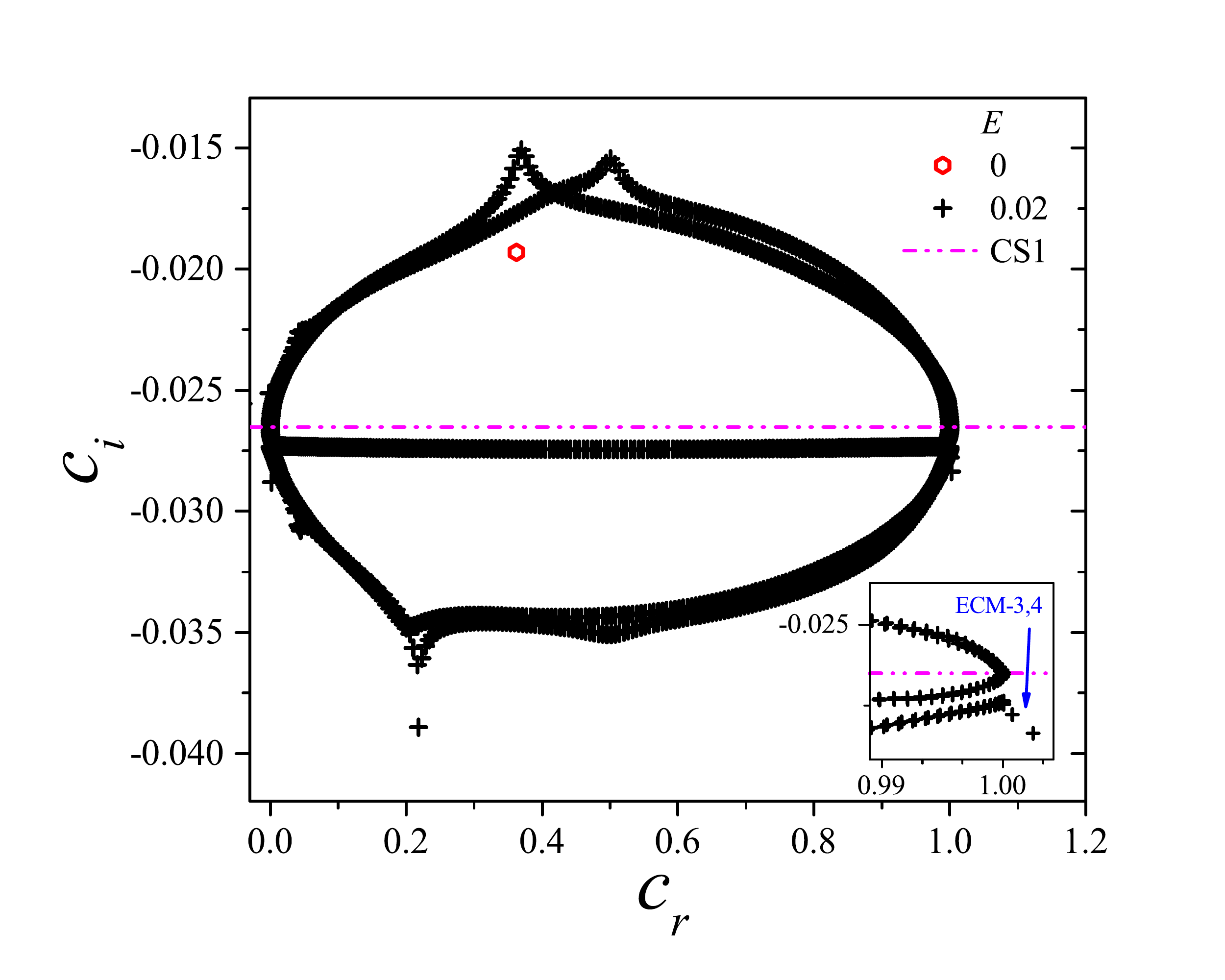}
		\caption{$ E = 0.02 $}
		\label{fig:E_0p02_Re1500}
	\end{subfigure}
	~ 
	\begin{subfigure}[b]{0.5\textwidth}
		\centering
		\includegraphics[width=\textwidth]{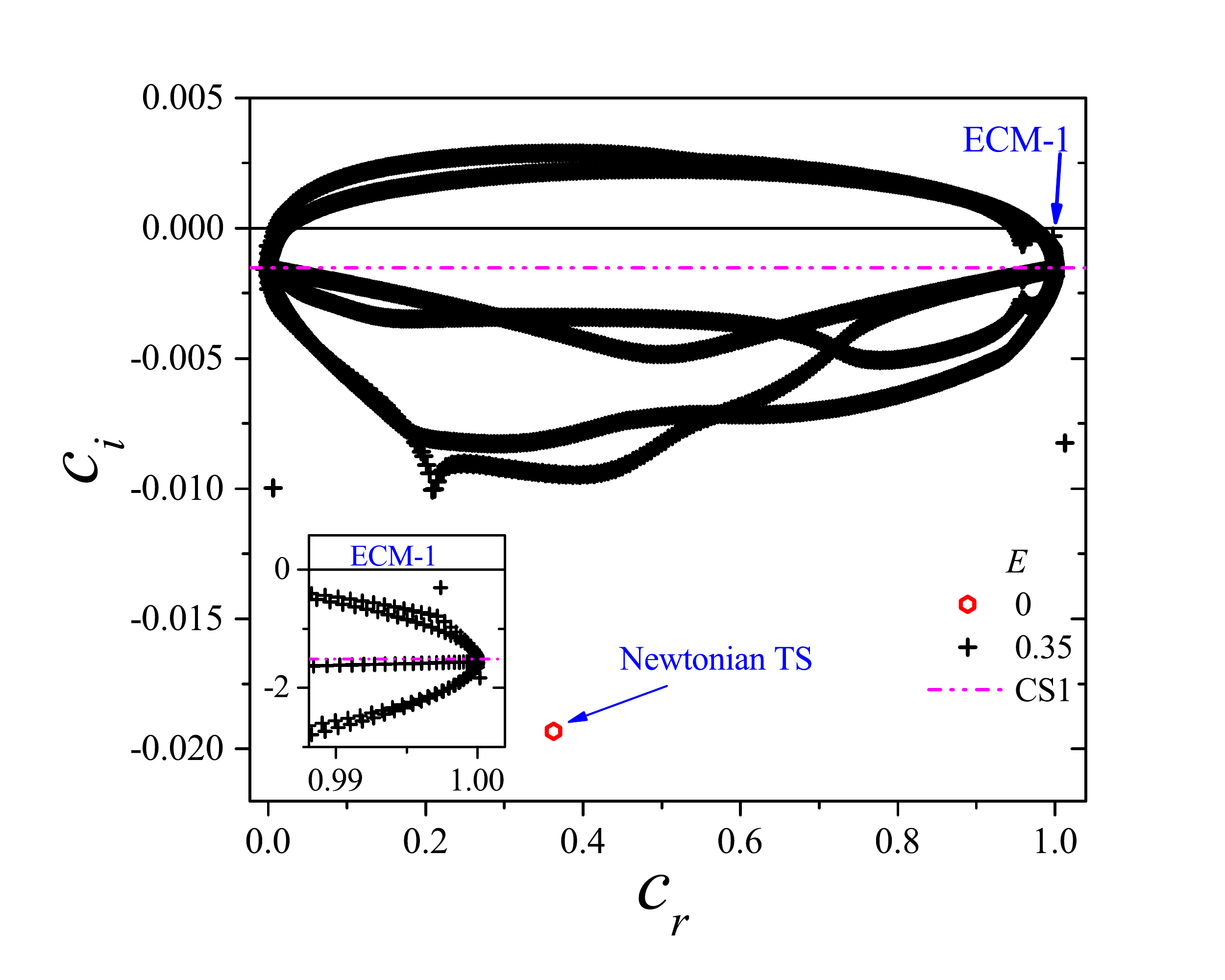}
		\caption{$ E = 0.35 $}
		\label{fig:Ep35_Re1500}
	\end{subfigure}%
	~
	\begin{subfigure}[b]{0.5\textwidth}
		\centering
		\includegraphics[width=\textwidth]{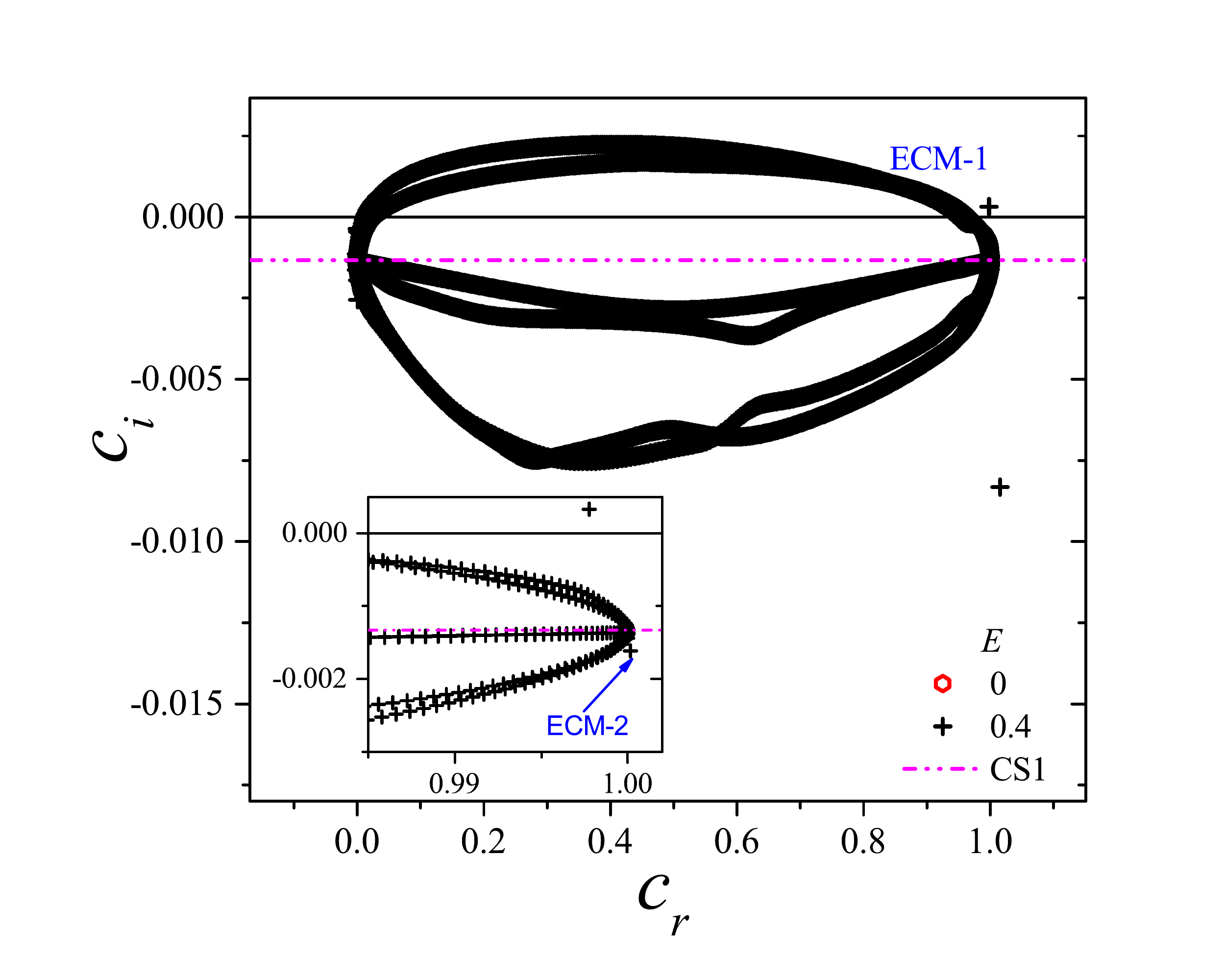}
		\caption{$ E = 0.4 $}
		\label{fig:Ep4_Re1500}
	\end{subfigure}
	\caption{\small Eigenspectra of viscoelastic channel flow for  $Re=1500$, $ k=0.4\pi, \beta=0.97 $ and varying $E$. For $ E \le  0.015$ (or, $W \leq 22$,  similar to the regime considered by \cite{Shekar2019}), the elastically modified TS mode is the least stable one. For $0.015 < E < 0.35$, there is no discrete mode above the CS.
	However, for $ E = 0.35 $, ECM-1 emerges above the CS to become the least stable mode, turning unstable at $ E \approx 0.4 $.  The corresponding Newtonian eigenspectrum ($E = 0$) for these set of parameters is shown for comparison. In panels (c)--(f), only the region near the CS is shown to illustrate the collapse and emergence of discrete modes from the CS.} 
	\label{fig:Re1500_k0.4pi_B0.97}
\end{figure*}

Figure~\ref{fig:ECM3_ECM4} shows the trajectories of the two leading Newtonian center modes (labelled NCM-1 and NCM-2; although these appear to emanate from the same point for $\beta = 1$, a closer examination reveals two distinct, but closely-spaced modes in the Newtonian spectrum) as  $\beta$ is gradually decreased from unity for a fixed $E$. 
Besides these Newtonian center modes (NCM-1 and NCM-2), four new modes emerge from the continuous spectrum (CS1). These modes (labelled ECM-1 to ECM-4 in Figs.~\ref{fig:ECM3_ECM4} and \ref{fig:ECM-1_ECM-2}) arise because of the combined effect of polymer elasticity and solvent viscosity at non-zero $ Re $, and hence do not have  counterparts in the Newtonian spectrum. The unstable center mode belongs to this class (ECM-1 in Fig.~\ref{fig:ECM-1_ECM-2}). 
Except ECM-1, however, all the other elasto-inertial center modes remain stable over the entire range of  $ \beta $, from the  Newtonian ($\beta = 1$) to the UCM ($\beta = 0$) limit, regardless of $ Re $ and $ E $. 
Figure~\ref{fig:ECM-1_ECM-2} depicts the trajectory of ECM-1 with decreasing $\beta$, starting from its emergence out of CS1 at $\beta \approx 0.95$, using the scaled growth rate $W kc_i$ (the continuous (red) line represents results from the shooting method).
Similar to the  trend exhibited by ECM-1 for varying $ E$ (at fixed $\beta$; see Sec.~\ref{sssec:Effect of elasticity}), wherein the instability existed only over a finite range of $E$, the mode is unstable only over a range of  $ \beta$  at fixed $E$ in Fig.~\ref{fig:ECM-1_ECM-2}), and becomes stable again below a critical $\beta$.
Thus, the trajectory of the unstable center mode with varying $ \beta  $, at a fixed $ E $, is similar in both pipe \citep[see Fig.~12 of][]{chaudharyetal_2020} and channel (Fig.~\ref{fig:Elastic_modes} of the present work) flows. 
However, in contrast to the pipe case, the  unstable center mode in channel flow persists even for $Re \sim O(1)$ in the limit $ \beta \rightarrow 1 $, albeit at high  $E$. We discuss this in detail in Sec.~\ref{ssec:Critical parameters and Scalings}.
   
In Fig.~\ref{fig:kWci_Beta_E0p1_and_E0p6}, we exclusively focus on the center mode ECM-1 to illustrate the importance of the solvent viscous contribution in rendering this mode unstable, by showing the variation of the scaled growth rate with $\beta$; Fig.~\ref{fig:cr_Beta_E0p1_and_E0p6} shows the variation of the corresponding phase speeds. At a fixed $ Re, E $ and $ k $, ECM-1 emerges from CS1 ($ c_i=-1/kW $) as $ \beta  $ is decreased  from the Newtonian limit ($ \beta \rightarrow 1 $). At a critical $ \beta $ (close to unity for higher $ E $) the elasto-inertial mode becomes unstable, and the range of $ \beta $ in which ECM-1 is unstable increases with decrease in $ E $. However,  the mode becomes stable again as $ \beta$ is decreased below a threshold. Crucially, for $\beta < 0.5$, we find that the center mode always remains stable in channel flow, at any $E, Re$. The absence of instability for $\beta < 0.5$ reinforces our predictions from the spectral analysis (in the previous section) that for  the center-mode instability, solvent viscosity is essential along with inertia and elasticity, again in agreement with the pipe flow results of \cite{Piyush_2018} and \cite{chaudharyetal_2020}. However, for pipe flow, the center mode becomes unstable even  as $\beta \approx 10^{-3}$, for sufficiently high $Re$. Intriguingly, this feature is not present in viscoelastic channel flows. 

\subsection{Relative stability of center and wall modes}
\label{ssec:Relative stability of center and wall modes}

\begin{figure*}
	\centering
	\begin{subfigure}[b]{0.5\textwidth}
		\centering
		\includegraphics[width=\textwidth]{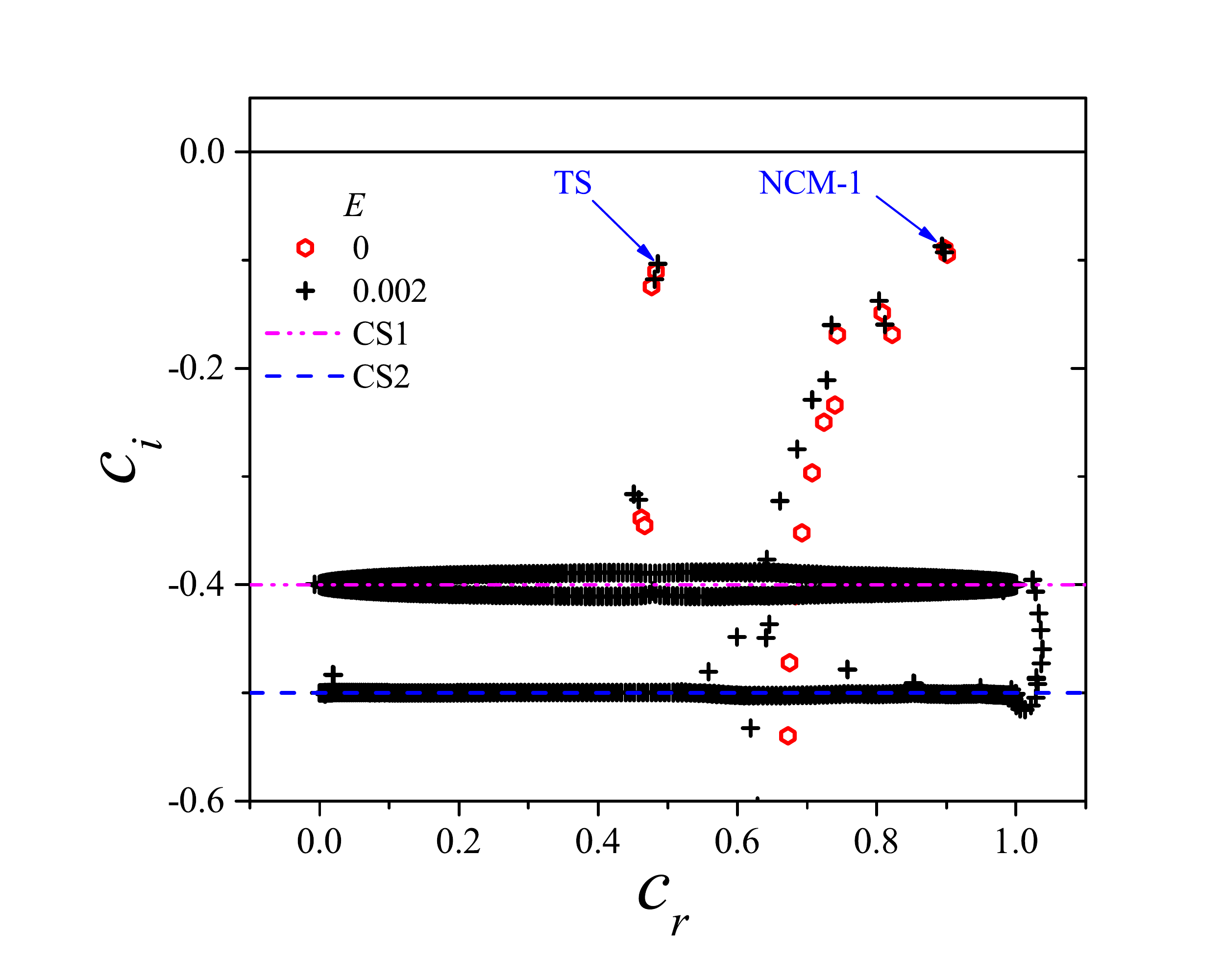}
		\caption{$ E = 0.002$}
		\label{fig:E0p002_Re500}
	\end{subfigure}%
	~ 
	\begin{subfigure}[b]{0.5\textwidth}
		\centering
		\includegraphics[width=\textwidth]{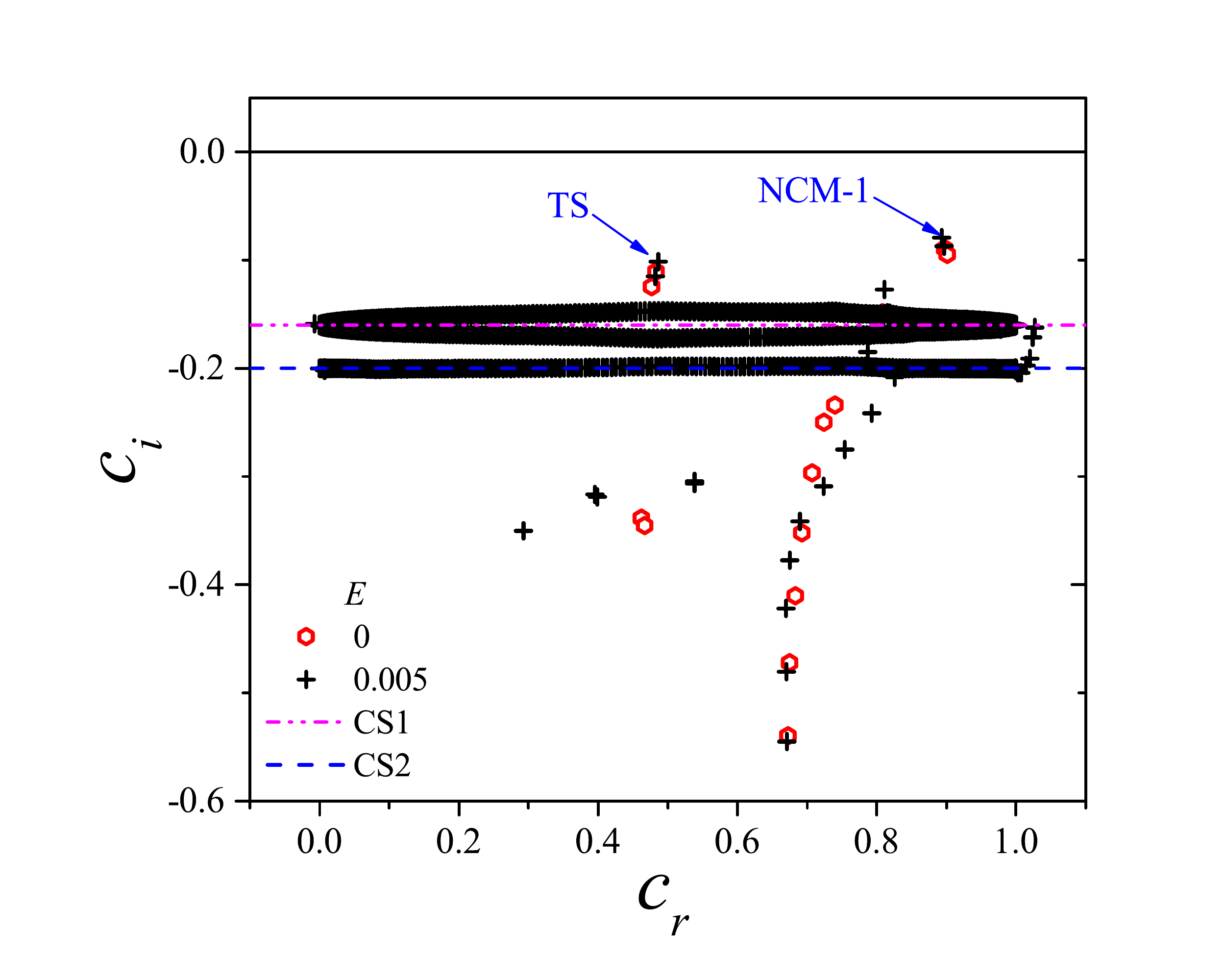}
		\caption{$ E = 0.005$}
		\label{fig:E0p005_Re500}
	\end{subfigure}
	~
	\begin{subfigure}[b]{0.5\textwidth}
		\centering
		\includegraphics[width=\textwidth]{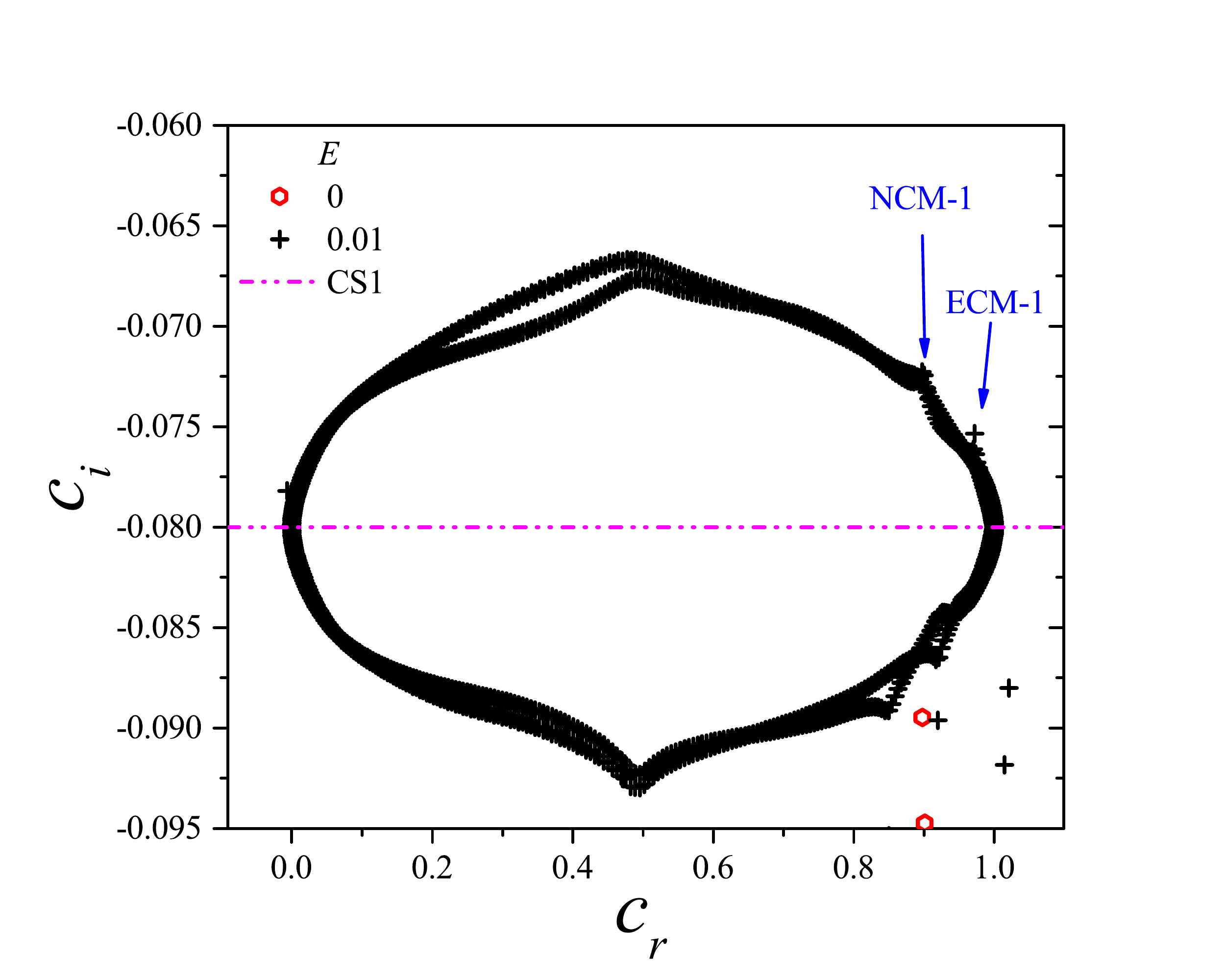}
		\caption{$ E = 0.01 $}
	 	\label{fig:E0p01_Re500}
	\end{subfigure}%
	~
	\begin{subfigure}[b]{0.5\textwidth}
		\centering
		\includegraphics[width=\textwidth]{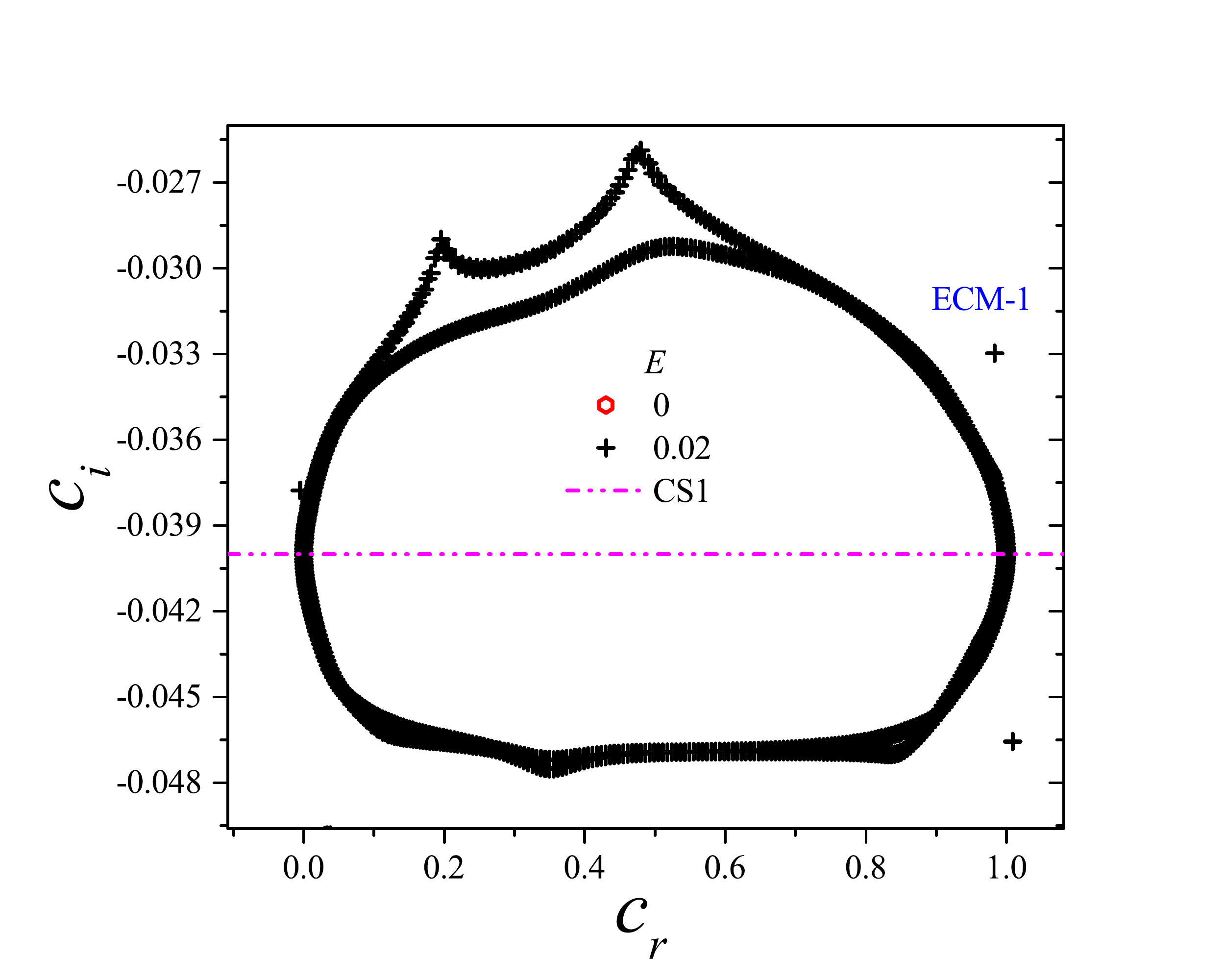}
		\caption{$ E = 0.02 $}
		\label{fig:E0p02_Re500}
	\end{subfigure} 
	
	\caption{\small Eigenspectra for viscoelastic channel flow for  $Re=500$, $ k=0.8\pi$ $(2.5)$, $ \beta=0.97 $ and varying $E$. For $k>2$, NCM-1 is the  least stable mode even in the Newtonian limit ($ E\rightarrow0 $). In panels (c) and (d), only the region near the CS is shown to illustrate the collapse and emergence of discrete modes from the CS.} 
	\label{fig:Re500_k2.5_B0.8}
\end{figure*}

\begin{figure*}
	\centering
	\begin{subfigure}[b]{0.5\textwidth}
		\centering
		\includegraphics[width=\textwidth]{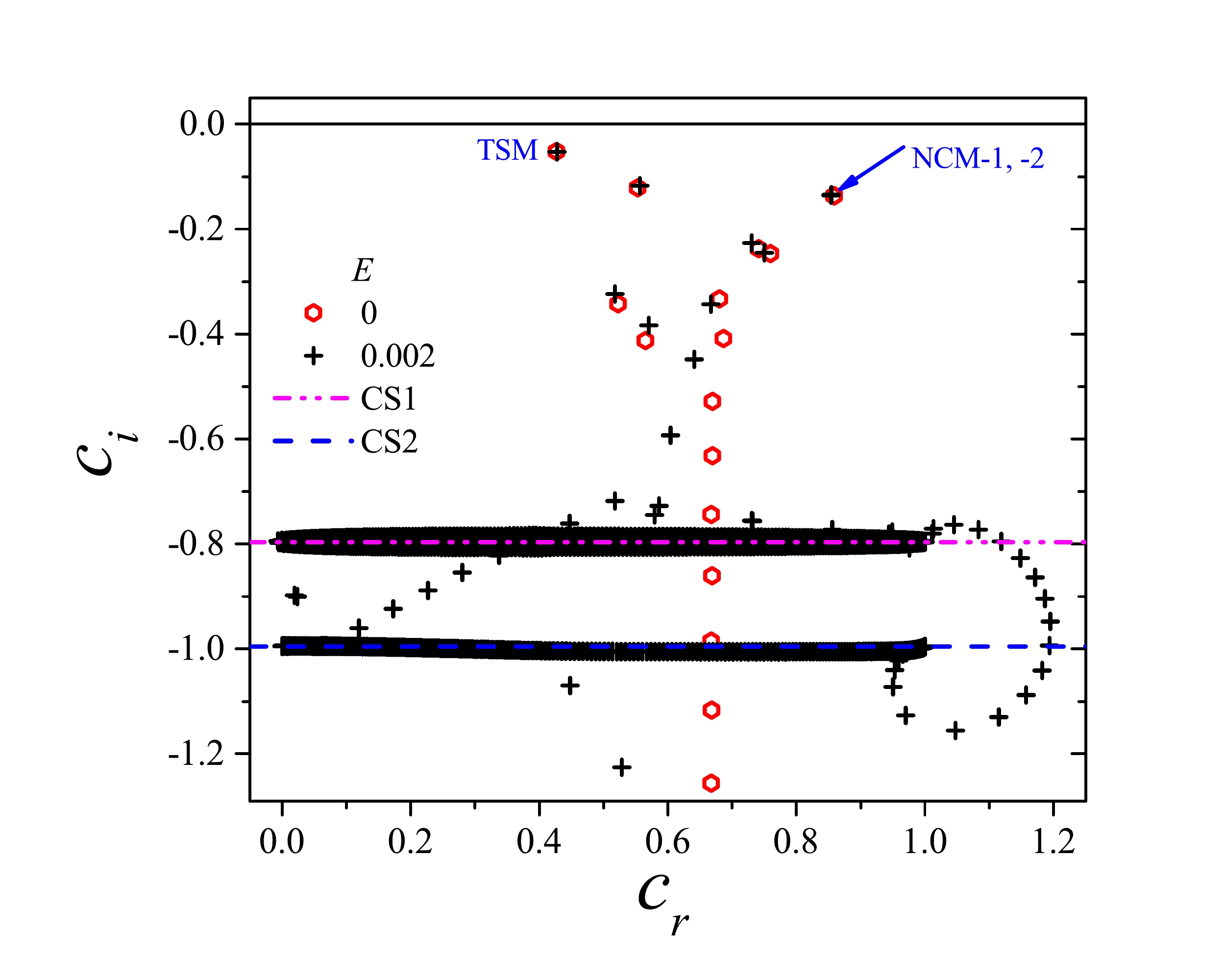}
		\caption{$  E = 0.002 $}
		\label{fig:Ep002_Re500_k0p4pi_B0p8}
	\end{subfigure}%
	~ 
	\begin{subfigure}[b]{0.5\textwidth}
		\centering
		\includegraphics[width=\textwidth]{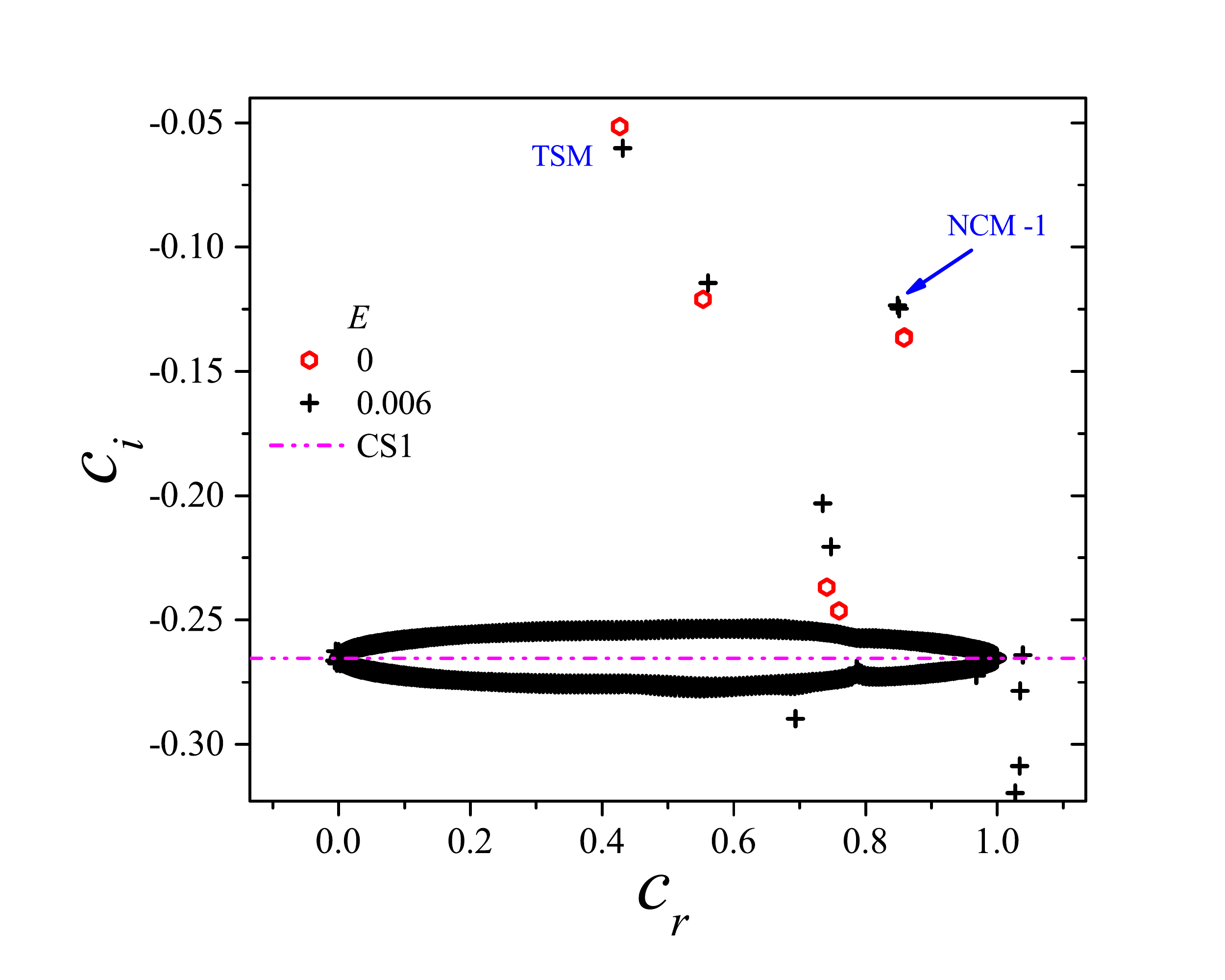}
		\caption{$ E = 0.006 $}
		\label{fig:Ep006_Re500_k0p4pi_B0p8}
	\end{subfigure}
	~
	\begin{subfigure}[b]{0.5\textwidth}
		\centering
		\includegraphics[width=\textwidth]{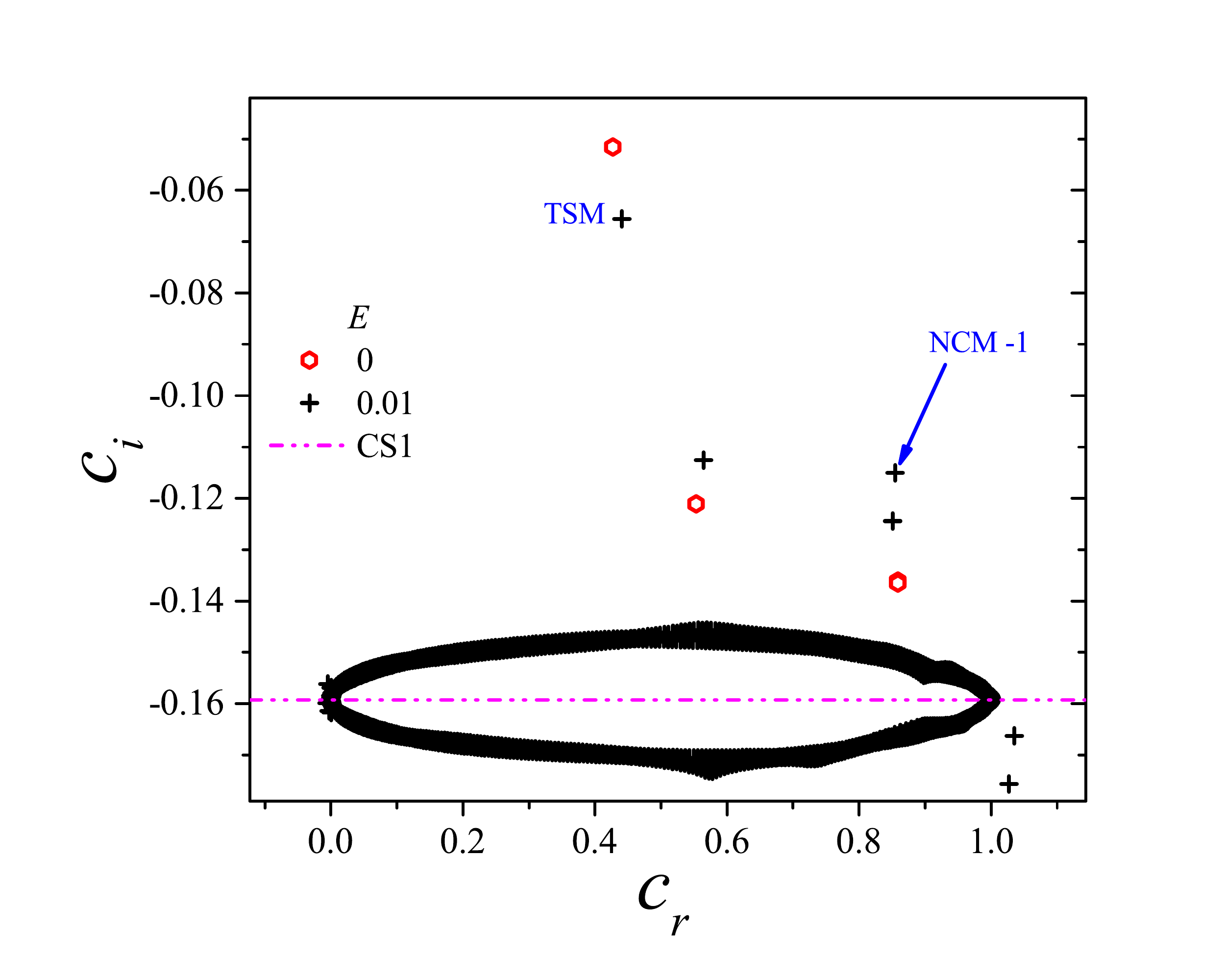}
		\caption{$  E = 0.01 $}
		\label{fig:Ep01_Re500_k0p4pi_B0p8}
	\end{subfigure}%
	~
	\begin{subfigure}[b]{0.5\textwidth}
		\centering
		\includegraphics[width=\textwidth]{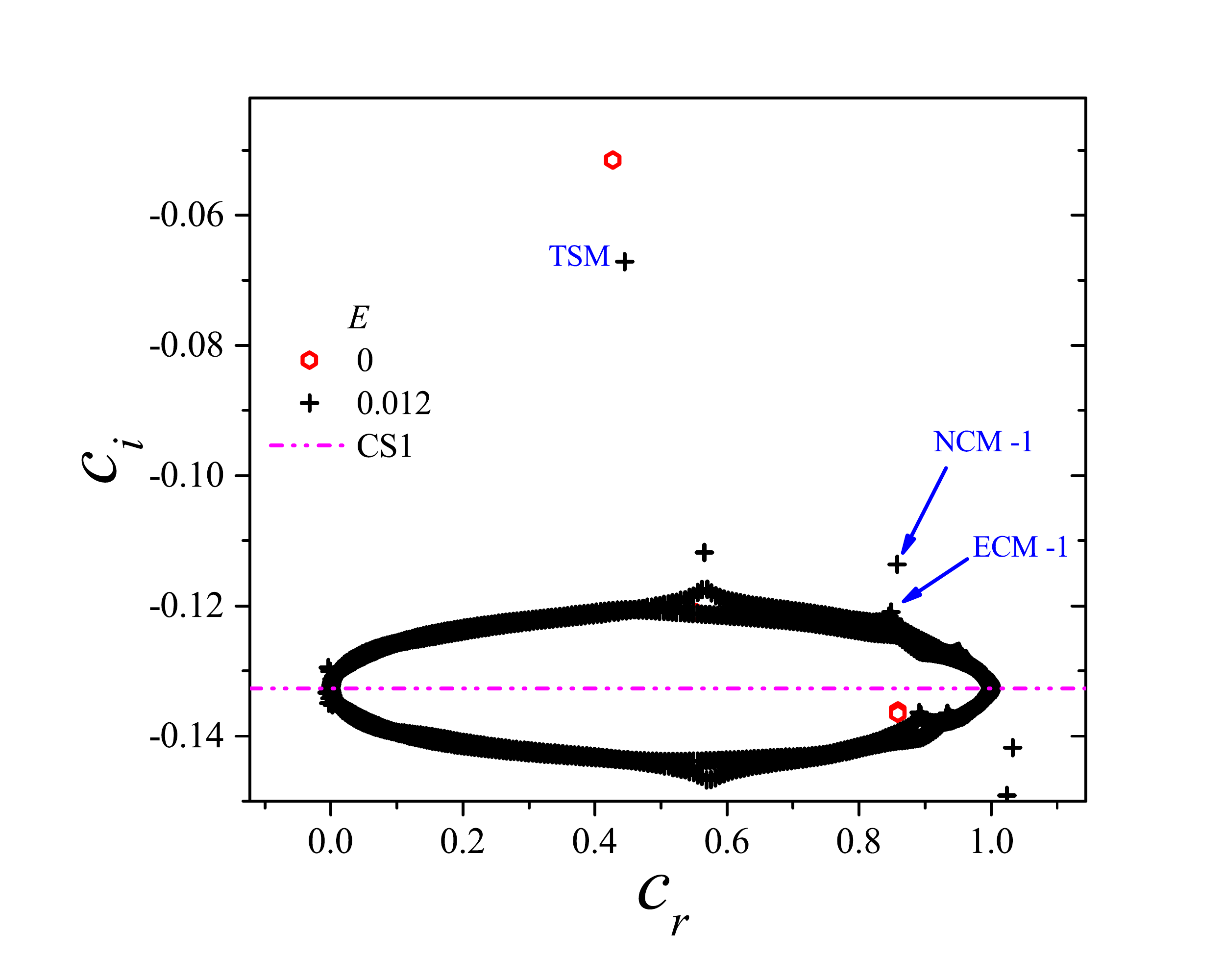}
		\caption{$ E = 0.012 $}
		\label{fig:Ep012_Re500_k0p4pi_B0p8}
	\end{subfigure} 
	~
	\begin{subfigure}[b]{0.5\textwidth}
		\centering
		\includegraphics[width=\textwidth]{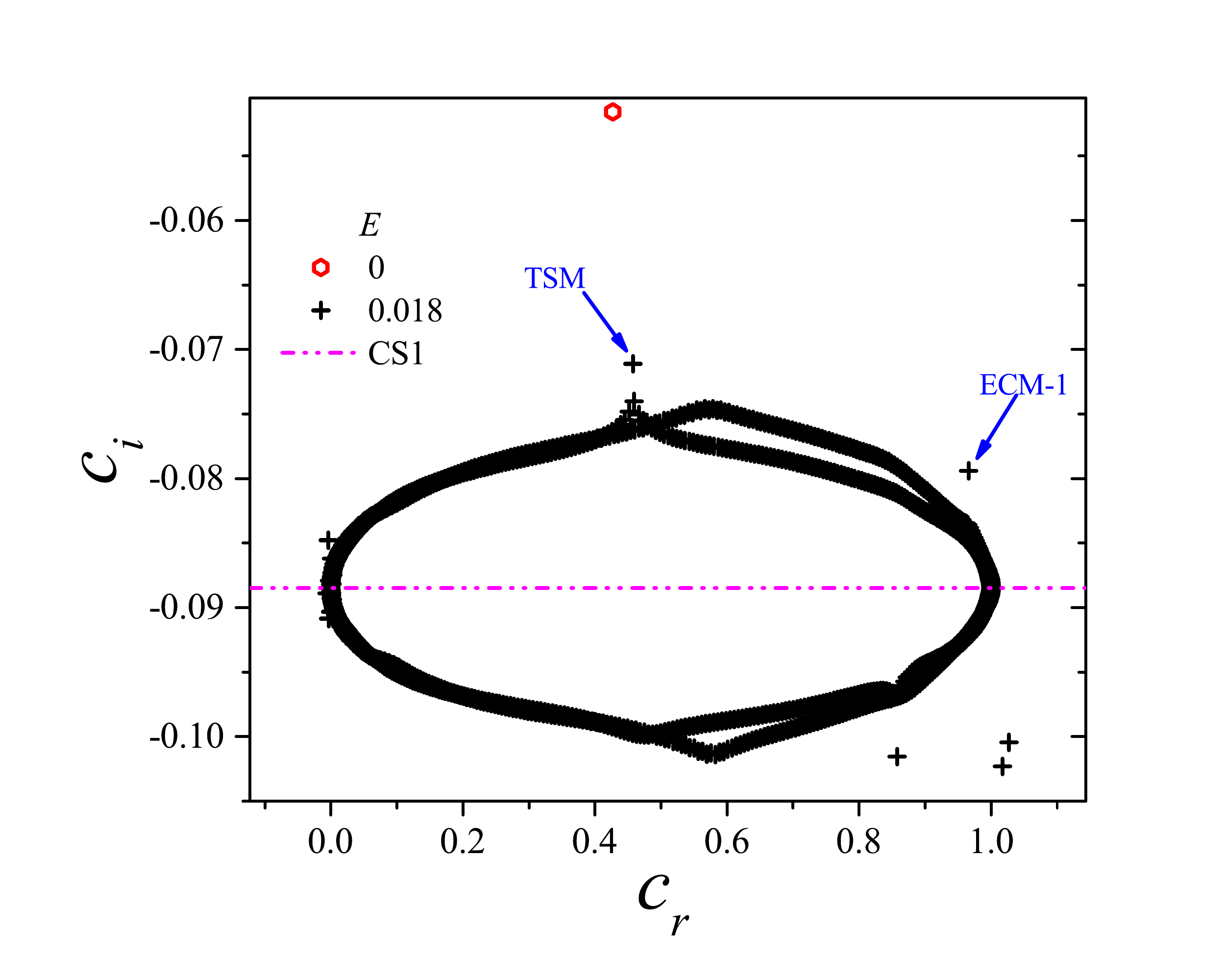}
		\caption{$ E = 0.018 $}
		\label{fig:Ep018_Re500_k0p4pi_B0p8}
	\end{subfigure}%
	~
	\begin{subfigure}[b]{0.5\textwidth}
		\centering
		\includegraphics[width=\textwidth]{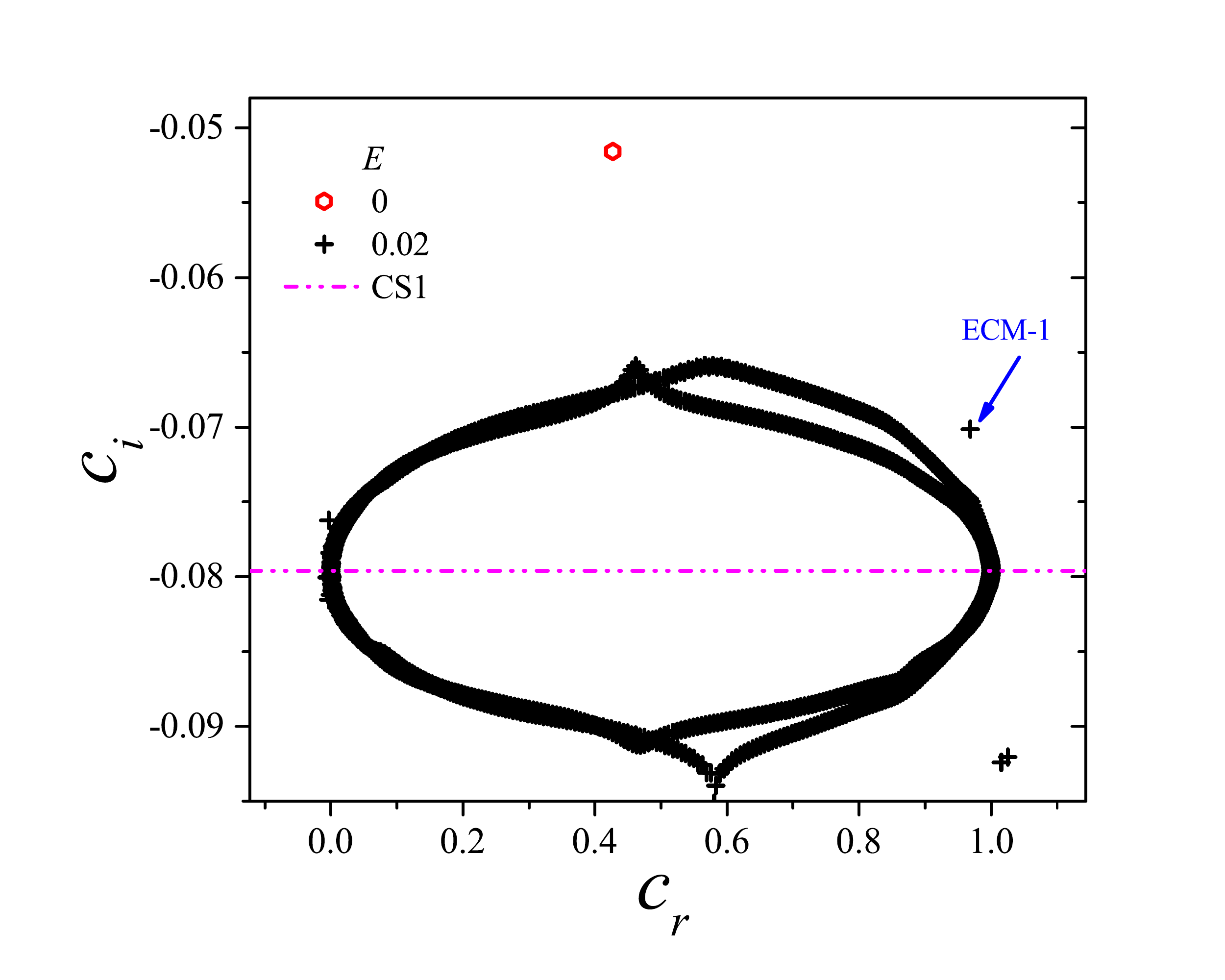}
		\caption{$ E = 0.02 $}
		\label{fig:Ep02_Re500_k0p4pi_B0p8}
	\end{subfigure}%

\caption{\small Eigenspectra of viscoelastic channel flow at the same $Re = 500$ and $\beta = 0.8$ as
in Fig.~\ref{fig:Re500_k2.5_B0.8} but with  $k = 0.4\pi$, and varying $E$. For $E < 0.02$, the elastically modified
	TS mode is the least stable, whereas  ECM-1 just emerges from CS1. However, for $E >0.02 $, the TS mode merges with CS1 and ECM-1 becomes the least stable mode dictating the stability of the system. The corresponding Newtonian eigenspectrum (with $E = 0$) is also shown for reference. In panels (c)--(f), only the region near the CS is magnified to illustrate the collapse and emergence of discrete modes from the CS.}
\label{fig:Re500_k0.4pi_B0.8}

\end{figure*}

\begin{figure}
        \centerline{\includegraphics[width=0.5\textwidth]{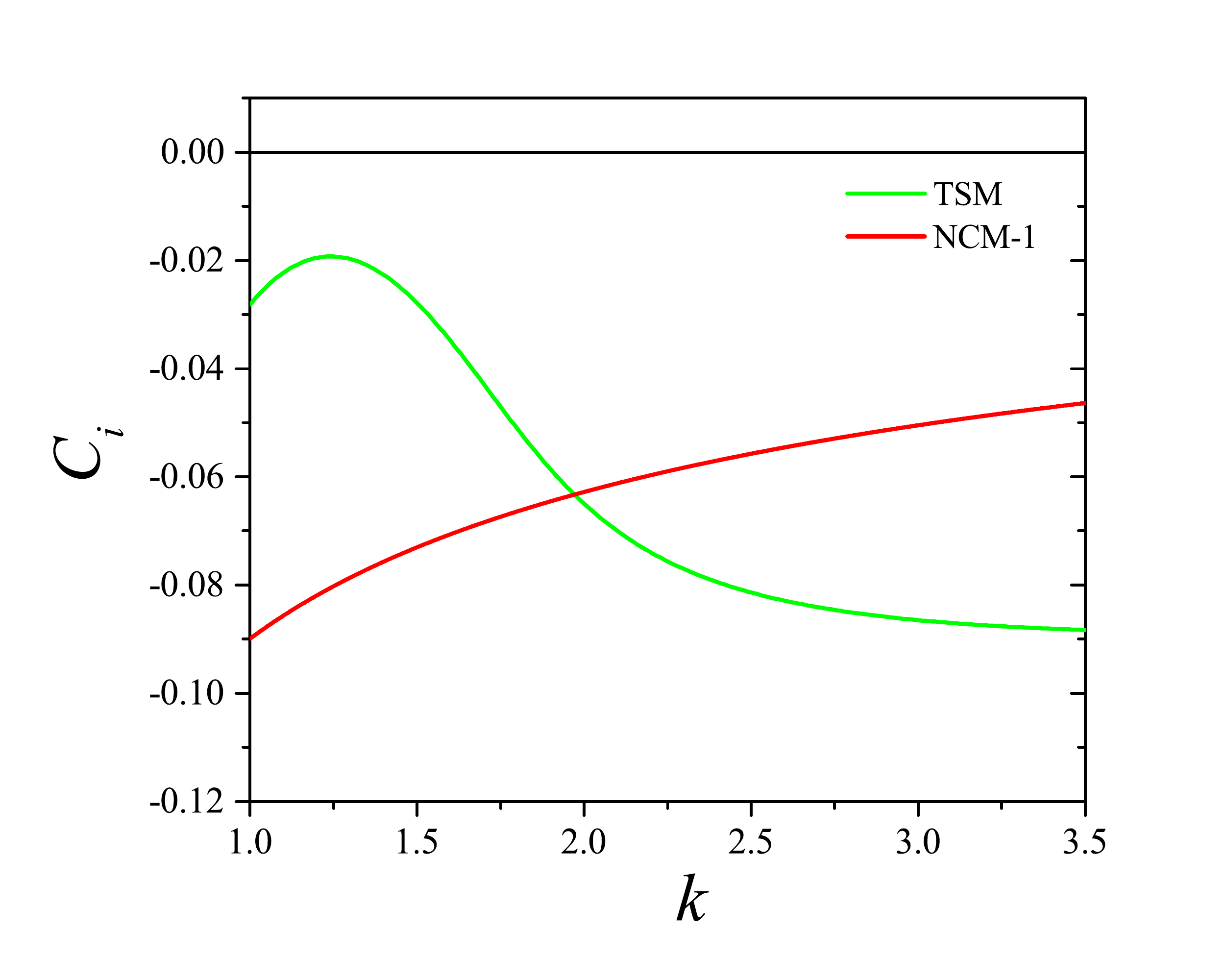}}
        \caption{\small Relative stability of center (NCM-1) and least-stable wall (TS) modes in Newtonian channel flow at $ Re=1500 $: variation of $c_i$ for these modes with  $k$. The wall mode is the least stable for $k < 2$, while the center mode becomes least stable for $ k \ge 2$.}
        \label{fig:Re1500_TS_vs_NCM}
\end{figure}

\begin{figure*}
	\centering
	\begin{subfigure}[b]{0.5\textwidth}
		\centering
		\includegraphics[width=\textwidth]{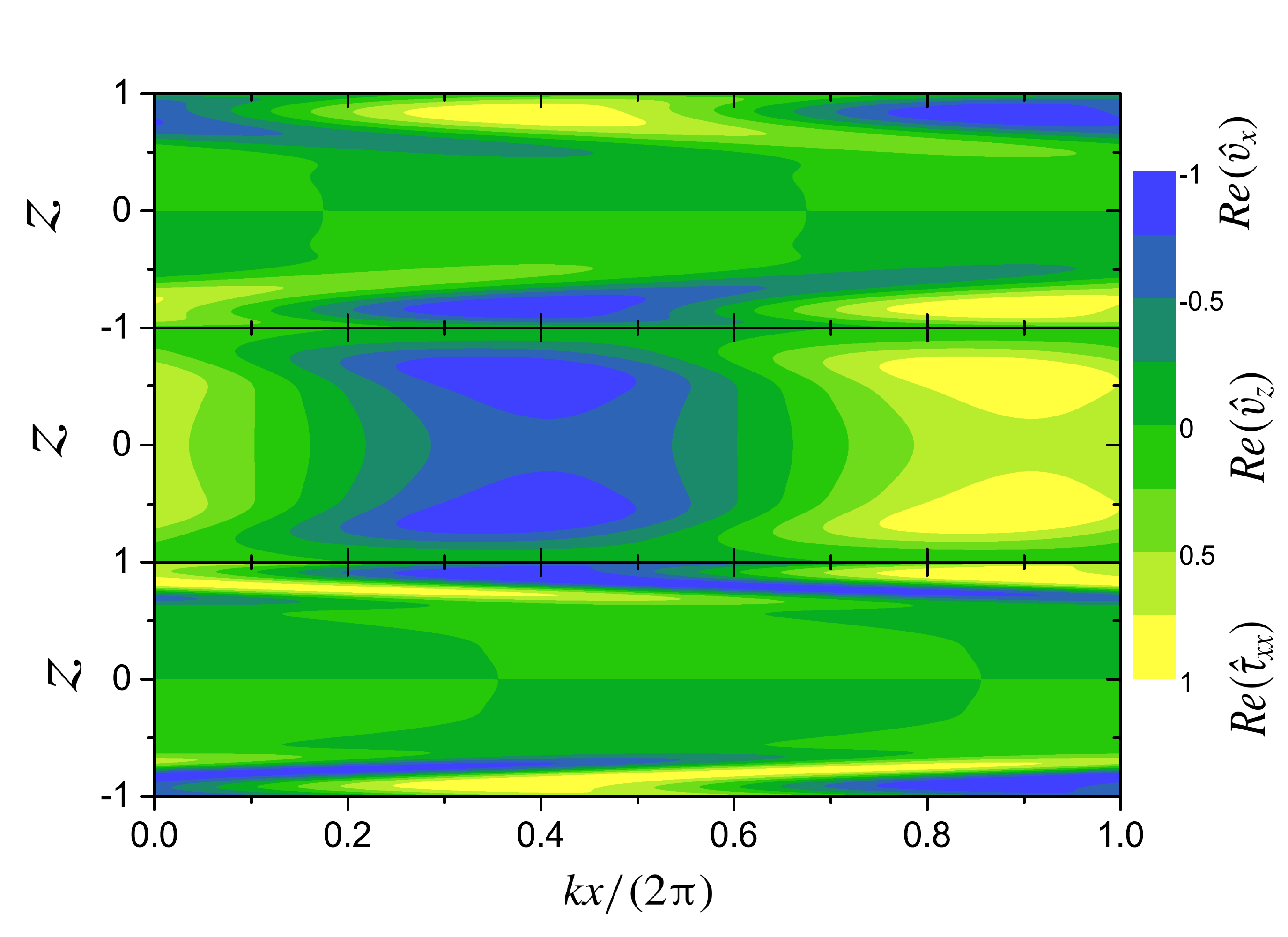}
		\caption{$ c = 0.484998 -0.10376i $}
		\label{fig:Cont_TSM-1}
	\end{subfigure}%
	~ 
	\begin{subfigure}[b]{0.5\textwidth}
		\centering
		\includegraphics[width=\textwidth]{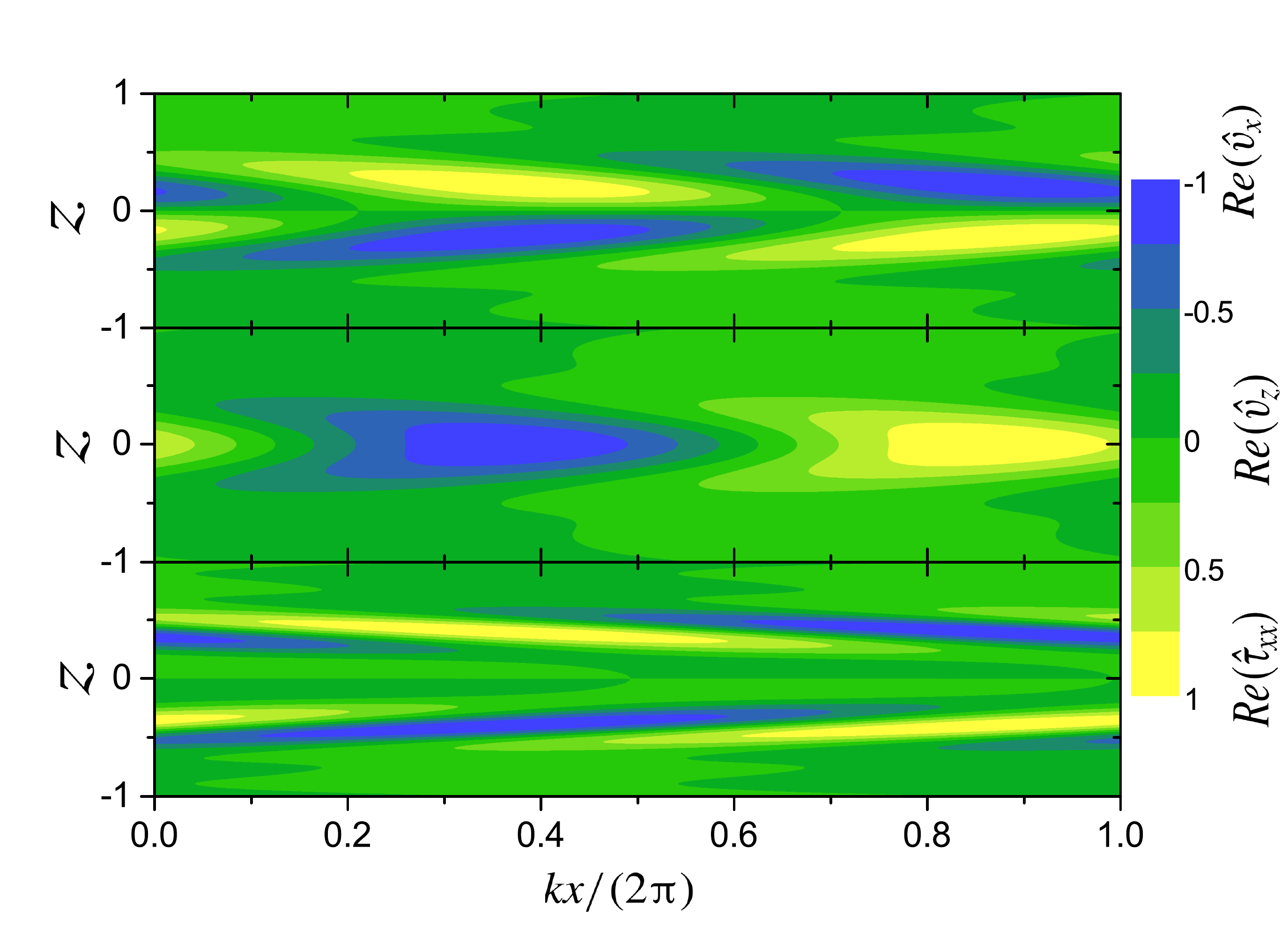}
		\caption{$ c =0.894151 -0.0865i $}
		\label{fig:Cont_NCM-1}
	\end{subfigure}
	\caption{\small Contours of the $v_x$, $v_z$, and $T_{xx}$ eigenfunctions for (a) wall (TS), and (b) center (NCM-1) modes in the $ x-z $ plane for $ Re=500, \beta=0.8, k=0.8\pi,  E=0.002$. }
	\label{fig:contours}
\end{figure*}

We have established above that the unstable ECM-1 in channel flow is not merely a continuation of the least stable Newtonian center mode (NCM-1), on account of their differing symmetries, but instead emerges out of  CS1 beyond a threshold $E$. In the present work, we propose that it is this unstable center mode that underlies the early transition to elastoinertial turbulence observed in both pipe and channel flow experiments, involving polymer solutions, discussed in Sec.~\ref{sec:Introduction}. 
In contrast, a recent DNS effort \citep{Shekar2019} has shown a resemblance between the phase-matched, ensemble-averaged structures of polymer stretch contours and the elastically-modified TS mode.
The authors carried out DNS for channel flow of a FENE-P fluid in the elasto-inertial turbulent regime ($Re = 1500$, $\beta = 0.97$; the Newtonian flow is turbulent at this $Re$),  and for $W$
in the range $0$--$50$, where the flow is linearly stable.
With increasing $W$, the simulations showed a reduction in drag from the Newtonian turbulent value,
eventually approaching the laminar value at $W \approx 10$,  suggesting complete relaminarization, in agreement with  observations  \citep{Choueiri2018}. For $W$ greater than $20$, 
the simulations again showed a weak increase in drag, and the authors attributed this mild increase to an
instability via a  two-dimensional non-linear mechanism. In this regime, simulation results showed very strong
and localized polymer stretch fluctuations similar to those  in the vicinity of the ‘critical
layer’ (the transverse location where the phase speed of the perturbation equals the base-flow velocity, in linear stability theory) of the elastically modified TS mode. Thus, the suggestion is that the fluctuating velocity field corresponding to the self-sustaining EIT state closely resembles the near-Newtonian velocity field of the TS (wall) mode for the small $E$'s under consideration ($0 < E < 0.03$), and that drives the polymer stretch, and the resulting large axial polymeric stresses, near the critical layer.

Thus, there are two qualitatively different mechanisms being put forward for transition (to elastoinertial turbulence) in viscoelastic channel flow, based on two different modes in the elasto-inertial spectrum:
the centermode (that has recently been shown, for a set of parameters, to continue subcritically to a novel EIT coherent structure; see \cite{page2020exact}), and the one advocated above by \cite{Shekar2019} based on the wall mode. A rigorous demonstration as to which mode would be dominant would require a weakly non-linear analysis leading to the determination of the first Landau coefficient; such an analysis, for the center mode, will be reported in a future communication.


\begin{figure*}
	\centering
		\includegraphics[width=0.5\textwidth]{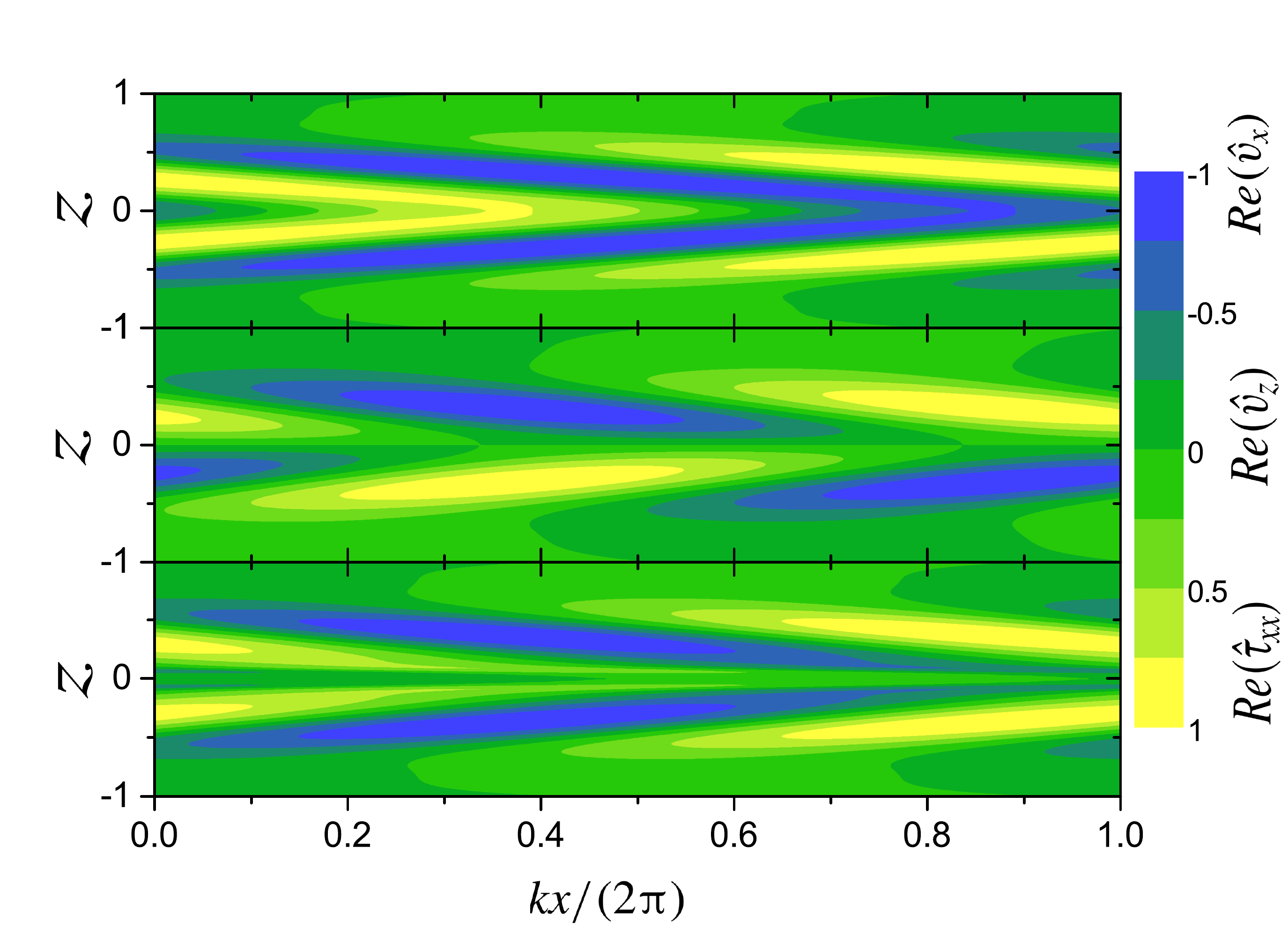}
\caption{\small Contours of $ v_x, v_z$ and $ T_{xx}$ for unstable (symmetric) center mode in the $ x-z $ plane for $ Re =500$, $\beta = 0.8$, $k=0.8\pi, E=0.12$. The unstable eigenvalue is $  c = 0.9995 +  2.3197 \times 10^{-4} i $.} 
\label{fig:contoursECM1}
\end{figure*} 
For the time being,
it is useful to examine, within the linear stability framework, the decay/growth rates of center (NCM-1 and ECM-1) and wall (TSM) modes as $E$ is varied (at fixed $Re$, $\beta$ and $k$), and demarcate the
$E$-intervals in which each of these modes is the most dominant one in the elasto-inertial spectrum.   
Figure~\ref{fig:Re1500_k0.4pi_B0.97} focuses on the relative stability of TSM and NCM-1  modes as $E$ is varied  (for the Oldroyd-B model), for $Re = 1500$, $k=0.4\pi$ and $\beta=0.97$, these parameter values being identical to those used by \cite{Shekar2019}  for the FENE-P model.  Recall from Sec.~\ref{ssec:origination_ECM} that, 
as $E$ is increased,  the NCM's merge with CS1, and new modes appear from it.
For $ 0 \leq E \leq 0.015$, which includes the range of $E$ considered by \cite{Shekar2019}, the elastically modified TS mode (i.e., TSM) is the least stable one  (see inset of Fig.~\ref{fig:E_0p015_Re1500}).  
For $E = 0.015$, NCM-1 has already collapsed onto CS1, and as $ E $ is increased  further to $0.02$, TSM also disappears into CS1 (Fig.~\ref{fig:E_0p02_Re1500}), and concomitantly new elasto-inertial center modes (ECM-3, 4; ECM-2 lies very close to the CS, and hence is not visible at this scale) appear from the lower side of CS1 (see inset of Fig.~\ref{fig:E_0p02_Re1500}). Although these new elasto-inertial center modes are not unstable at this parameter range, nonetheless, these are the least stable discrete modes at this value of $ E $. Importantly, there are no discrete modes above the CS for $0.02 < E < 0.35$, and thus the CS modes are the least stable in this range.
It is only at a much higher $E \approx 0.35$ that ECM-1 emerges above CS1. Subsequently, ECM-1 becomes unstable at $E \approx 0.4$, and thereby, dictates the stability for all higher $E$'s
(see inset of Figs.~\ref{fig:Ep35_Re1500} and \ref{fig:Ep4_Re1500}).

In Figs.~\ref{fig:Re500_k2.5_B0.8} and \ref{fig:Re500_k0.4pi_B0.8}, we investigate the relative stability of TSM and the center modes at a lower $Re = 500$, $\beta = 0.8$, and for two different  $k =0.8\pi$ and $k = 0.4\pi$ respectively. Surprisingly, for the larger $k$ (Figs.~\ref{fig:E0p002_Re500} and \ref{fig:E0p005_Re500}),  NCM-1 is less stable than TSM (red circles) even in the Newtonian limit. 
Figure~\ref{fig:E0p01_Re500} shows that TSM has already collapsed into CS1, while NCM-1 lies just above it, in contrast to the behaviour seen in Fig.~\ref{fig:E_0p015_Re1500}.
 As soon as both the TSM and NCM-1  merge into CS1, the new elasto-inertial center mode (ECM-1) emerges above  CS1  (Fig.~\ref{fig:E0p01_Re500}), eventually becoming unstable at higher $E$.   The spectra at the lower $k = 0.4\pi$ (Fig.~\ref{fig:Re500_k0.4pi_B0.8}) but at the same $Re$ and $\beta$ as in Fig.~\ref{fig:Re500_k2.5_B0.8}, however, show that the TS mode remains the least stable for $E < 0.02$ before merging into the CS.  The ECM-1 mode emerges above the CS for $E > 0.02$, as the least stable in the spectrum.

Thus, at sufficiently high $E$'s, the center mode ECM-1 is always the least stable/unstable mode in the elasto-inertial spectrum, 
but even for smaller $E$'s (where ECM-1 has not yet emerged from the CS), one  could have the original Newtonian center mode (NCM-1) be less stable than the wall mode (TSM) depending on $k$.
 In light of this, the relative stability of the wall (TS) and center (NCM) modes
in Newtonian channel flow at different  $ k $, for $Re = 1500$
(Figure~\ref{fig:Re1500_TS_vs_NCM}), reveals that increasing  $ k $ changes the relative stability of TS mode and NCM-1, with the latter being the least stable for $ k\geq 2$. 
 An important inference from Figs.~\ref{fig:Re1500_k0.4pi_B0.97}--\ref{fig:Re1500_TS_vs_NCM} is that, even in parameter regimes where channel flow is linearly stable, there are intervals where the center mode (ECM-1 or NCM-1) or the continuous spectrum is the least stable, and are likely to influence the (subcritical) nonlinear dynamics of the transition.  Indeed, in Fig.~\ref{fig:Re1500_k0.4pi_B0.97} alone, there is a significant range of $E$ for which there is no discrete mode above the CS, a fact that might be attributed to the near-unity $\beta$ ($=0.97$) considered. Thus, the connection between the least stable wall (TSM) mode in Newtonian channel flow and the (2D) elasto-inertial turbulent structures noted by \cite{Shekar2019} may not be generic in the $Re$-$E$-$\beta$ space.  We return to the question involving the relative magnitudes of the growth rates of th different modes in Sec.~\ref{ssec:ECS_vs_EIT}.
%
%

The contours corresponding to the velocity ($\hat{v}_x(x,z), \hat{v}_z(x,z)$) and streamwise component of the polymeric stress ($\hat{\tau}_{xx}(x,z)$) eigenfunctions of the TS and NCM-1 modes are shown in Figs.~\ref{fig:Cont_TSM-1} and \ref{fig:Cont_NCM-1}. While both these modes are antisymmetric about the channel centerline, the structures of the TS mode are confined near the wall, while the NCM-1 structures display maximum variation away from the walls; in both cases, the confinement is prominent in the tangential velocity and  streamwise polymer stress eigenfunctions. For the small $ E $ considered, the velocity contours are quite  reminiscent of their Newtonian counterparts (not shown). For the higher $E = 0.12$, the elasto-inertial center mode has become unstable, and the 2D contour plots of $ \hat{v}_x(x,z), \hat{v}_z(x,z)$ and  $\hat{\tau}_{xx}(x,z)$ corresponding to this mode is shown in Fig.~\ref{fig:contoursECM1}. In contrast to the TS mode, ECM-1 is a symmetric mode, with both the velocity and stress perturbations being relatively less confined. The proposal of the center mode underlying EIT dynamics seems to have
support from the recent finding of a novel EIT structure \citep{page2020exact} 
that bifurcates subcritically from the center-mode instability, and has 
the same symmetry about the channel centerline.

\section{Neutral stability curves}
\label{sec:Neutral stability curves for unstable elasto-inertial center mode}

\begin{figure*}
	\centering
	\begin{subfigure}[b]{0.5\textwidth}
		\centering
		\includegraphics[width=\textwidth]{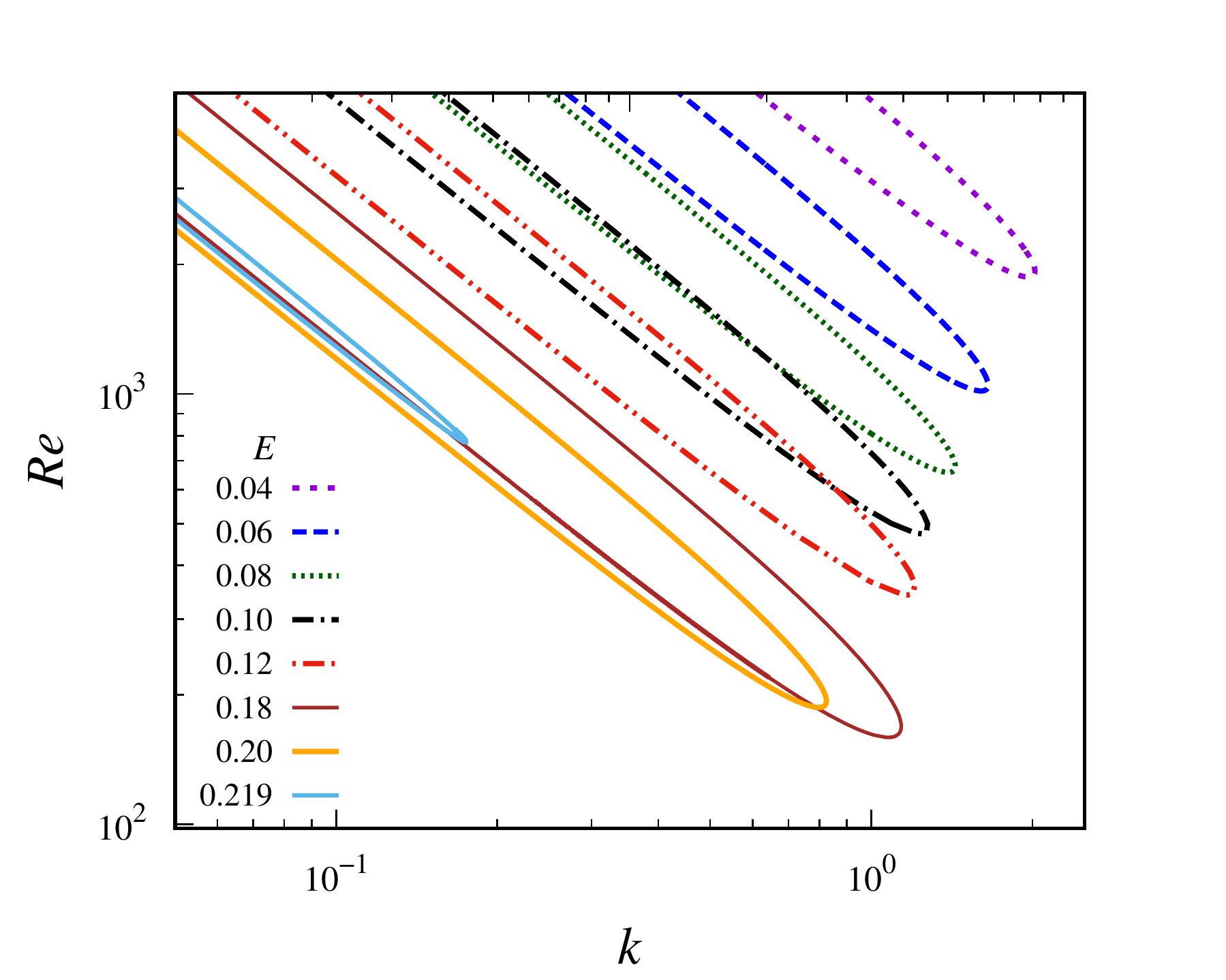}
		\caption{$\beta =0.65$}
		\label{fig:NC_Re-k-Beta_p8_log-log}
	\end{subfigure}%
	~ 
	\begin{subfigure}[b]{0.5\textwidth}
		\centering
		\includegraphics[width=\textwidth]{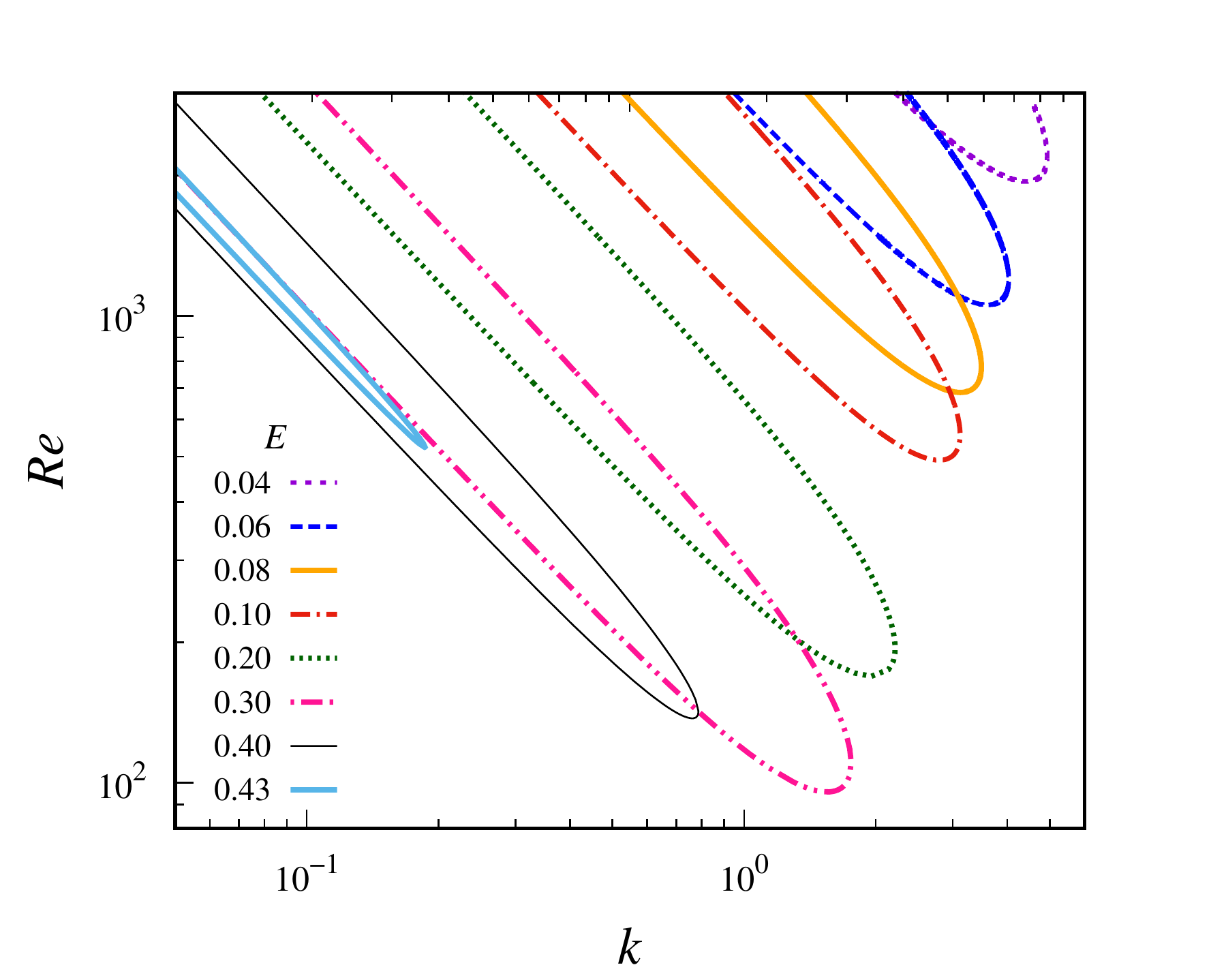}
		\caption{$\beta =0.8$}
		\label{fig:NC_Re-k-Beta_p9_log-log}
	\end{subfigure}
    \caption{\small Neutral stability curves in  the  $Re$--$k$ plane for different $ E $ at: (a) $ \beta =0.65$, and (b) $ \beta =0.8$.}
	\label{fig:NC_log-log_Re-k}
\end{figure*}

\begin{figure*}
	\centering
		~ 
	\begin{subfigure}[b]{0.5\textwidth}
		\centering
		\includegraphics[width=\textwidth]{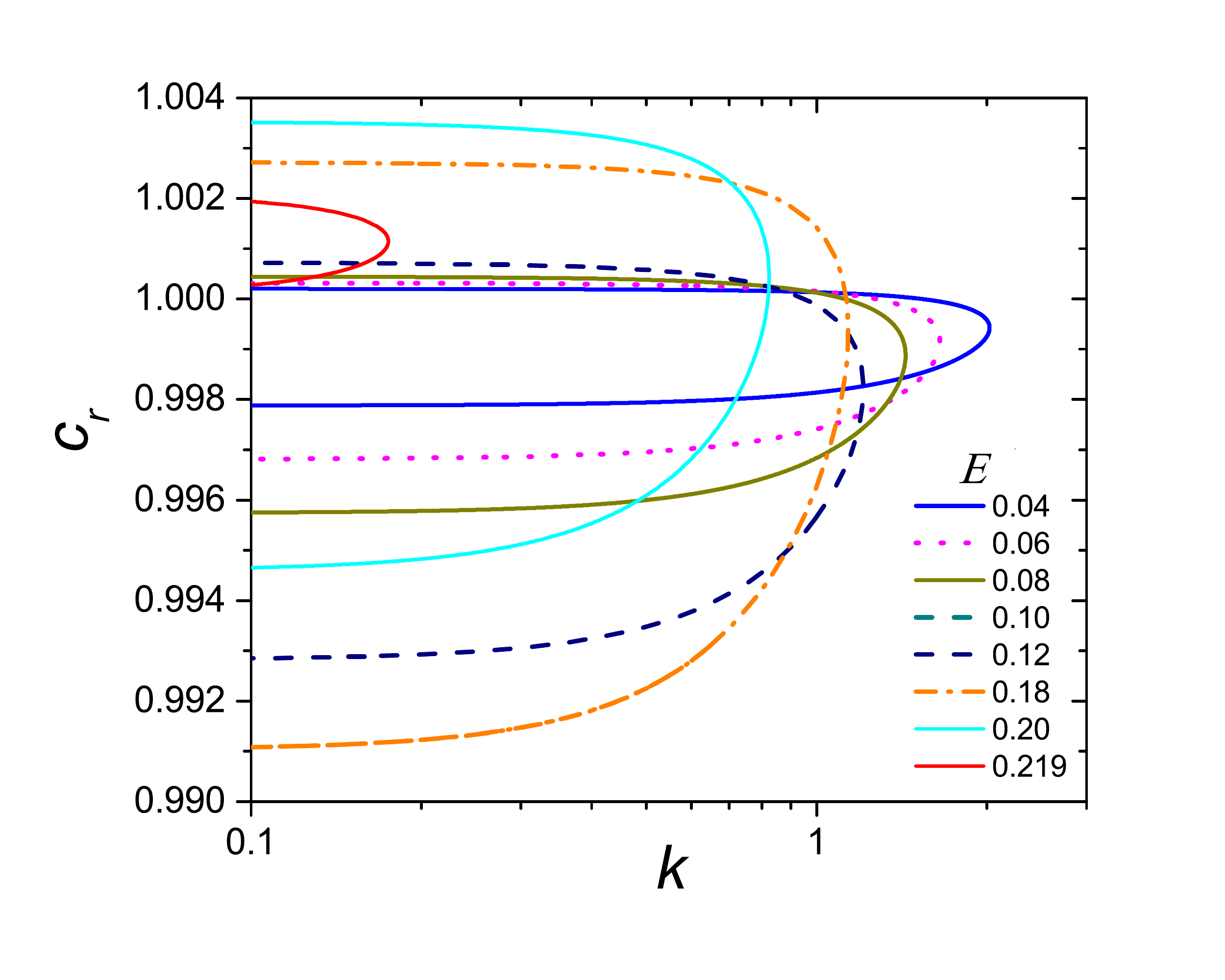}
		\caption{$\beta =0.65$}
		\label{fig:NC_cr-k-Beta_p8_log-log}
	\end{subfigure}%
	~ 
	\begin{subfigure}[b]{0.5\textwidth}
		\centering
		\includegraphics[width=\textwidth]{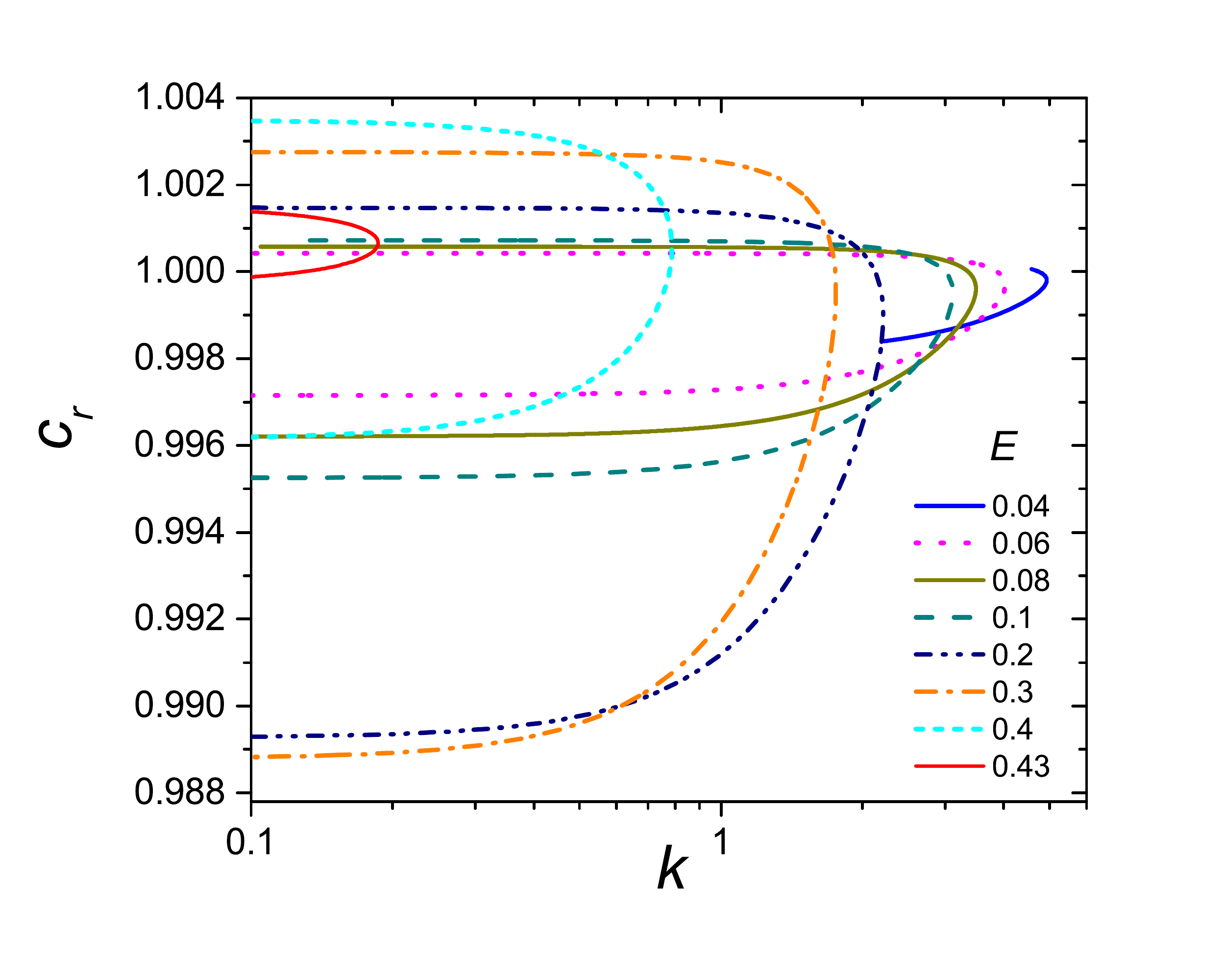}
		\caption{$\beta =0.8$}
		\label{fig:NC_cr-k-Beta_p9_log-log}
	\end{subfigure}
\caption{\small The variation of phase speed as a function of $ k $ corresponding to the  neutral stability curves at different $ E $ shown in Fig.~\ref{fig:NC_log-log_Re-k} for $\beta = 0.65, 0.8$.}
	\label{fig:NC_log-log_cr-k}
    \end{figure*}
    
In Fig.~\ref{fig:NC_log-log_Re-k}, we present neutral stability curves (at fixed $\beta$, and with varying $E$) for the channel-flow center mode, which are in the form of loops in the $Re-k$ plane, with the region inside each neutral loop being unstable.  
For $k \ll 1$, we find $Re \sim k^{-1}$ along both the upper and lower branches 
of the loops for $\beta = 0.65$ and $0.8$ in Fig.~\ref{fig:NC_log-log_Re-k}, and for other $\beta$'s (not shown). In contrast, for pipe flow, this scaling is valid along the lower branch \citep[regardless of $\beta$; see][]{Piyush_2018,chaudharyetal_2020}, with the upper branch conforming to this scaling only for $\beta < 0.9$. While the neutral loops for channel flow shown in Fig.~\ref{fig:NC_log-log_Re-k} remain single-lobed for any $\beta$, those for pipe flow display instead a two-lobed structure for $\beta > 0.9$ \citep{chaudharyetal_2020}. For a fixed $\beta$ and $E$, the
critical Reynolds number $(Re_c)$ is the minimum of the $Re$-$k$ curve, and from Figs.~\ref{fig:NC_Re-k-Beta_p8_log-log} and \ref{fig:NC_Re-k-Beta_p9_log-log}, is seen to
exhibit a non-monotonic variation with increasing $E$.  For sufficiently high $E$, increasing $E$ is accompanied by a shrinking of the $Re$-$k$ loop, leading to its disappearance beyond a critical $E$. Thus,  similar to pipe flow \citep{Piyush_2018,chaudharyetal_2020}, the center-mode instability ceases to exist at sufficiently high $E$. The phase speeds corresponding to the neutral curves in Fig.~\ref{fig:NC_log-log_Re-k} are shown in Fig.~\ref{fig:NC_log-log_cr-k}, and remain close to unity, with the range of $c_r$'s, for any given $k$, again exhibiting a non-monotonic dependence on $E$. Importantly, and in sharp contrast to pipe flow, the phase speeds of the neutral modes along the upper branch exceed unity.


\begin{figure*}
	\centering
	\begin{subfigure}[b]{0.5\textwidth}
		\centering
		\includegraphics[width=\textwidth]{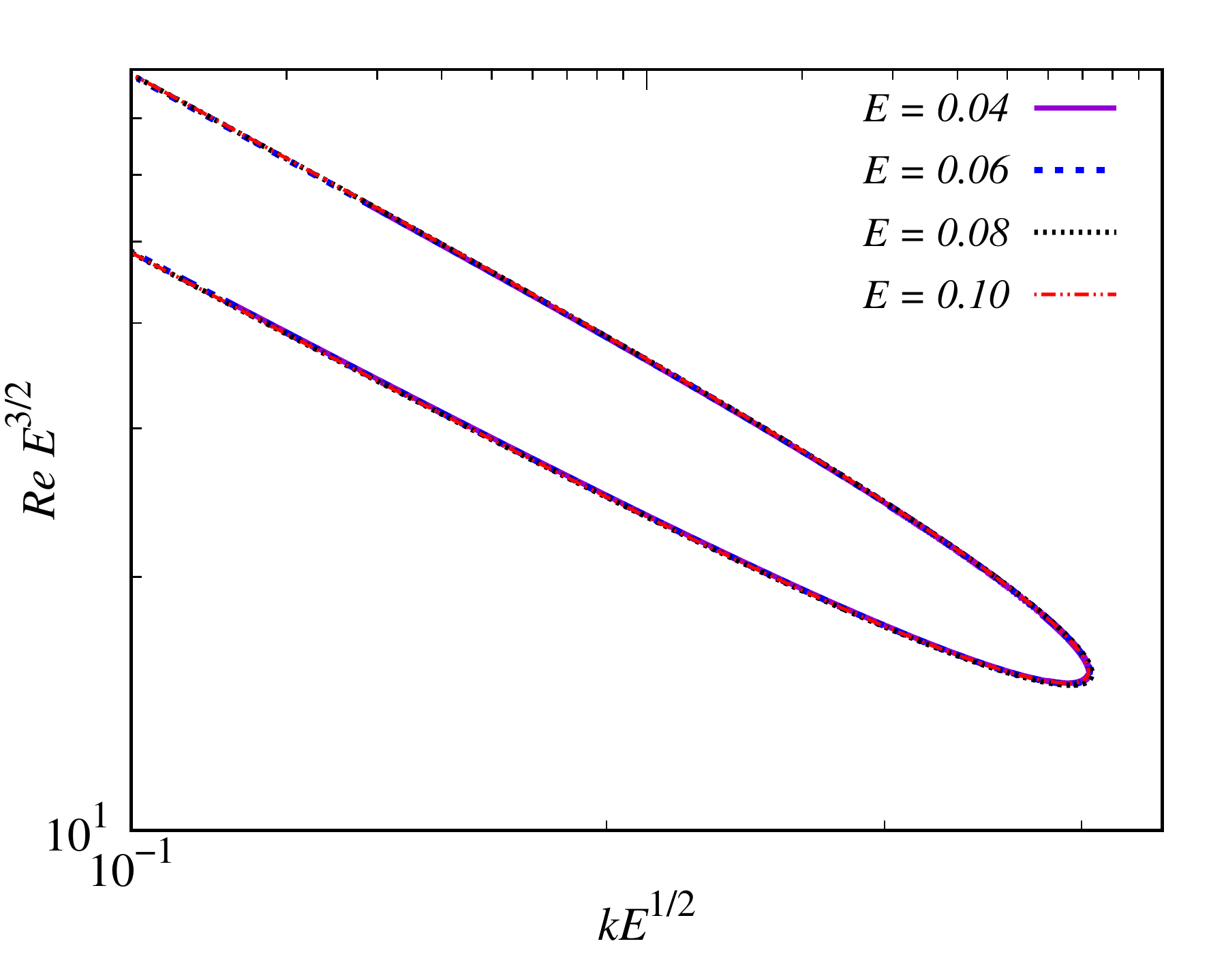}
		\caption{$\beta =0.65$}
		\label{fig:NC_collapsing_Re-k_Beta0p65}
	\end{subfigure}%
	~ 
	\begin{subfigure}[b]{0.5\textwidth}
		\centering
		\includegraphics[width=\textwidth]{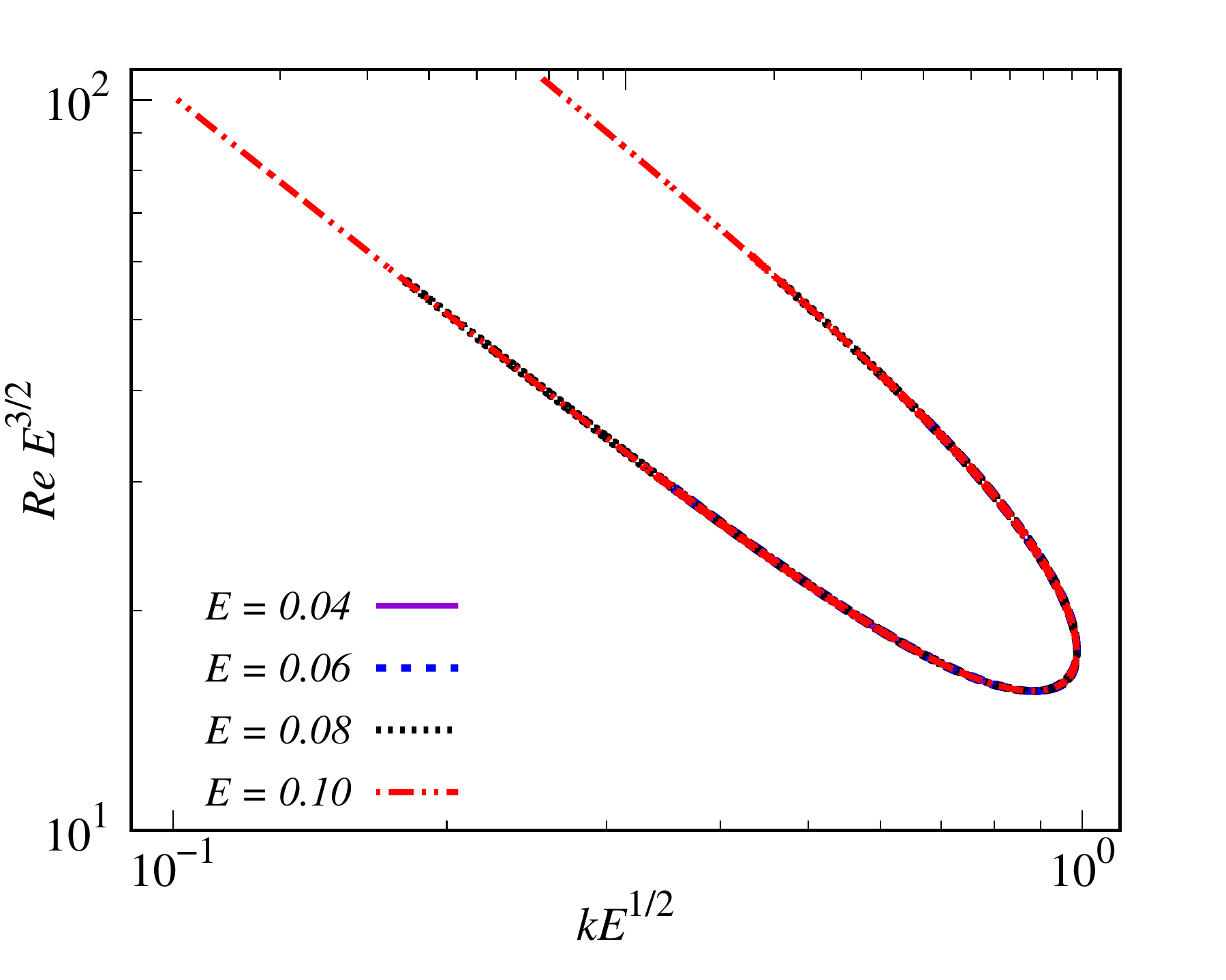}
		\caption{$\beta =0.8$}
		\label{fig:NC_collapsing_Re-k_Beta0p8}
	\end{subfigure}
	~ 
	\begin{subfigure}[b]{0.5\textwidth}
		\centering
		\includegraphics[width=\textwidth]{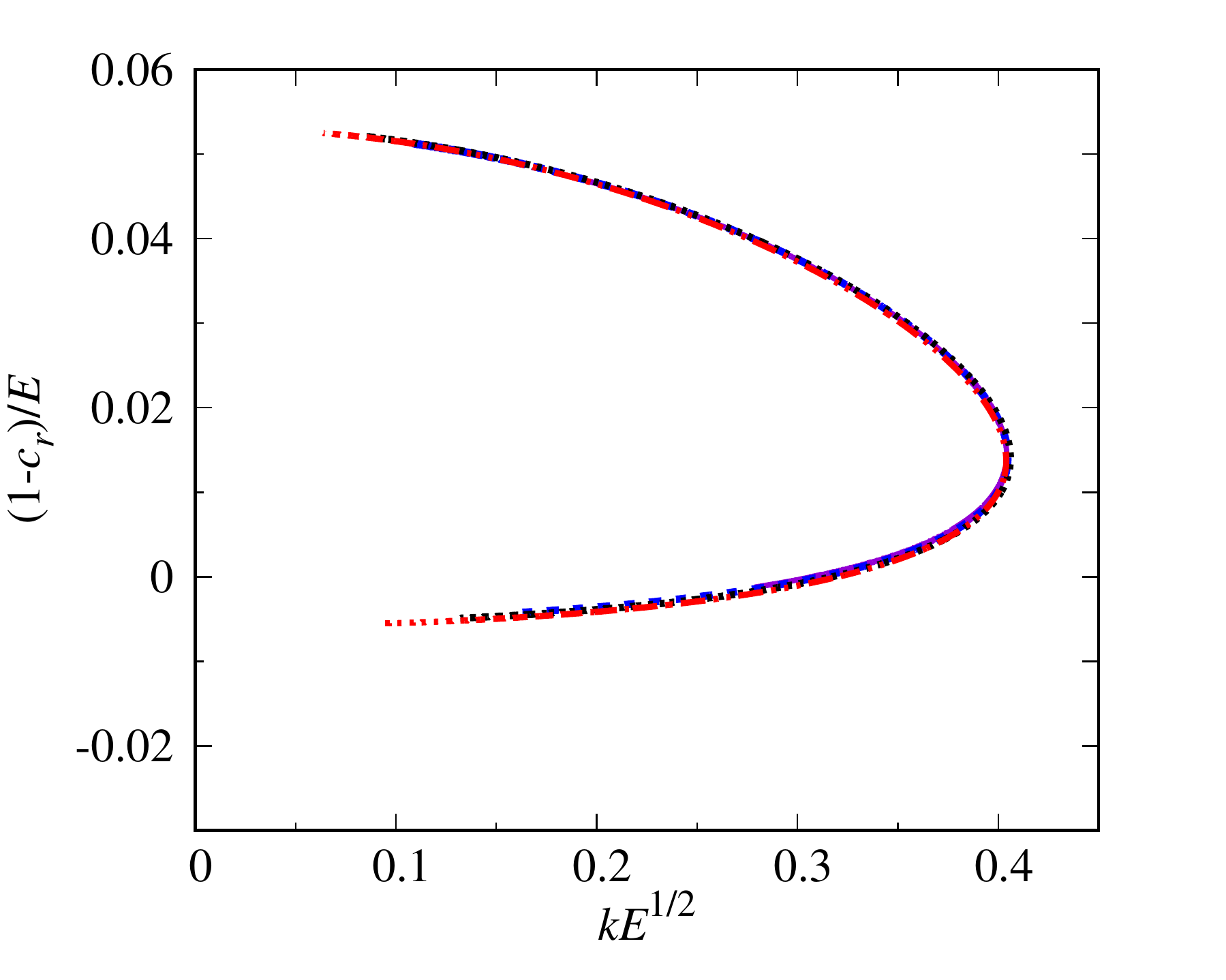}
		\caption{$\beta =0.65$}
		\label{fig:NC_collapsing_cr-k_Beta0p65}
	\end{subfigure}%
	~ 
	\begin{subfigure}[b]{0.5\textwidth}
		\centering
		\includegraphics[width=\textwidth]{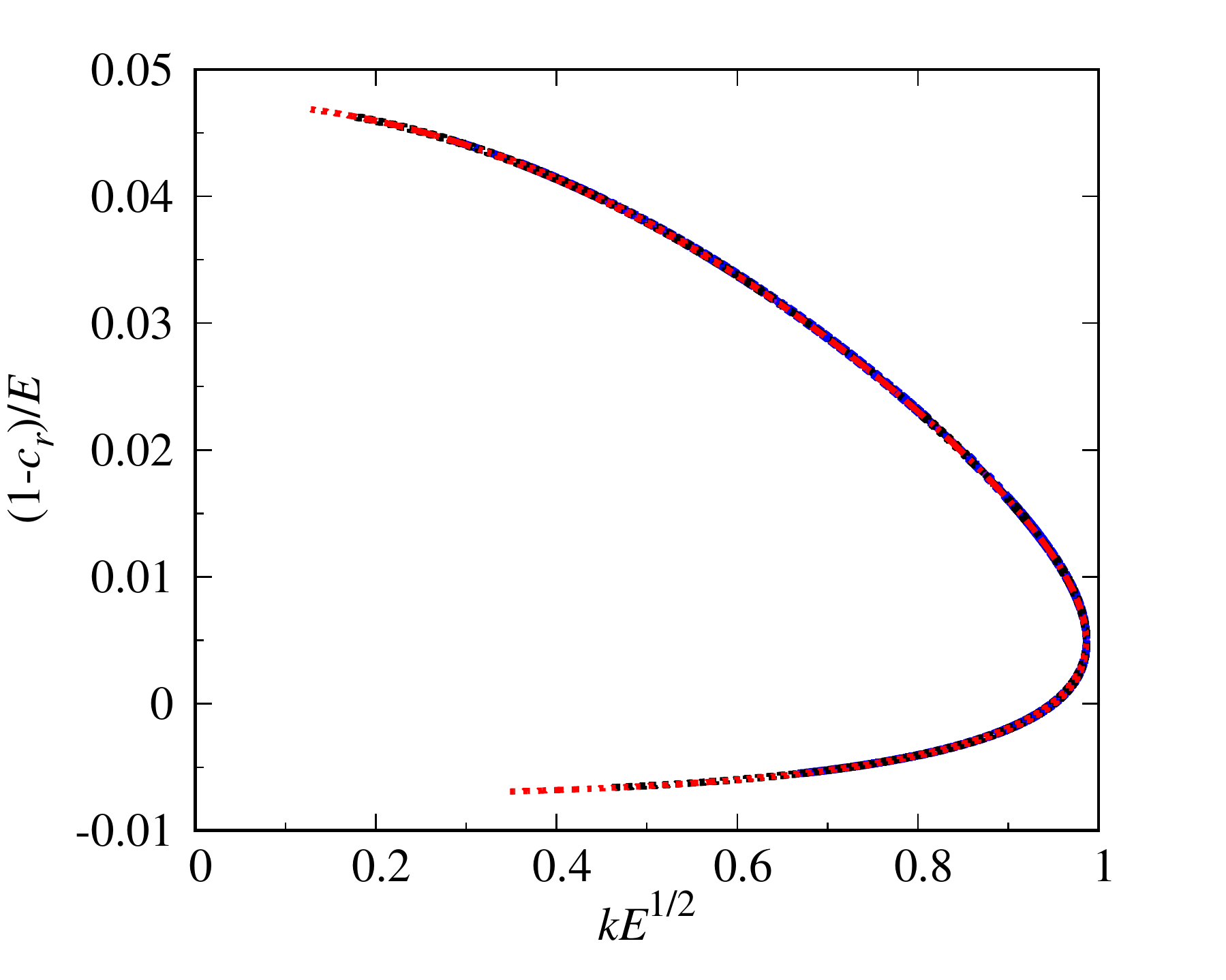}
		\caption{$\beta =0.8$}
		\label{fig:NC_collapsing_cr-k_Beta0p8}
	\end{subfigure}
	\caption{\small Collapse of the neutral curves for small $ E $ and for $\beta$'s shown in Figs.~\ref{fig:NC_log-log_Re-k}
	and \ref{fig:NC_log-log_cr-k}. Panels (a) and (b): rescaled neutral stability curves in the $Re E^{3/2}$--$k E^{1/2}$ plane; panels (c) and (d): corresponding rescaled phase speeds  in the $ (1-c_r)/E$--$kE^{1/2}$ plane.}
	\label{fig:Data_collapsing}
\end{figure*}

\subsection{Scaled neutral curves}
\label{ssec:Data collapse}

Figures~\ref{fig:NC_log-log_Re-k} and \ref{fig:NC_log-log_cr-k} are strongly suggestive of a collapse of  neutral curves and the corresponding phase speeds, especially for the smaller $E$'s, on suitable rescaling. 
Figures~\ref{fig:NC_collapsing_Re-k_Beta0p65} and \ref{fig:NC_collapsing_Re-k_Beta0p8} show a collapse of the different small-$E$ neutral loops onto a single master curve in the $Re E^{3/2}$--$k E^{1/2}$ plane, for the $\beta$'s chosen in the aforementioned figures, implying that the threshold Reynolds number diverges as $E^{-3/2}$ as one approaches the Newtonian limit $E = 0$. In Figs.~\ref{fig:NC_collapsing_cr-k_Beta0p65} and
\ref{fig:NC_collapsing_cr-k_Beta0p8},  the phase speeds along the neutral curve exhibit a similar collapse when plotted as $(1-c_r)/E$ vs. $kE^{1/2}$, suggesting that $(1-c_r) \sim O(E)$ along the neutral curve.
 A similar collapse was also reported for pipe flow \citep{Piyush_2018,chaudharyetal_2020}.
An alternate route to the Newtonian limit, that of $\beta$ approaching unity for a fixed $E$, also
appears to yield a collapse of the neutral curves when plotted in terms of $ Re[E(1-\beta)]^{3/2} $ and $ k[E(1-\beta)]^{1/2} $, in the limit $ [E(1-\beta)]\ll 1$ (Fig.~\ref{fig:NC_Double_scaling}). However, this collapse is not as perfect as the one obtained above for small $E$, even in the limit $\beta \rightarrow 1$. In particular, the upper branch of the $Re$--$k$ curves collapses very well for $\beta \approx 0.99$, but the collapse is not perfect in the lower branches and near the minimum of the neutral curves.
Figure \ref{fig:Dual Scaling} shows the rescaled critical Reynolds number, $Re_c E^{3/2}$, and the corresponding rescaled critical wavenumber,  $k_c E^{1/2}$, as a function of $(1-\beta)$. This plot suggests that 
$Re_c$ and $k_c$ begin to approach the scalings $Re_c \propto (E(1-\beta))^{-3/2}$, $k_c \propto (E(1-\beta))^{-1/2}$ only for $\beta \approx 0.99$.

\begin{figure*}
	\centering
	\begin{subfigure}[b]{0.5\textwidth}
		\centering
		\includegraphics[width=\textwidth]{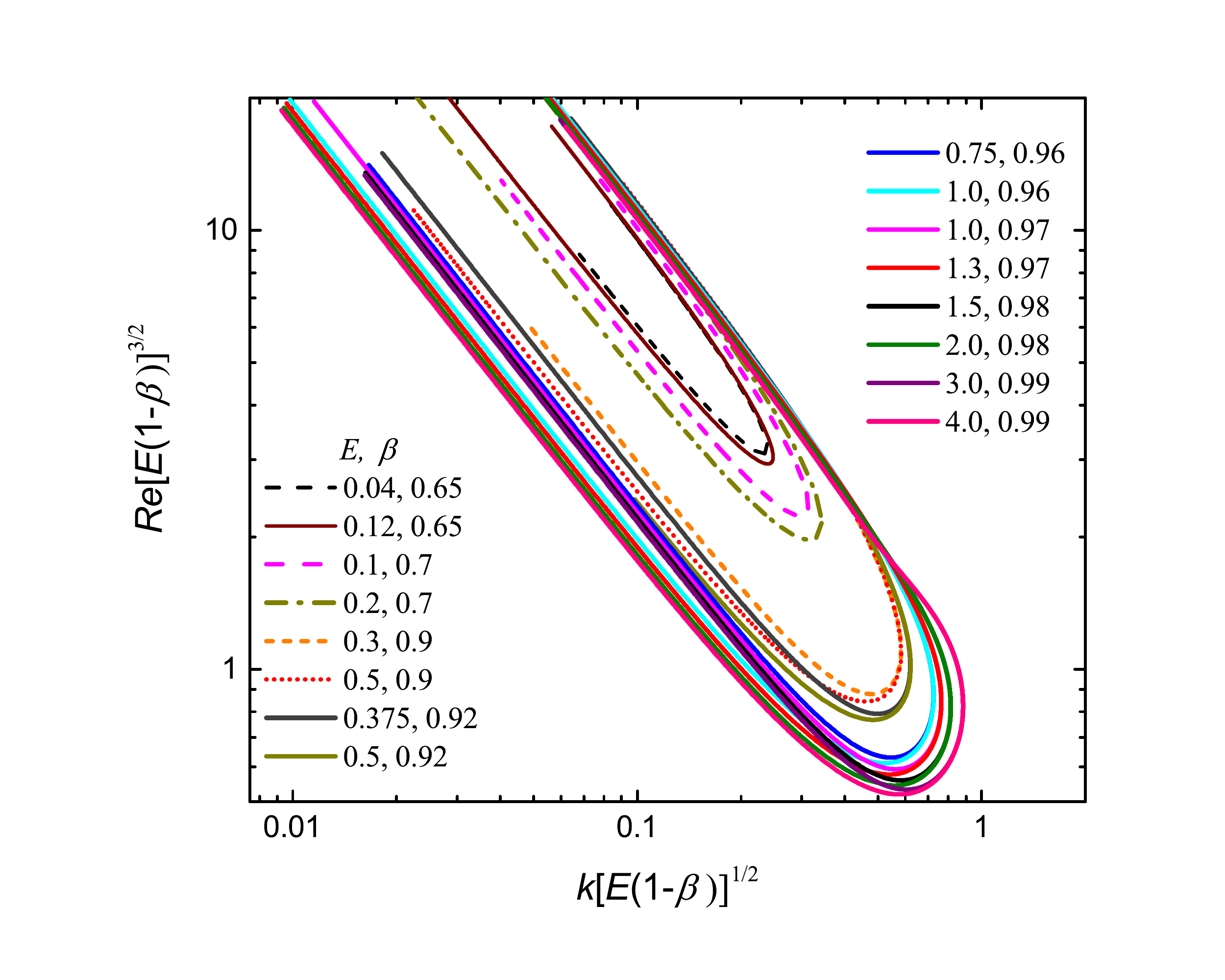}
		\caption{Rescaled neutral curves}
		\label{fig:NC_Double_scaling}
	\end{subfigure}%
	~ 
	\begin{subfigure}[b]{0.5\textwidth}
		\centering
		\includegraphics[width=\textwidth]{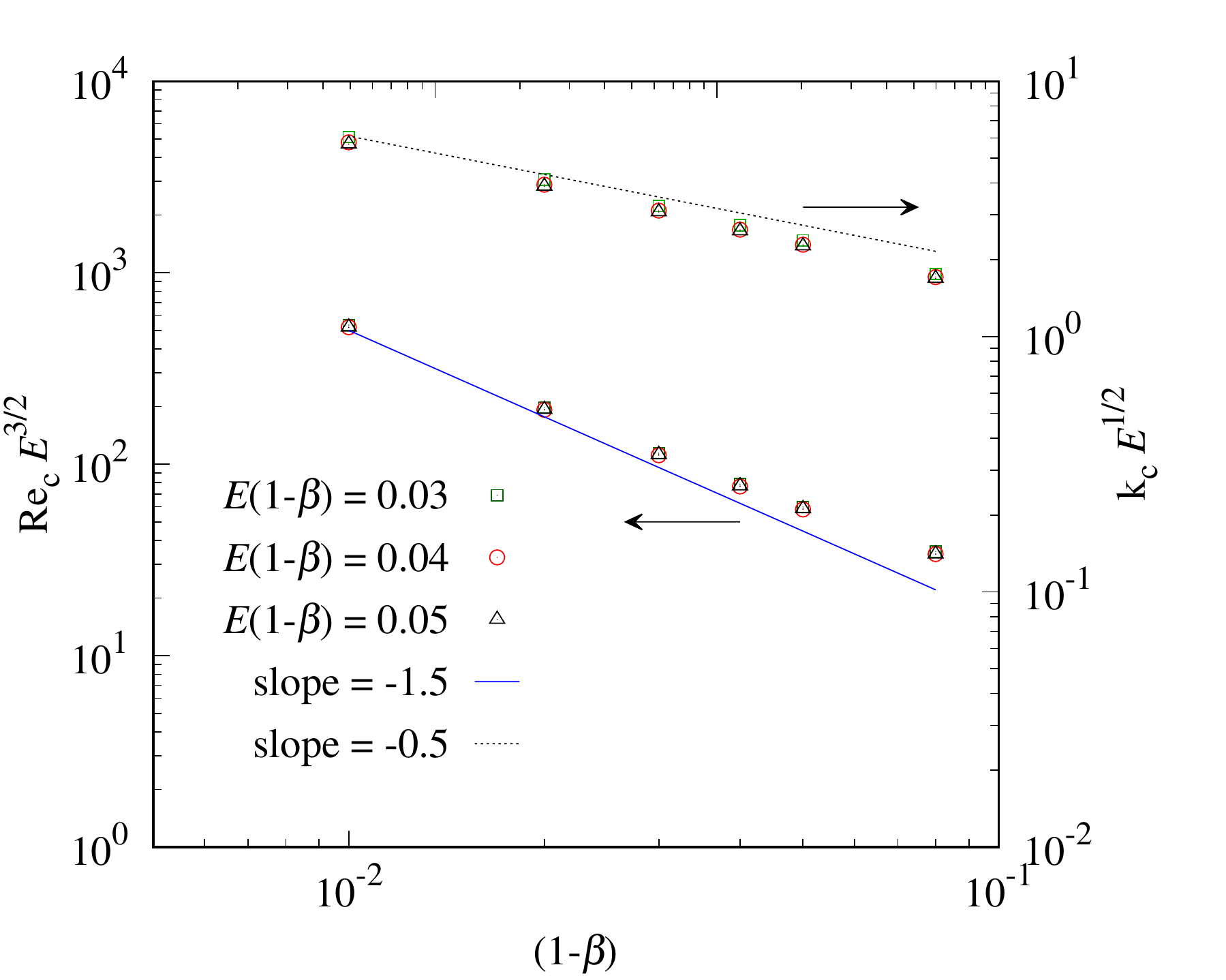}
		\caption{Critical parameters}
		\label{fig:Dual Scaling}
	\end{subfigure}
	\caption{\small Collapse in the limit $(1-\beta) \ll 1$ and $E$ fixed: In panel~(a), neutral stability curves at different $E$ and $\beta$ plotted in terms of the scaled Reynolds number $ Re[E(1-\beta)^{-3/2}]$ and wavenumber  $k[E(1-\beta)]^{-1/2}$. For $ \beta \rightarrow 1 $, the rescaled neutral curves exhibit a data collapse. In panel~(b), rescaled critical parameters at different $E$ and $\beta$ plotted as $Re_c E^{3/2}$, $k_c E^{1/2}$ vs. $(1-\beta)$ fall on lines of slopes $-3/2$ and $-1/2$ respectively, indicating again that, $Re_c \propto [E(1-\beta)]^{-3/2}$ and $k_c \propto [E(1-\beta)]^{-1/2}$.}
	\label{fig:doublescalingbetaunity}
\end{figure*}

%
%

\begin{figure*}
    \centering
    \begin{subfigure}[b]{0.5\textwidth}
        \centering
        \includegraphics[width=\textwidth]{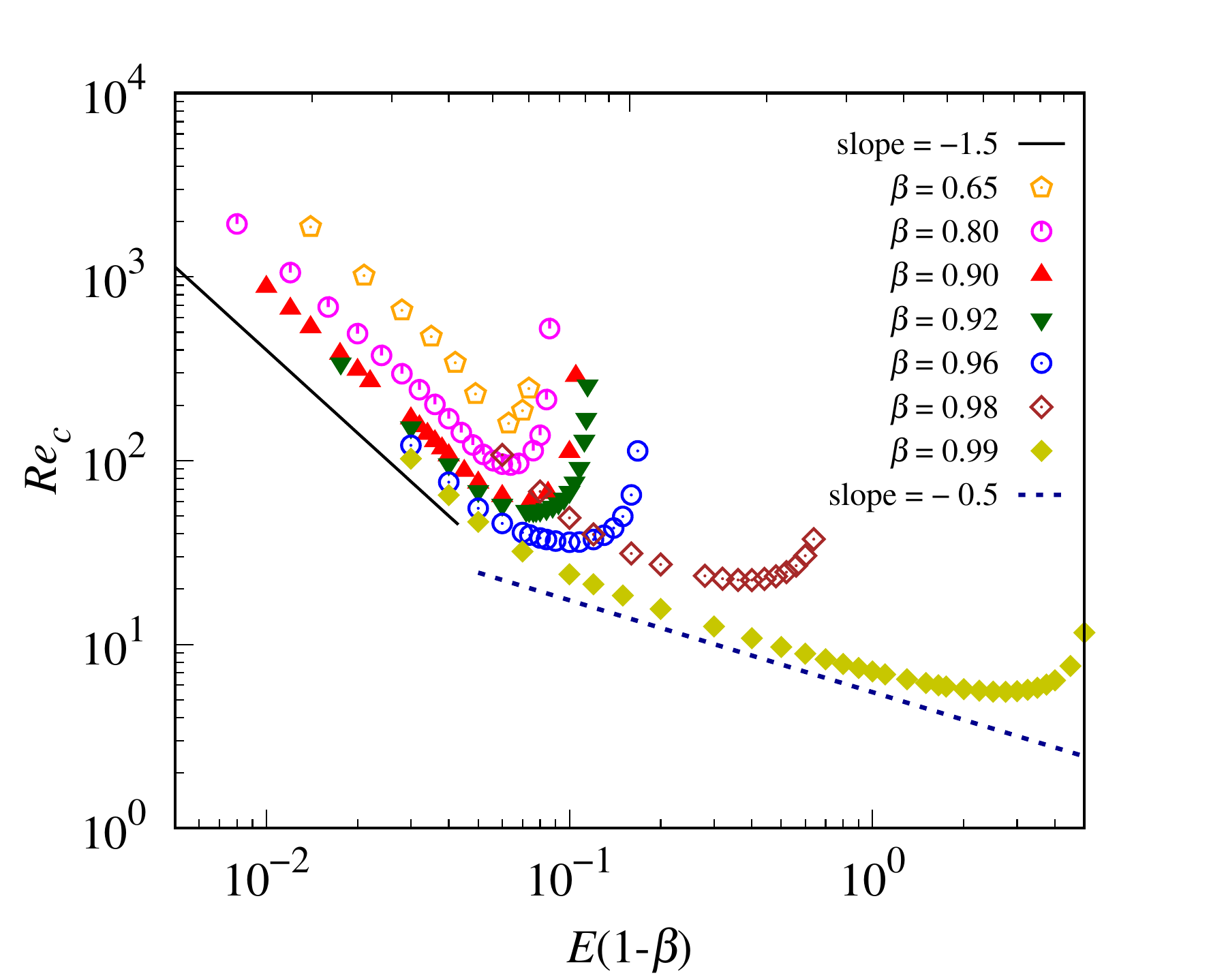}
        \caption{}
        \label{fig:NC_Scaling_RevsE}
    \end{subfigure}~
    \begin{subfigure}[b]{0.5\textwidth}
	\centering
	\includegraphics[width=\textwidth]{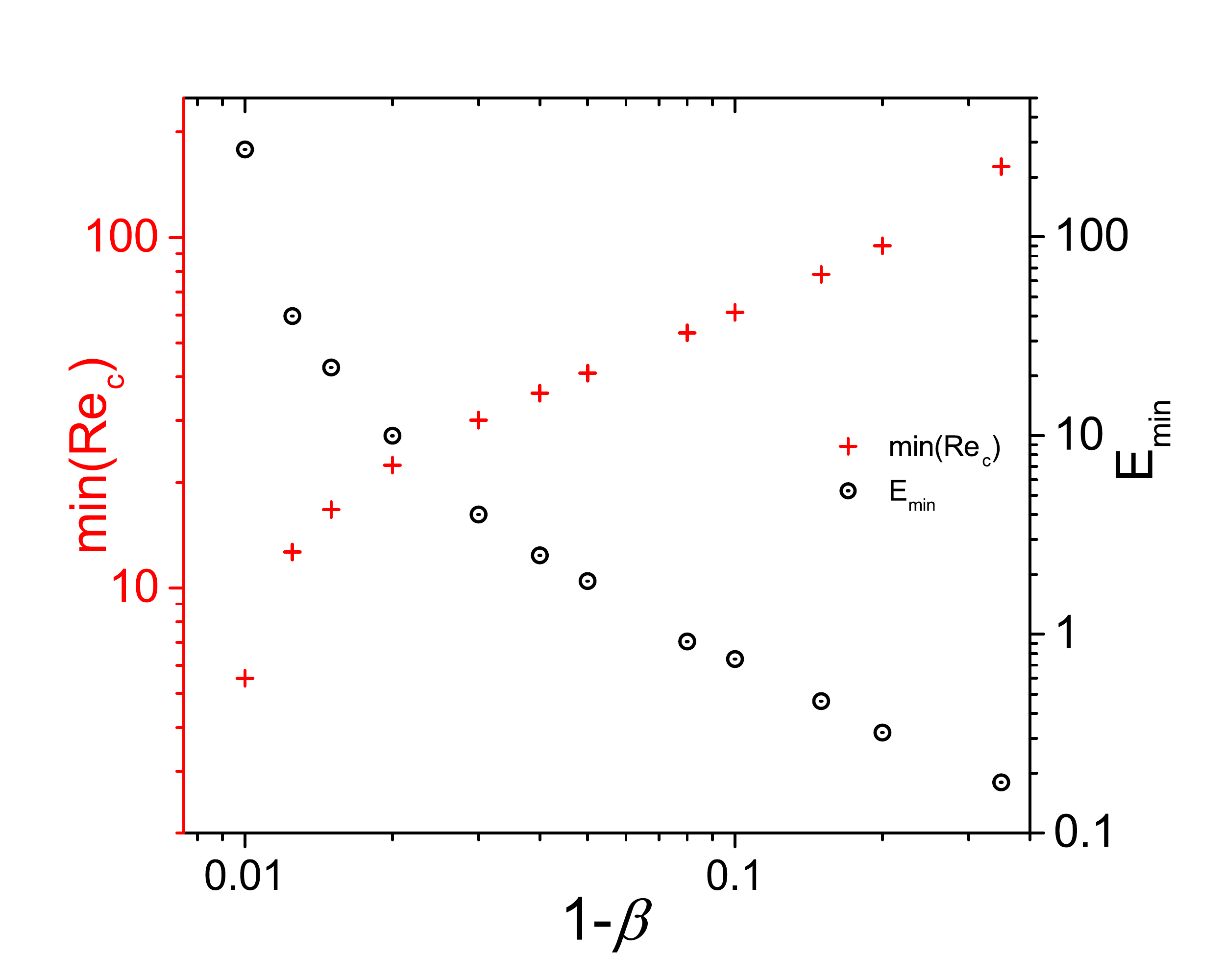}
	\caption{}
	\label{fig:Re_min-Beta-E}
    \end{subfigure} ~
    \begin{subfigure}[b]{0.5\textwidth}
        \centering
        \includegraphics[width=\textwidth]{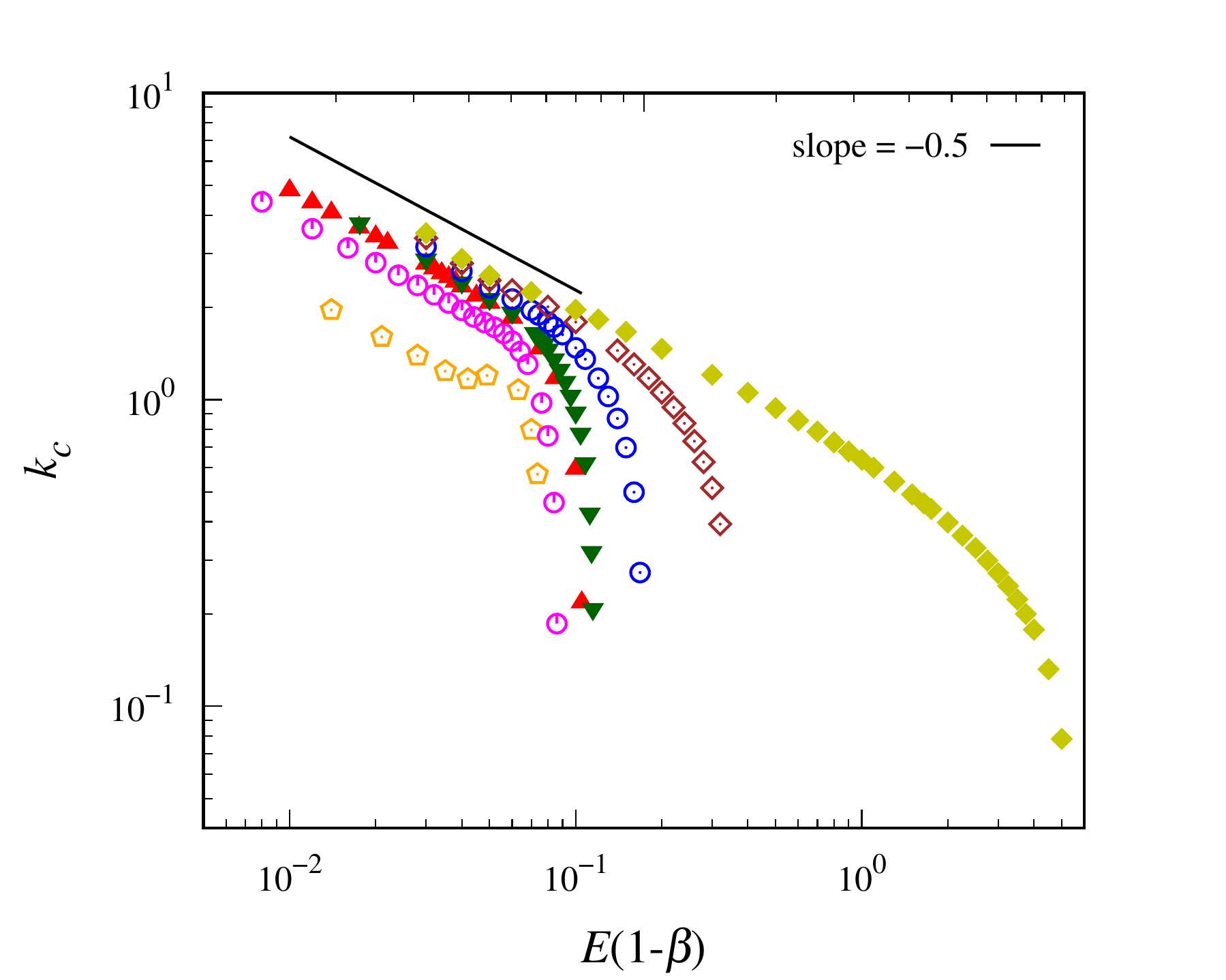}
        \caption{}
          \label{fig:NC_Scaling_kvsE}
    \end{subfigure}~
    \begin{subfigure}[b]{0.5\textwidth}
	\centering
	\includegraphics[width=\textwidth]{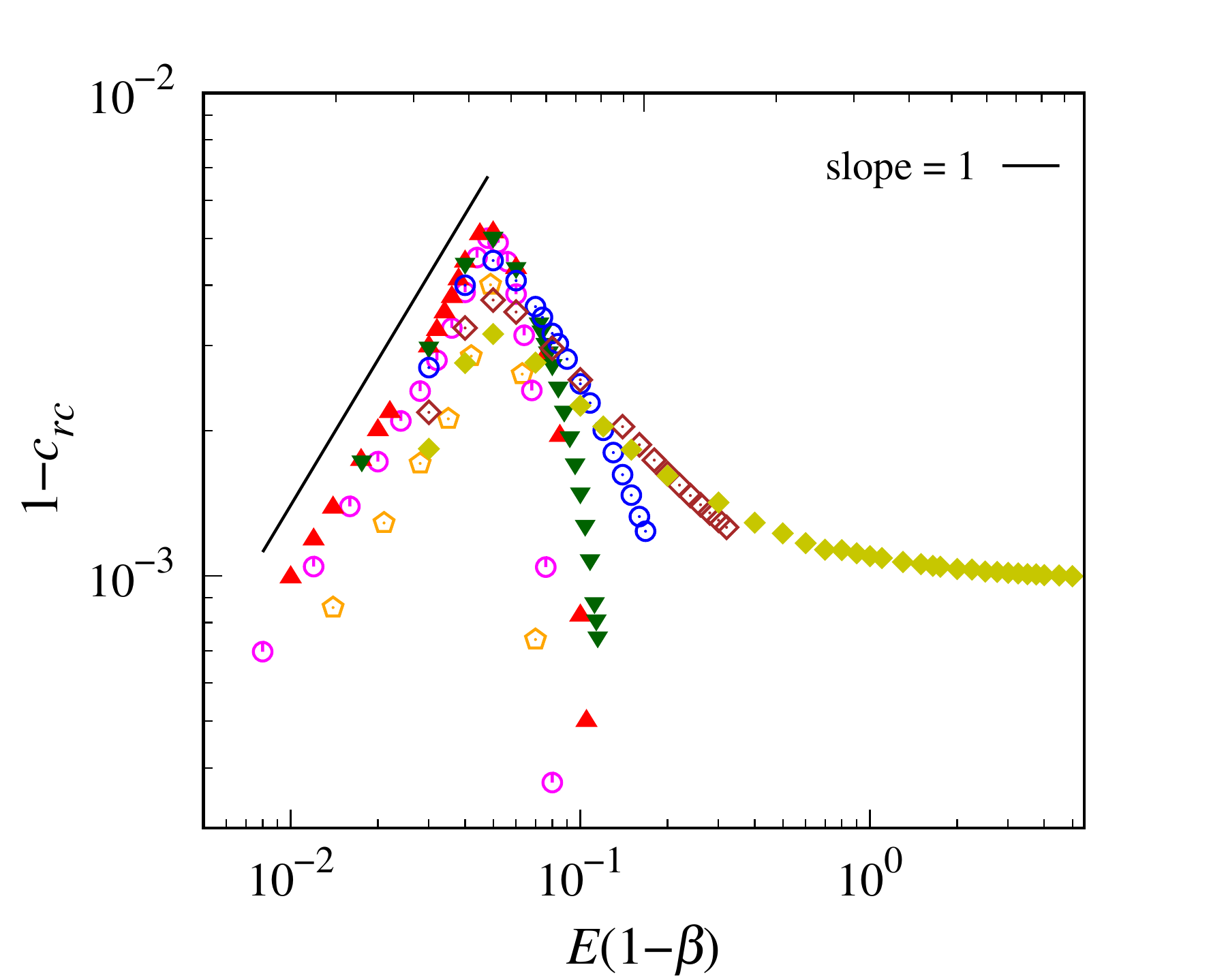}
	\caption{}
	\label{fig:NC_Scaling_crvsE}
    \end{subfigure}%
\caption{\small (a) Variation of critical parameters with $ E (1-\beta) $: (a) The critical Reynolds number  scales as $Re_c \propto [E(1-\beta)]^{-3/2}$ for $E(1-\beta) \ll 1$; (b) the minimum $Re_c $ in panel~(a) and the corresponding $E_{min}$; (c) critical wavenumber $ k_c \propto [E(1-\beta)]^{-1/2}$; and (d) phase speed, $ (1- c_r) \propto [E(1-\beta)] $. As shown in panel~(b), the center-mode instability persists in channel flow  up to $ Re \approx 5$ for very high $ E \sim 10^4  $ and for $\beta \approx 0.99 $. } 
\label{fig:Scaling_Re_and_k_cr_vsE}
\end{figure*}

\subsection{Critical parameters and Scalings}
\label{ssec:Critical parameters and Scalings}
The critical parameters ($ Re_c, k_c$ and $ c_{rc} $) are plotted as a function of $ E(1-\beta) $ in Fig.~\ref{fig:Scaling_Re_and_k_cr_vsE}.  The variation of $Re_c$ (Fig.~\ref{fig:NC_Scaling_RevsE}) is non-monotonic with $E(1-\beta)$, with $Re_c$ scaling as $(E(1-\beta))^{-3/2}$ for $E(1-\beta) \ll 1$, but showing a nearly vertical rise beyond a threshold $E$, denoted $E_{min}$, in a manner very similar to pipe flow \citep{Piyush_2018,chaudharyetal_2020}. 
A similar non-monotonic behaviour of $ Re_c $ with $ E $ has been obtained for elasto-inertial wall mode instabilities in plane Poiseuille flow of Oldroyd-B  \citep{sadanandan_sureshkumar_2002,Brandi2019} and FENE-P \citep{Zaki2013} fluids. However, since wall modes in channel flow are strongly stabilized by solvent viscous effects, the minima  in $Re_c - E $ curves  shift towards  higher $ Re_c $ with increase in  $ \beta $ for a fixed $ E $ \citep[see, for example, Fig.~1a of][]{sadanandan_sureshkumar_2002}.  In stark contrast, for the unstable center modes (Fig.~\ref{fig:NC_Scaling_RevsE}),  the $Re_c$'s shift towards lower values as $\beta$ approaches unity, thereby illustrating the contrasting roles played by solvent viscous effects on the center- and wall-mode instabilities.
Figure~\ref{fig:Re_min-Beta-E}  further reinforces the effect of $\beta$ by showing the variation of the minimum $Re_c$ (obtained from Fig.~\ref{fig:NC_Scaling_RevsE}) and the corresponding $E_{min}$ with $(1-\beta)$.
 Unlike pipe flow, where the center-mode instability ceases
to exist below a $Re_c \approx 60$, the instability in channel flow persists down to  $Re_c \sim O(1)$  for $ \beta \rightarrow 1$, albeit at very high $E$.
Figure~\ref{fig:NC_Scaling_kvsE} shows that the critical wavenumber scales as $ k_c \propto [E(1-\beta)]^{-1/2}$ for $E (1-\beta) \ll 1$,  while Fig.~\ref{fig:NC_Scaling_crvsE} shows that 
the critical phase speed scales as  ($ 1-c_r $) $ \propto $ $[E(1-\beta)]$, both similar to pipe flow.


Similar to the collapse of the neutral curves for $E \ll 1$, a collapse is also exhibited by the eigenfunctions when plotted using a suitably rescaled wall-normal coordinate for $Re \gg 1$, $E \ll 1$.  In this regard, there are two possible asymptotic regimes: one in which ($k, \beta$) are fixed and $Re$ and $E$ are varied so as to remain in the unstable region, and the other in which 
$\beta$ is fixed, and 
 the eigenfunctions are tracked along different sets of critical parameters
 ($Re_c, k_c$) for different $E$. For the latter case, Fig.~\ref{fig:Eigenfunctions_critical_Re_and_k} shows
 that the tangential and normal
velocity  eigenfunctions are increasingly localized in the vicinity of the channel centerline, 
within a boundary layer of thickness of $O(Re^{-1/3 })$; the $Re$-dependence of this boundary layer thickness may be obtained using a scaling analysis, as outlined in \cite{chaudharyetal_2020}.
Instead, if one considers a fixed $k$, and the limit $ Re,W \rightarrow \infty $, such that the ratio  $W/Re^{1/2} \sim O(1)$ in order to be in the unstable region, the  eigenfunctions become localized in a  boundary layer of thickness of $\mathcal{O}(Re^{-1/4})$  in the vicinity of channel centerline.


\begin{figure*}
    \centering
    \begin{subfigure}[b]{0.5\textwidth}
        \centering
        \includegraphics[width=\textwidth]{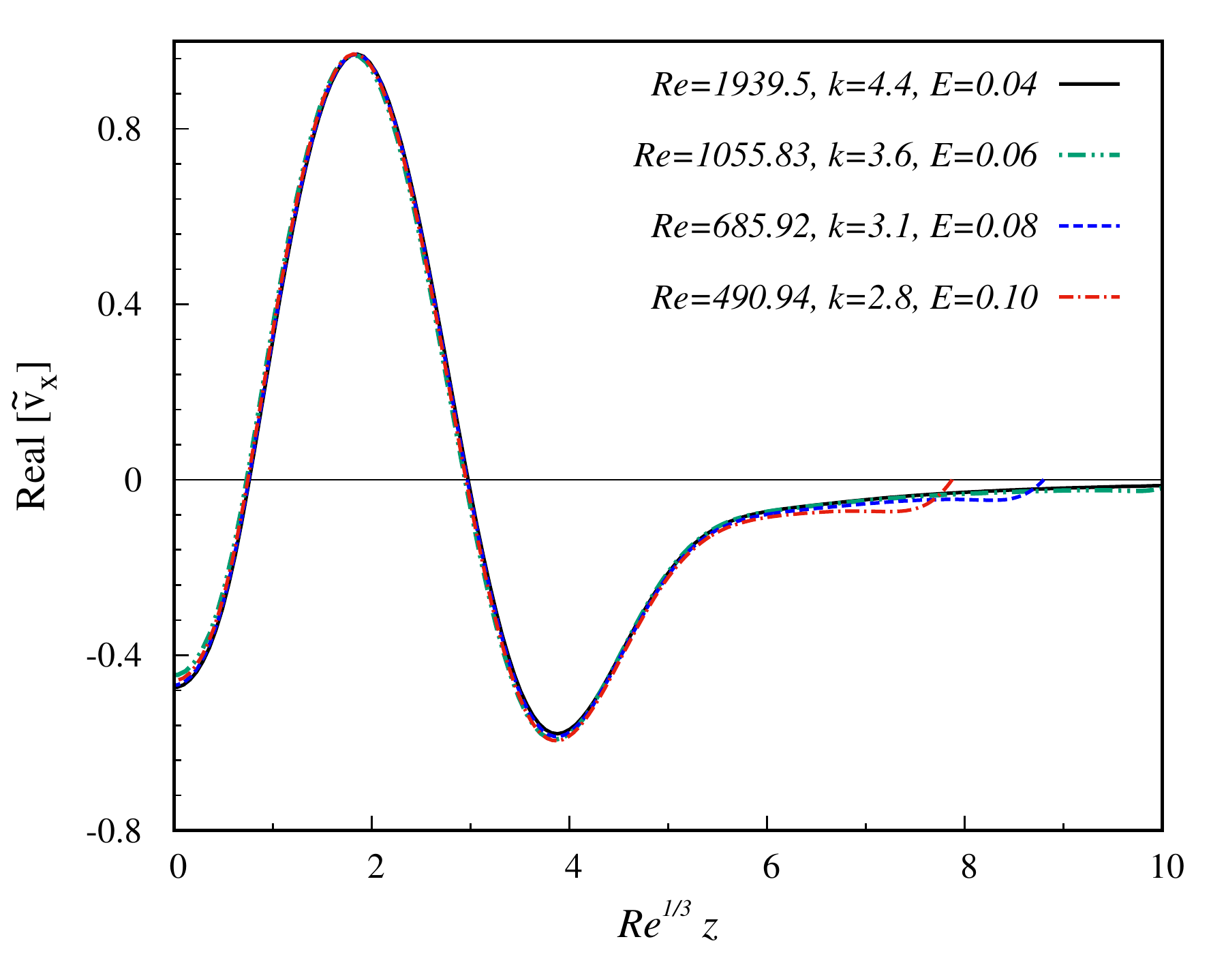}
        \caption{axial}
        \label{fig:V_x_Eigfn_for_fixed_Re_critical}
    \end{subfigure}%
    ~ 
    \begin{subfigure}[b]{0.5\textwidth}
        \centering
        \includegraphics[width=\textwidth]{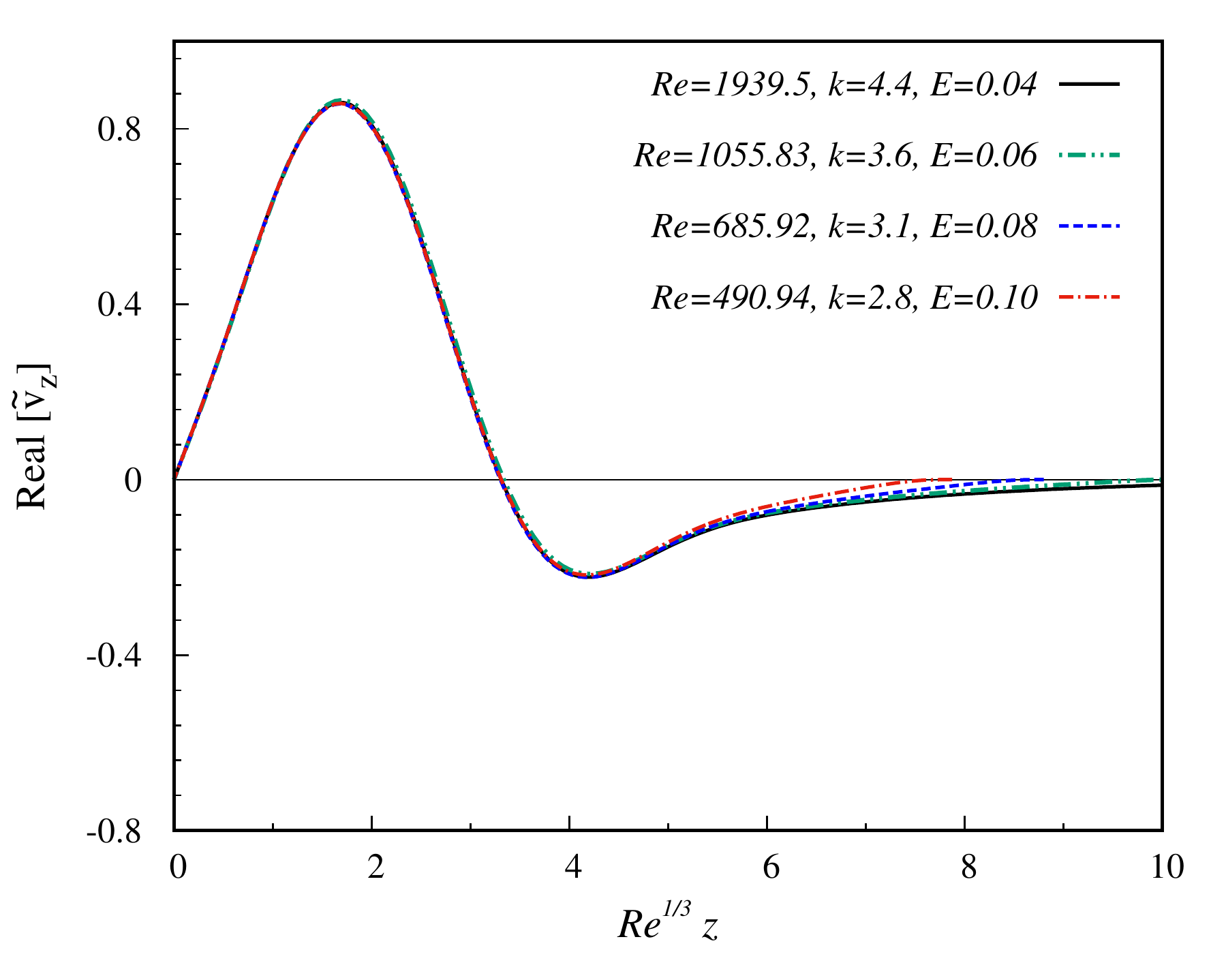}
        \caption{wall-normal}
          \label{fig:V_z_Eigfn_for_fixed_Re_critical}
    \end{subfigure}
\caption{\small The collapse of stream-wise and wall-normal eigenfunctions  corresponding to $Re_c$ and $k_c$ (at $\beta=0.8$ and different $E$) when plotted against the rescaled wall-normal coordinate scaled using the  viscous layer thickness of $\mathcal{O}(Re^{-1/3})$.}
\label{fig:Eigenfunctions_critical_Re_and_k}
\end{figure*}

\begin{figure*}
    \centering
    \begin{subfigure}[b]{0.5\textwidth}
        \centering
        \includegraphics[width=\textwidth]{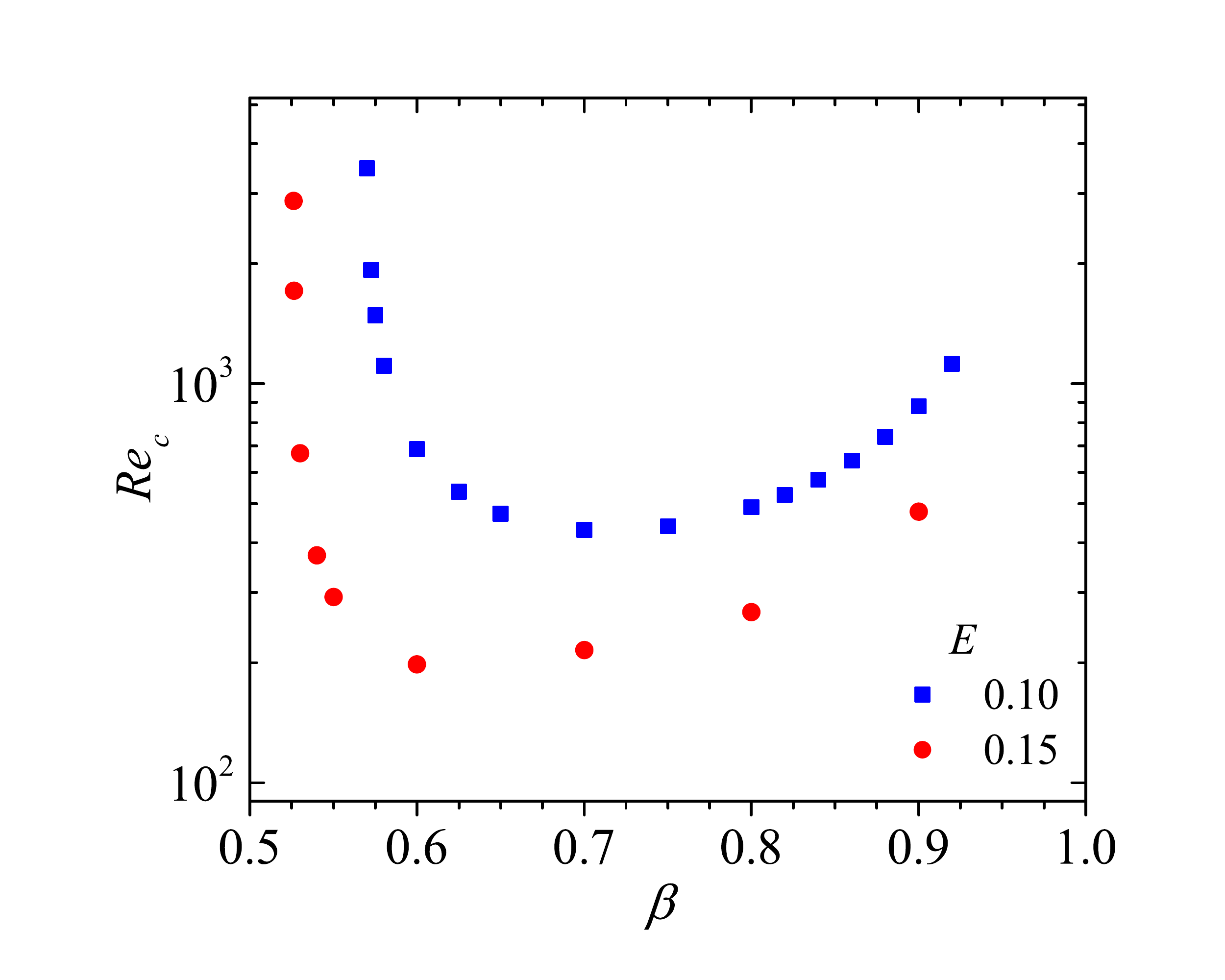}
        \caption{}
        \label{fig:RevsBeta}
    \end{subfigure}%
    ~ 
    \begin{subfigure}[b]{0.5\textwidth}
        \centering
        \includegraphics[width=\textwidth]{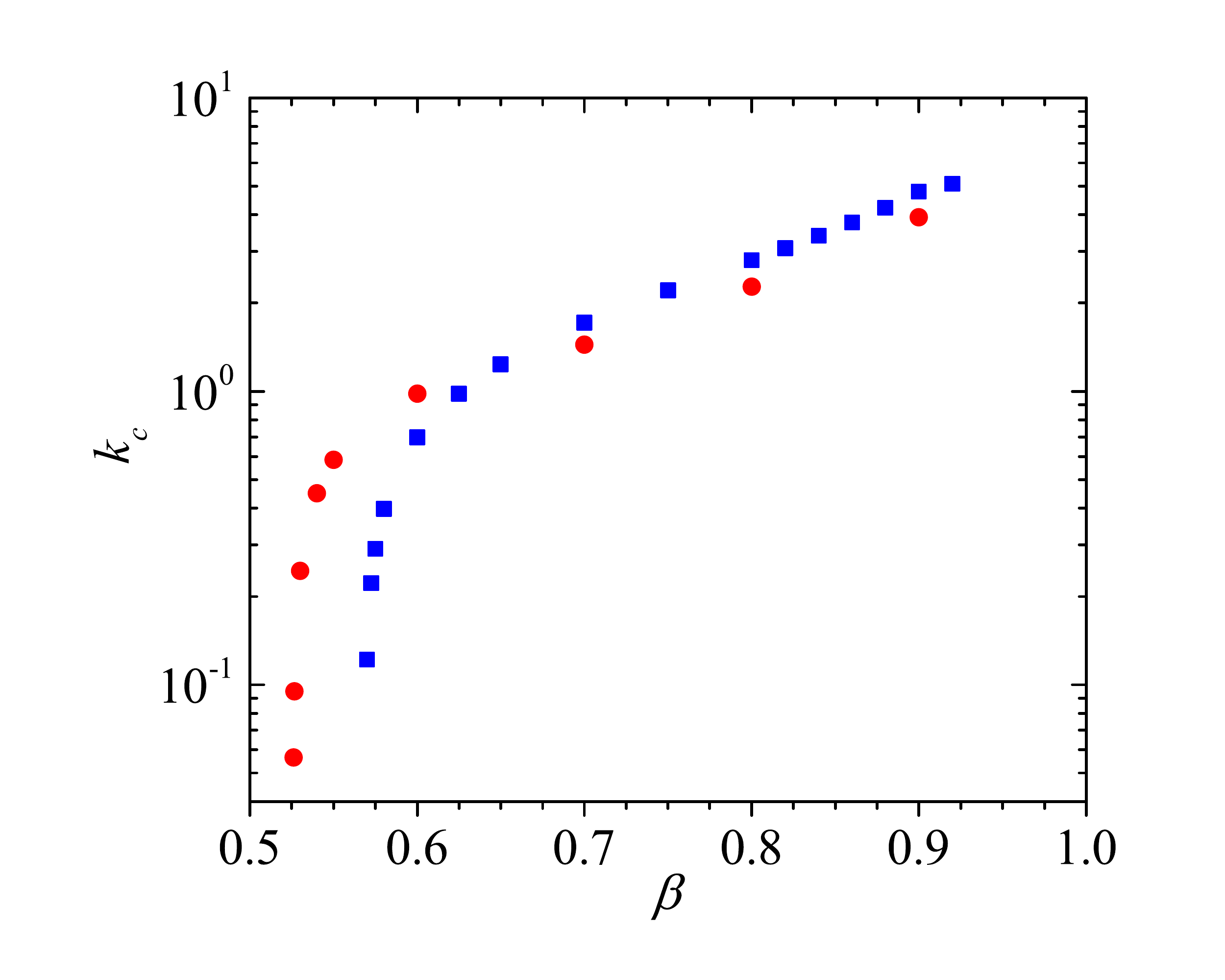}
        \caption{}
          \label{fig:kvsBeta}
    \end{subfigure}
\caption{Variation of (a)  $ Re_c $ and (b) critical wavenumber $ k_c $ as a function of the viscosity ratio $ \beta $ at fixed $ E $.  The minimum $ \beta $ required to sustain the center-mode instability in  channel flow is  $  \approx 0.5$. } 
\label{fig:Re_and_k_vs_Beta}
\end{figure*}

\subsection{Effect of solvent viscosity  on critical parameters} 
\label{ssec:Role of beta}
  
The center-mode instability in pipe Poiseuille flow discussed in our earlier works  \citep{Piyush_2018,chaudharyetal_2020}, rather counter-intuitively, required the presence of solvent viscous effects, with the flow being stable in the UCM limit.
Nevertheless, the pipe-flow instability does continue to exist for very low $\beta \sim 0.001$, with $Re_c$ exhibiting a weak divergence for $\beta \rightarrow 0$. 
In marked contrast, a finite solvent viscous threshold is required for the channel flow instability, with
the instability ceasing to exist  below $ \beta \approx 0.5$ at  $ E =0.01 $ (Fig.~\ref{fig:RevsBeta}). We have further verified that this is, in fact, the lowest $ \beta $ for which the instability is present for any $E$.  Figure~\ref{fig:RevsBeta} also shows a non-monotonic  behaviour of $ Re_c $ with $ \beta  $, at fixed $ E \sim O(1)$, rather similar to the variation of $Re_c$ with $E $ (at fixed $\beta$). 
In the limit of $\beta \rightarrow 1$, $Re_c$ does diverge for channel flow, in a manner similar to that seen in pipe flow \citep[see Fig.~5 of][]{Piyush_2018}.
The divergence of $Re_c$ for $\beta \rightarrow 1$ appears, at first sight, to contradict the results shown in Fig.~\ref{fig:Re_min-Beta-E},
where $Re_c$ decreases in the same limit. There is no inconsistency, however, since the parameters kept constant differ in the two cases. 
In Fig.~\ref{fig:RevsBeta}, $E$ is fixed at $0.1$, while in Fig.~\ref{fig:Re_min-Beta-E}, $E$ is allowed to vary, and increases to very high values for $\beta \rightarrow 1$.
%
The eigenfunctions at the lowest $\beta$'s for which the center-mode instability is present are shown in Fig.~\ref{fig:eigfnRe800k1pt5E0pt1beta0pt6}. Interestingly, the eigenfunctions at $\beta = 0.6$  (and $Re = 800$, $k = 1.5$) are qualitatively similar to the eigenfunctions at a much higher $\beta = 0.96$
(and $Re = 650$, $k = 1$) shown in 
Fig.~\ref{fig:Eigenfunction_vx_vz}, suggesting that the shape of the center-mode eigenfunctions is rather robust over the entire unstable range of $\beta$'s.

\begin{figure*}
	\centering
	\begin{subfigure}[b]{0.5\textwidth}
		\centering
		\includegraphics[width=\textwidth]{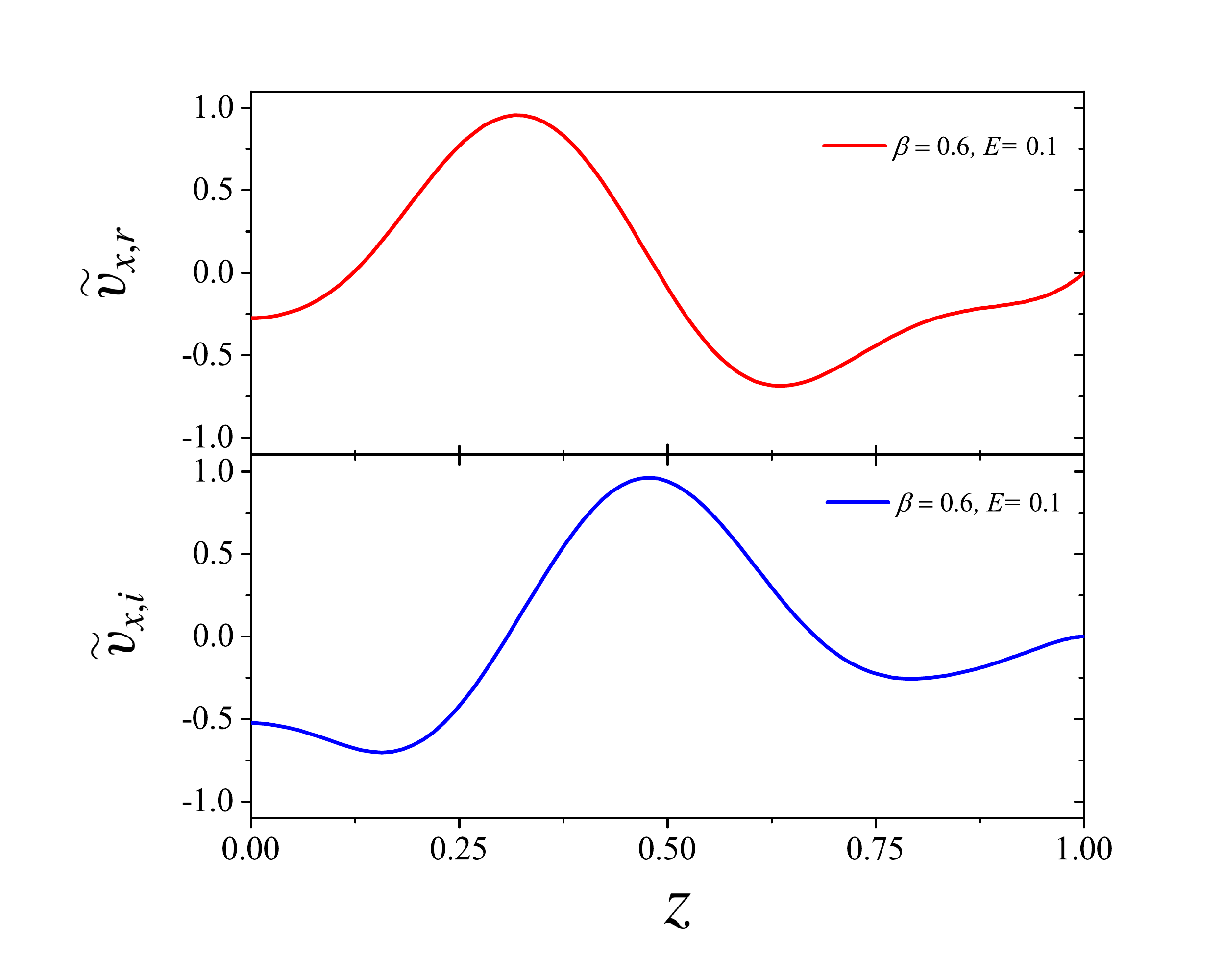}
		\caption{$\tilde{v}_x$}
		\label{fig:Ep1_Re800_k0p5pi_B0p8_vx}
	\end{subfigure}%
	~ 
	\begin{subfigure}[b]{0.5\textwidth}
		\centering
		\includegraphics[width=\textwidth]{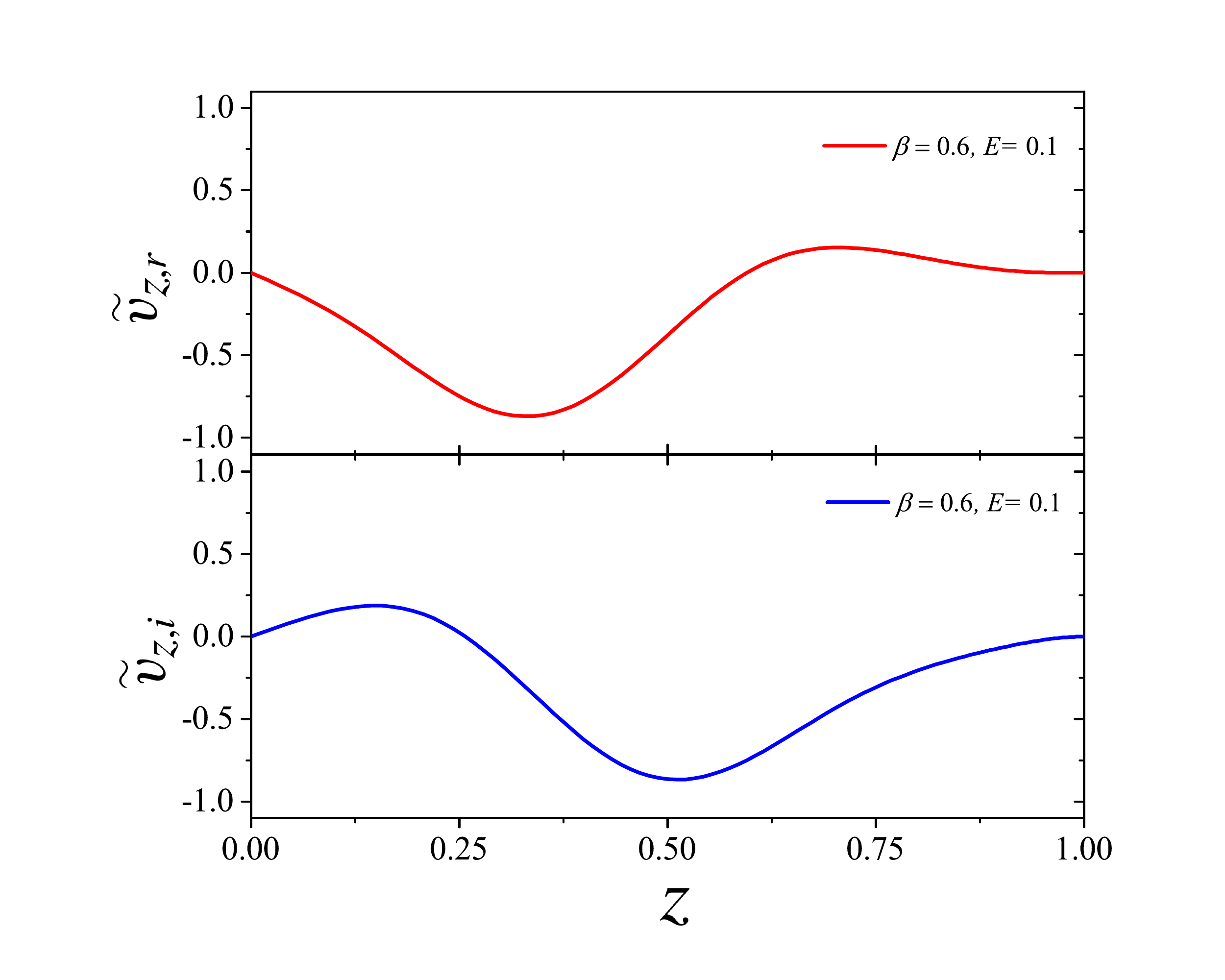}
		\caption{$\tilde{v}_z$}
		\label{fig:Ep1_Re800_k0p4pi_B0p8_vz}
	\end{subfigure}

\caption{\small Normalized eigenfunctions for the streamwise (a) and wall normal (b) perturbation velocities near the lowest value of $ \beta$'s for which center-mode instability exists in viscoelastic channel flow. Data shown for the eigenvalue $ c = 0.99778 + 5.78112 \times 10^{-5} i $ at $ Re =800, k=0.6, E=0.1,\beta =0.6$.}
\label{fig:eigfnRe800k1pt5E0pt1beta0pt6}

\end{figure*}

%

\begin{figure}
        \centerline{\includegraphics[width=0.5\textwidth]{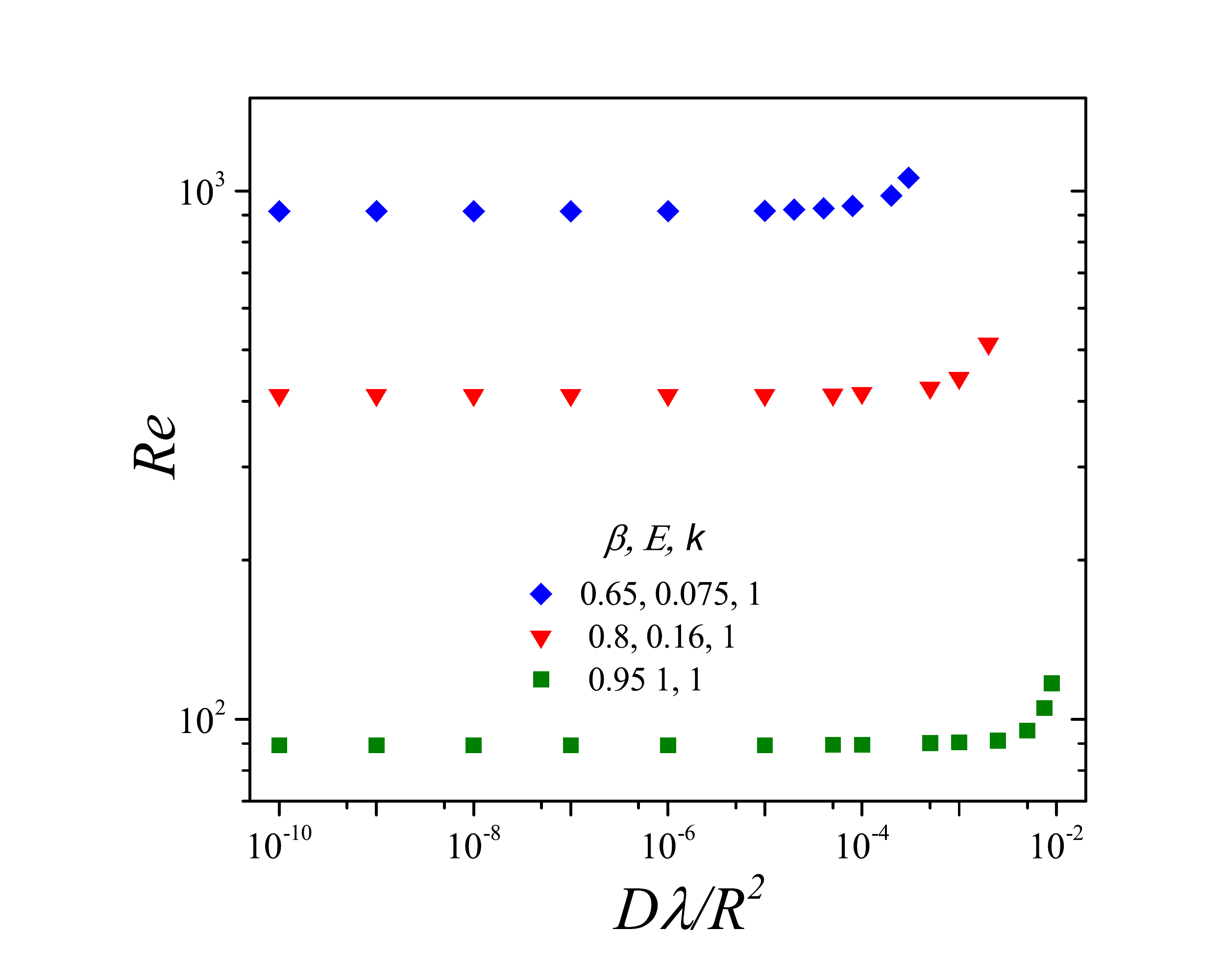}}
        \caption{\small The effect of stress diffusion coefficient $D \lambda/H^2$ on the threshold $Re$ required for center-mode instability at different $E$, $\beta$, $k$.}
        \label{fig:diffusion}
\end{figure} 

\subsection{Role of diffusion on the center-mode instability}
\label{ssec:diffusion}
In this section, we explore the role of stress diffusion on the center-mode instability. The underlying microscopic origin of stress diffusion is the Brownian (translational) diffusion of the polymer molecules, with a diffusivity $D \sim 10^{-12}$ m$^2$/s, and a corresponding Schmidt number $Sc = \nu/D \sim 10^6$, with $\nu$ being the kinematic viscosity of the polymer solution. To this end, the 
Oldroyd-B constitutive equation is now augmented with 
 a stress diffusion term, whose importance, in dimensionless terms, is characterized by $D \lambda/H^2 $  \citep{chaudharyetal_2020}.
While many older \citep{sureshkumar1995diffusive,sureshkumar_1997} and a few recent \citep{lopez_choueiri_hof_2019} DNS studies have incorporated an artificially large diffusion coefficient with $Sc \sim O(1)$, the work of \cite{Sid_2018_PRF} has demonstrated that the 2D EIT structures are suppressed for $Sc < 9$. It therefore behooves us to examine whether stress diffusion has a similar effect on the center-mode instability analyzed in this study, especially because of our premise that the center-mode instability is the mechanism underlying the onset of EIT. 
Based on the $D$ given above, a typical relaxation time $\lambda \sim 10^{-3}$s, and with channel half-width $H \sim 1$mm, 
the dimensionless diffusivity $D \lambda/H^2 \sim 10^{-9}$. 
Note that, with the stress diffusion term included, boundary conditions need to be
prescribed for the polymeric stress.  Following earlier efforts \citep{sureshkumar1995diffusive}, these are obtained by using the constitutive equation without diffusion at the two boundaries.  Figure~\ref{fig:diffusion} shows the threshold $Re$ for the center-mode instability as a function of $D \lambda/H^2$, for fixed sets of $E$, $\beta$ and $k$. 
For $D\lambda/H^2 \rightarrow 0$, the threshold $Re$ for instability for the model with stress diffusion approaches that of the Oldroyd-B model without diffusion; importantly, $Re_c$ remains virtually unaltered for the aforementioned estimate of $D\lambda/H^2 \sim O(10^{-9})$. However, similar to pipe flow \citep{chaudharyetal_2020}, $Re_c$ increases steeply for $D\lambda/H^2$ greater than a threshold that is a function of $E$ and $\beta$. For $(\beta,E,k) \equiv (0.8, 0.16, 1)$, this threshold is $O(10^{-3})$, corresponding to $Sc = E/(D \lambda/H^2) \sim 100$ for $E = 0.1$. This stabilization of the linear center-mode instability beyond a threshold stress diffusivity is broadly consistent with the disappearance of the span-wise structures in the fully nonlinear simulations of \cite{Sid_2018_PRF} discussed above.
%

\subsection{Comparison with  experiments}
\label{ssec: Comparison}
We compare our theoretical predictions with the experiments of \cite{Srinivas-Kumaran2017}, who studied the flow of 
 30 and 50 ppm  polyacrylamide (PAAm) solutions (molecular weight  $ 5 \times 10^6 $) through rectangular microchannels with a gap width of $160\mu$m and a cross-sectional aspect ratio of $1:10$. The rather high aspect ratio used in these experiments justifies 
a comparison of their results with the present linear stability results obtained using the plane-Poiseuille flow approximation. 
 The transition was characterized by an increase in the standard deviation of velocity fluctuations, as inferred using particle image velocimetry. We estimated
 the elasticity numbers ($E \equiv \lambda \nu /\rho H^2$, $\lambda$ being the longest relaxation time of polymer while $\nu,$ and $H$ are respectively the kinematic viscosity of the solution and channel half width) for these experiments using Zimm relaxation times. 
The $Re_c$'s from our stability analysis  are in very good agreement with the threshold $Re_t$ inferred from experiments (Table~\ref{table:theory vs experiments}).

A point, made on more than one occasion in the manuscript, is that viscoelastic channel flow continues to be linearly unstable even at $Re \sim O(1)$, provided the elasticity number is sufficiently large. In Fig.~\ref{fig:Re_min-Beta-E}, $Re_c$ dips down to about $5$ at an $E$ of $O(200)$ (with $\beta = 0.99$). In this regard, it is worth mentioning the recent experiments of Steinberg and co-workers \citep{Varshney_2017_wake,Varshney_2018,Varshney_2018_mixing}, which demonstrate the feasibility of achieving very high $E$'s with  dilute polymer solutions. The experiments involve a channel flow setup, although the focus is entirely different; the authors analyze elasticity-induced transitions in the free-shear flow set up between a pair of cylindrical obstacles embedded in the imposed pressure-driven flow. Importantly, the experiments access $W$'s in excess of $10^3$ with $Re$ still being substantially smaller than unity. While the authors' interpretation of their results are based on the instability of the elastic shear layer between the pair of cylinders, and motivated by the  elastic turbulence paradigm, it is worth noting that the small-radii cylinders might also act as a trigger for exciting the elastoinertial center-mode instability discussed here. Note that the polymer concentration in the above experiments is quite low ($c = 80$ppm, with the overlap concentration $c^* \approx 200$ppm), and shear thinning effects are therefore negligible. In contrast, there have been other reports of  instabilities \citep{bodiguel-et-al-2015,poole-2016,picaut-et-al-2017,Bidhan_PRF} in channel/tube flows of highly shear-thinning  concentrated  solutions ($\beta < 0.2$), but these observations cannot be explained  by the center-mode instability which is absent for $\beta \leq 0.5$.

%

\begin{table}
	\begin{center}
		\def~{\hphantom{0}}
		\begin{tabular}{lcccc} \hline
			$E$ & $\beta$ & $C_p(ppm)$ & $Re_c$ (theory) & $Re_t$ (experiment) \\ 
			\hline 
			0.22 &  $0.915$ & $50$ & $289$ & $267-311$ \\
			0.22 &  $0.92$ & $30$ & $333$ & $311-355$ \\
			\hline
		\end{tabular}
		\caption{\small Comparison of present theoretical predictions for $Re_c$ with the  experimentally inferred transition Reynolds number $Re_t$ of  \cite{Srinivas-Kumaran2017} for the flow of polyacrylamide solutions in rectangular microchannels. Here, $C_p$ denotes the concentration of the polymer solutions used.}
		\label{table:theory vs experiments}
	\end{center}
\end{table}



\subsection{Linear  vs. nonlinear transition scenarios in  viscoelastic channel flow}
\label{ssec:ECS_vs_EIT}

As mentioned in the Introduction, transition to turbulence in  canonical parallel shear flows of Newtonian fluids has a subcritical character, being preceded by the  emergence and proliferation of nonlinear three-dimensional solutions (including travelling waves), termed `exact coherent states' (ECS), in an appropriate phase space.  Motivated by this Newtonian picture, \cite{Graham2007} studied the effect of viscoelasticity (using a FENE-P model) on the simplest  ECS solutions in plane Poiseuille flow, \textit{viz.}, the nonlinear travelling waves  originally found for the Newtonian case by \cite{waleffe_2001}, with the aim of inferring the effect of viscoelasticity on transition.
The results from Fig.~2 of \cite{Graham2007} for the Reynolds number $Re_{min}$ required for the existence of the travelling-wave ECS  are shown in Fig.~\ref{fig:ECS_vs_EIT_Beta_0.90} for
$\beta = 0.9$ and  in Fig.~\ref{fig:ECS_vs_EIT_Beta_0.97} for $\beta = 0.97$; the results have been replotted as a function of $E$, rather than $W$ used by those authors. The first effects of viscoelasticity, extending up to 
  $ E \le 0.01 $,  manifest as  a slight decrease (not visible on the scale of the plot) in $Re_{min}$ from the Newtonian value; for $ E > 0.03 $, however, $ Re_{min} $ increases abruptly, implying a rapid shrinking (and subsequent disappearance) of the regime of existence of the simplest ECS. 
Assuming this stabilizing effect to hold for the other ECS's with a non-trivial time dependence (for instance, relative periodic orbits), one may infer that viscoelasticity  tends to suppress the subcritical Newtonian transition.   Figure~\ref{fig:ECS_vs_EIT} also shows  the threshold Reynolds number, $Re_c$,  for the onset of the center-mode instability. For completeness, we  show, in addition, the $Re_c$ for the elastically modified TS mode (recall that $Re_c$ in this case equals $5772$ for $E = 0$). Note that while the results of \cite{Graham2007} are for a FENE-P fluid and the present results have been obtained using the Oldroyd-B model, our preliminary stability calculations for a FENE-P fluid  show that the present results are not qualitatively altered by finite extensibility.

Figure~\ref{fig:ECS_vs_EIT} allows one to rationally infer the transition scenario pertinent to a given viscoelastic channel flow configuration, and should serve as a guide for future experimental efforts probing transition in the flow of polymer solutions through rectangular channels. 
Note that two types of transition experiments  have been carried out in the literature: the `forced transition', wherein the inlet was subjected to a disturbance of fixed finite amplitude 
\citep[for instance, a commonly used forcing mechanism is via fluid injection at the walls; see][]{darbyshire_mullin_1995,hof_etal_2003}, and the `natural transition' that ensues in the absence of any
imposed disturbances.  
Based on the above, one may clearly differentiate between two extreme scenarios for channel-flow transition. The first is that of a `noisy' experimental set-up, where the sub-critical 
forced transition occurs at an $Re_c \approx 1000$ in the Newtonian limit (correlated to the emergence of the ECS's at a slightly lower $Re$). The viscoelasticity-induced suppression of the ECS's then leads to a steep increase in $Re_c$ with increasing $E$, and finally, at much higher $E$'s, a rapid decrease in $Re_c$ results corresponding to the onset of the linear center-mode instability. At the other extreme, for a sufficiently refined setup, the Newtonian transition would be the natural one, occurring at $Re_c = 5772$ for $E = 0$, with $Re_c$ exhibiting a relatively gentle increase with $E$ thereafter, along the TS-wall mode branch, until the point of intersection with the centermode branch. This intersection corresponds to a fairly modest $E$ of $O(10^{-2})$ for $\beta = 0.9$ (see Fig.~\ref{fig:ECS_vs_EIT_Beta_0.90}), after which $Re_c$ begins to decrease due to the center-mode instability, similar to the forced transition above. For intermediate noise levels, one expects the transition scenario to interpolate between these two extremes.

Interestingly, Fig.~\ref{fig:ECS_vs_EIT} bears a qualitative resemblance to that obtained by \cite{Samanta2013} for their pipe-flow experiments (see Fig.~3a therein). Note that $E$ in Fig.~\ref{fig:ECS_vs_EIT} may be treated as a surrogate for the polymer concentration used in \cite{Samanta2013}; in either case, a given experiment corresponds to a vertical line in Fig.~\ref{fig:ECS_vs_EIT}. 
For Newtonian pipe flow, the forced transition is again subcritical (and related to the emergence of ECS's similar to those for channel flow), 
and in the experiments of \cite{Samanta2013}, this transition occurred at $Re_c \approx 2000$ (an exact critical point of $Re_c = 2040 \pm 10$  has been identified in this regard based on the emergence and subsequent splitting of the ECS's -- see \cite{avila2011}). However, the linear stability of pipe flow implies that the natural transition, although at a higher $Re_c$,  is again sub-critical,
and therefore, in contrast to the channel flow case. Thus, while the natural transition in the Newtonian limit can, in principle, be delayed to very high Reynolds numbers in suitably refined setups \citep{pfenniger1961boundary}, it occurred at $Re_c \approx 6500$ for \cite{Samanta2013}.
For the forced transition, \cite{Samanta2013} did observe an increase in $Re_c$ with polymer concentration, similar to the role played by $E$ in the subcritical channel-flow transition discussed above, and that may be rationalized based on the elasticity-induced suppression of the underlying ECS solutions. However, the $Re_c$ for the natural transition decreased from $6500$ with increasing $E$ (although the authors explicitly state the Newtonian threshold, as is also evident from their Fig.~2a, their Fig.~3a nevertheless does not connect to this Newtonian threshold, and is instead suggestive of an apparent divergence of the threshold
$Re$ in the limit of zero concentration).
As mentioned in \cite{chaudharyetal_2020}, this runs counter to the stabilizing role of elasticity on the simplest ECS's predicted by \cite{Graham2007}, and implies a differing role of elasticity on the more complex set of ECS's that presumably determine the turbulent trajectory at the higher $Re$. This behavior for pipe flow above suggests that the effect of an increasing $E$ on the channel flow transition, in cases where the transition occurs at $Re$'s greater than $O(1000)$ (and until close to the linear TS-mode threshold), might depend on the relative influences of the TS wall-mode vis-a-vis the ECS solutions which in turn might depend both on the $Re$ and on the detailed nature of the induced disturbance. When the ECS solutions play a dominant role for small $E$, similar to \cite{Samanta2013}, one expects the $Re_c$ to decrease with increasing $E$ to begin with, with a subsequent more rapid decrease at higher $E$ arising due to the center-mode instability.

\begin{figure}
	\centering
	\begin{subfigure}[t]{0.5\textwidth}
		\centering
		\includegraphics[width=\textwidth]{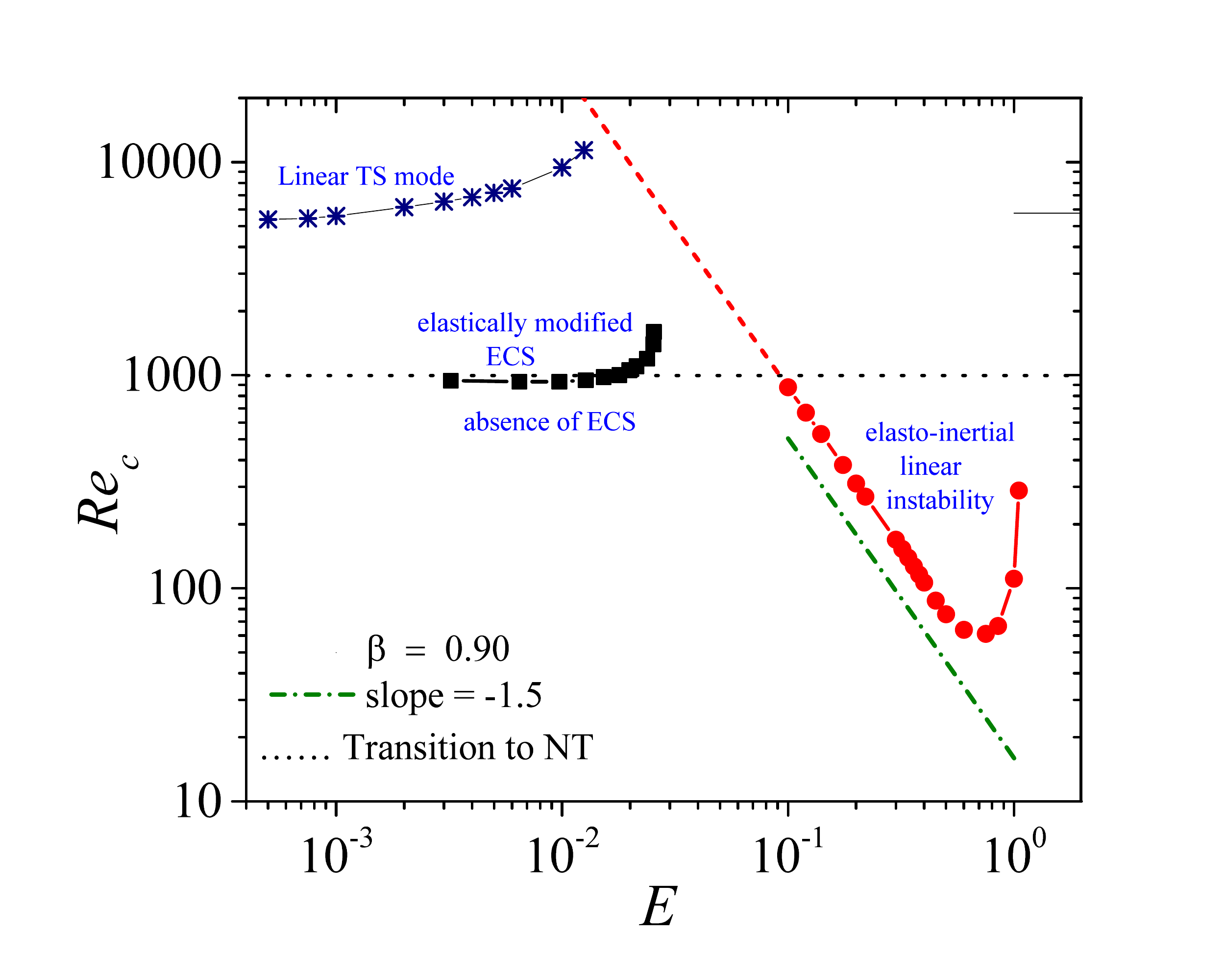}
		\caption{$\beta =0.9$}
		\label{fig:ECS_vs_EIT_Beta_0.90}
	\end{subfigure}%
	~ 
	\begin{subfigure}[t]{0.5\textwidth}
		\centering
		\includegraphics[width=\textwidth]{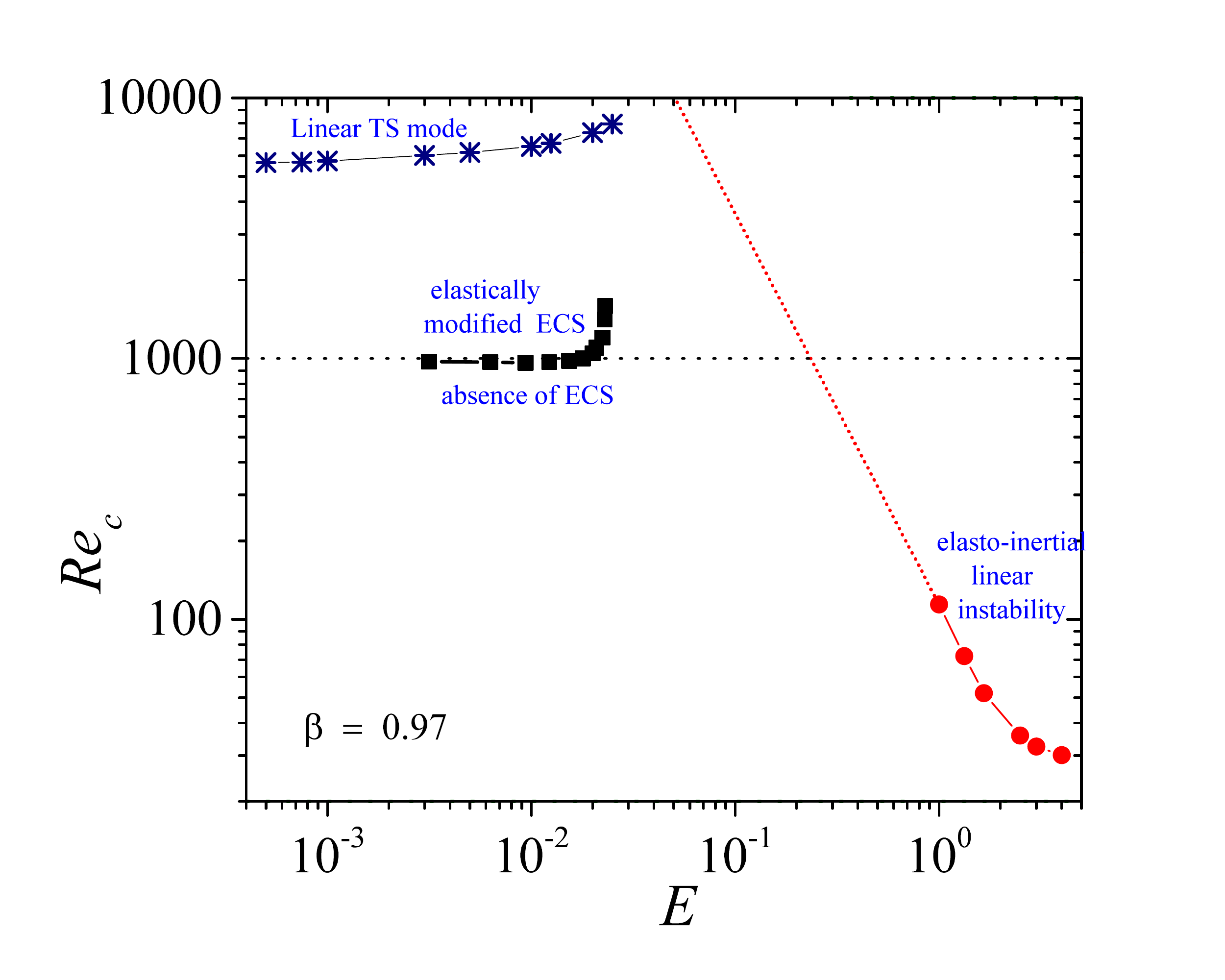}
		\caption{$\beta =0.97$}
		\label{fig:ECS_vs_EIT_Beta_0.97}
	\end{subfigure}
	\caption{\small Boundaries demarcating the existence of elastically-modified ECS solutions \citep[black squares;][]{Graham2007}, the elastically-modified linear TS mode (blue stars; present study) and the linear center-mode  instabilities (red circles; present study) in the $Re$--$E$ plane for $ \beta = 0.9$, $0.97$.  The lack of points on the center-mode threshold curve is only an apparent one, since the numerics have begun conforming to the small-$E$ asymptote (The red dashed line represents the $Re_c \propto E^{-3/2}$ scaling for the center-mode, extrapolated down to $E \sim 0.01$). The black dotted line represents the experimental threshold for Newtonian turbulence (NT).}
	\label{fig:ECS_vs_EIT}
\end{figure}

  \begin{figure}
	\centering
	\begin{subfigure}[t]{0.5\textwidth}
		\centering
		\includegraphics[width=\textwidth]{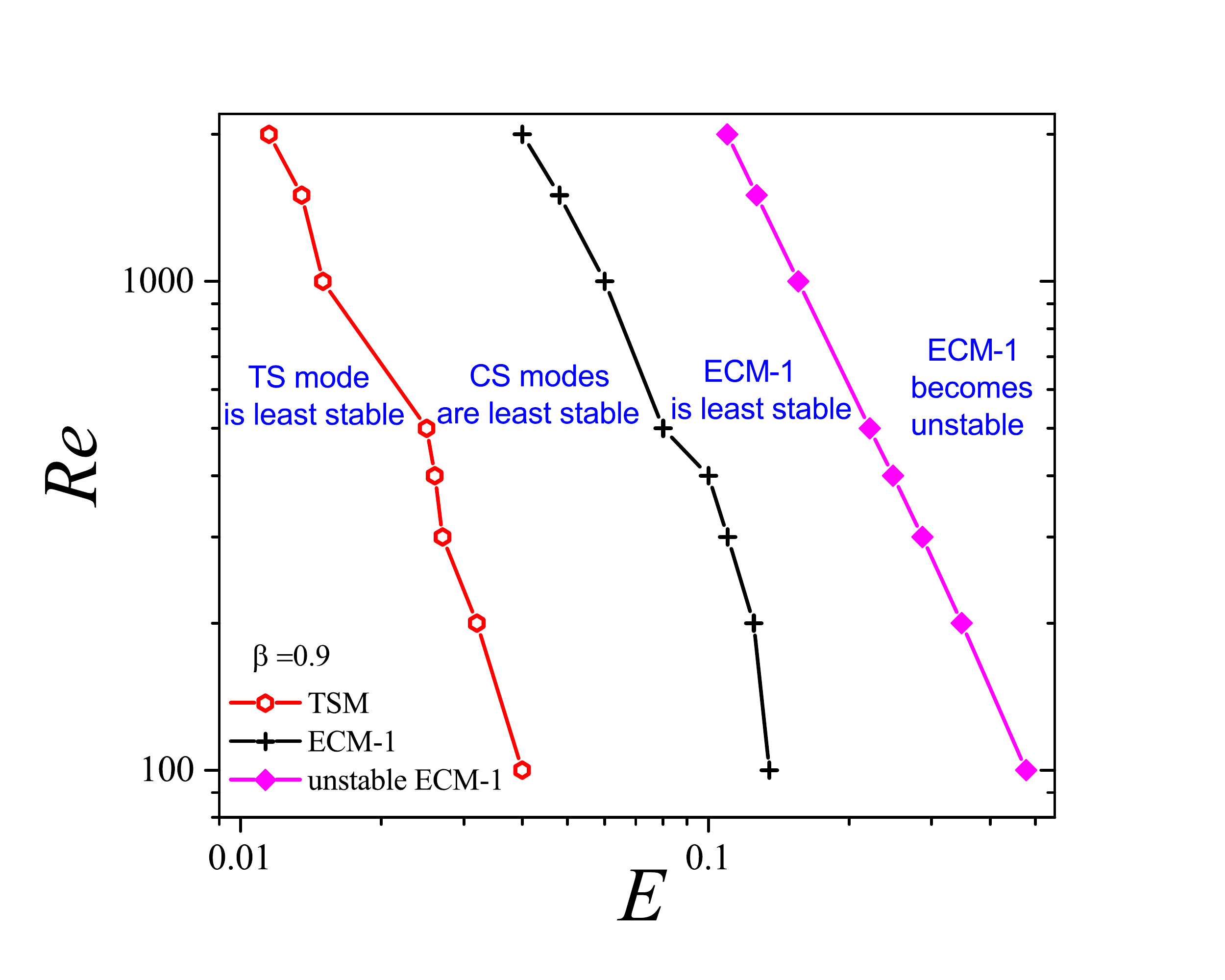}
		\caption{$\beta =0.9$}
		\label{fig:tsandcmbeta0.9}
	\end{subfigure}%
	~ 
	\begin{subfigure}[t]{0.5\textwidth}
		\centering
		\includegraphics[width=\textwidth]{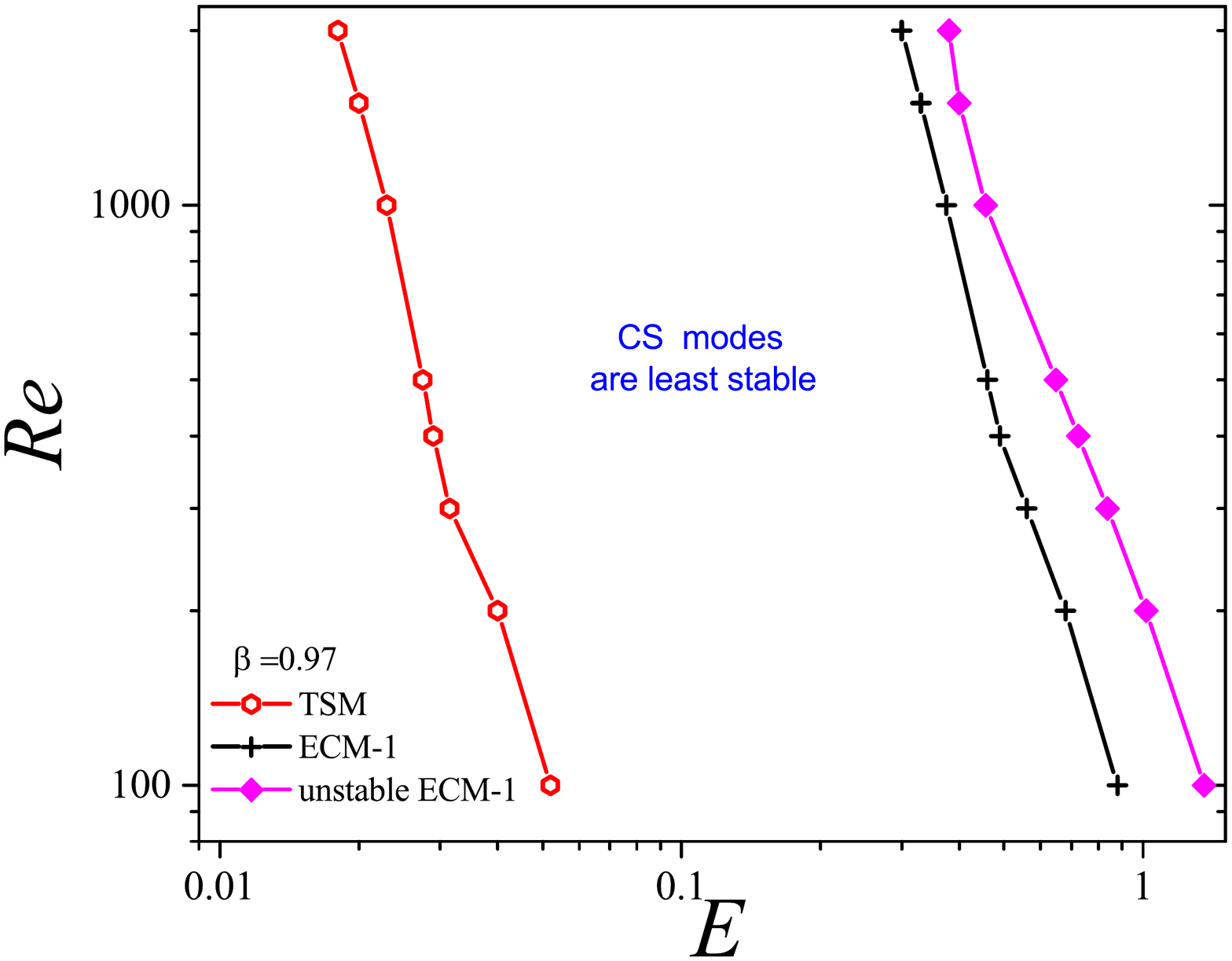}
		\caption{$\beta =0.97$}
		\label{fig:tsandcmbeta0.97}
	\end{subfigure}
	\caption{\small Regions in the $Re$--$E$ plane where the TS, CS, and ECM-1 are the least stable/unstable for $k = 0.4\pi$ and $ \beta = 0.9$, $0.97$. The TS mode is the least stable in the region to the left of the red curve, while the CS modes are the least stable in the region between the red and black curves. 
 The center mode (ECM-1) is the least stable in the region between the black and pink curves, and is unstable in the region to the right of the pink curves.}
	\label{fig:tsandcentermodes}
\end{figure}

In the context of the forced transition scenario above, we mentioned the suppression of the ECS's at a fairly modest $E$, and the emergence of the center-mode-mediated transition only at higher $E$'s, implying the existence of an intermediate $E$-interval where neither mechanism might be operative. For instance, considering a fixed-$Re$ path, with $Re \approx 1500$ in Fig~\ref{fig:ECS_vs_EIT_Beta_0.90} for $\beta = 0.9$, the ECS solutions are restricted to $E$ below an (approximate) threshold of $0.04$; in contrast, the 2D center-mode instability is only operative for $E > 0.09$. Thus, there is the possibility of transition in the interval $0.04 < E < 0.09$ being controlled by novel subcritical mechanisms. In this regard, as briefly mentioned in the Introduction and discussed below, two very different mechanisms, with their origins in the center and wall modes of the elasto-inertial spectrum, have recently been proposed.

The first proposal, by \cite{Shekar2019}, is rooted in the least stable TS wall mode, as already discussed in
Sec.~\ref{ssec:Relative stability of center and wall modes}. However, it was demonstrated therein that the continuation of the TS mode is no longer present in the elasto-inertial spectrum as $E$ is increased. Indeed, it was shown that there is a range of $E$'s for which there are no discrete stable modes above the CS, with the CS being the least stable in this range. 
The center mode eventually emerges above the CS at higher $E$'s, and 
is the least stable  or unstable mode in viscoelastic channel flows, implying that, beyond the
smallest $E$'s,  even a nonlinear (subcritical) mechanism underlying the transition must necessarily involve the signatures, either of the least-stable center mode or the stable CS. 
This scenario is further illustrated in Fig.~\ref{fig:tsandcentermodes}, where we demarcate regions in the $Re$-$E$ plane for a fixed $k = 0.4\pi$ (and for $\beta = 0.9$ and $0.97$) where the TS, CS and the center modes are least stable or unstable. 
For $k = 0.4\pi$, the TS mode is the least stable only for sufficiently small $E$'s (e.g., for $E < 0.015$  for $\beta = 0.9$ and $E < 0.02$ for $\beta = 0.97$ in Fig.~\ref{fig:tsandcentermodes}); for an intermediate range of $E$'s (a range that increases in extent as $\beta$ approaches unity),  there are no discrete modes above the CS in the elastoinertial spectrum, with the CS dominating the dynamics. At higher $E$'s, the center mode emerges above the CS, and is either the least stable or unstable mode. The least stable nature of the TS mode  at the lowest $E$'s (for $k = 0.4 \pi$) in Fig.~\ref{fig:tsandcentermodes}  is, however, sensitive to the wavenumber chosen, and as already seen in
Sec.~\ref{ssec:Relative stability of center and wall modes}, for $k > 2$, the center mode is the least stable even in the Newtonian limit.

The second mechanism, proposed by \cite{page2020exact},  is based on a novel elasto-inertial coherent state that bifurcates subcritically from the center-mode instability,  therefore continues to exist even in regimes where the centermode is stable (thereby being relevant to the aforementioned intermediate range of $E$'s). In particular, \cite{page2020exact} carried out DNS using the FENE-P model, and used an arc-length procedure to continue the center-mode eigenfunction to the subcritical regime. Their study identified a structure with polymer stretch contours resembling an `arrow head' configuration,  and shares similarities with the structures seen transiently in DNS of the EIT regime \citep{dubief2020coherent}.
These 2D elasto-inertial coherent states owe their origin to both inertia and elasticity, and thus are absent in the Newtonian limit, unlike the elastically modified 3D ECS's analyzed by Graham and co-workers which are, essentially, of a Newtonian origin.

\section{Conclusions}
\label{sec:conclusions}
The present study provides a comprehensive account of the linear stability of plane Poiseuille flow of an Oldroyd-B fluid, and shows that in the limit of sufficiently elastic ($E \sim 0.01$ and higher) and moderate-to-highly dilute ($\beta$  $\sim 0.6$ and higher) solutions, the flow becomes unstable to a two-dimensional center mode with phase speed close to the maximum base-flow velocity, and at a critical Reynolds number, $Re_c$, much lower than the typical Newtonian threshold of $\sim 1100$. We also provide a detailed account of the  emergence of the unstable center mode in the elasto-inertial spectrum. 
Several features of the instability predicted here for channel flow are analogous to those for viscoelastic pipe flow \citep{Piyush_2018,chaudharyetal_2020}, including the scaling of critical Reynolds $Re_c \propto (E (1-\beta))^{-3/2}$ and wavenumbers $k_c \propto (E (1-\beta))^{-1/2}$ in the limit $E(1-\beta) \ll 1$, fixed $E$. Although the disturbances in the aforementioned asymptotic limit are strongly localized near the channel centerline, this is no longer true for experimentally relevant values of $\beta$ and $E$. 
In fact, our theoretical predictions for $Re_c$ are in very good agreement with the observations of \cite{Srinivas-Kumaran2017} for transition in rectangular microchannels.

There are a  few crucial differences between the center-mode stability characteristics of viscoelastic channel and pipe flows, the most important being the absence of the center-mode instability for $\beta < 0.5$ in channel flow, in contrast to
 its persistence down to $\beta \sim 10^{-3}$ in pipe flow. 
In either case, the destabilizing role of solvent viscous effects on the center-mode instability is in contrast to their stabilizing role for wall-mode instabilities \citep{sadanandan_sureshkumar_2002,khalid_solvent}. 
 In the opposite limit of $\beta \rightarrow 1$, the instability persists down to $Re \approx 5$ for channel flow, while being restricted to $Re > 63$ in pipe flow. Thus, while the channel center-mode instability requires a finite solvent viscous threshold, the pipe center-mode instability requires a finite inertial threshold for its existence. It is also worth noting that the prediction of a linear instability for $Re \sim O(1)$, for channel flow, is a significant departure from the prevailing viewpoint of such rectilinear shearing flows being linearly stable at low $Re$, wherein a nonlinear subcritical mechanism was hitherto considered to be the only route to instability \citep{meulenbroek_sarloos2004,morozov_saarloos2005,Pan_2012_PRL}.



Despite the differences for $\beta < 0.5$ and $\beta \rightarrow 1$, for the intermediate range of $\beta$'s, there appears to be a universal linear mechanism underlying the onset of elasto-inertial turbulence in both viscoelastic channel and pipe flows. Thus, the viscoelastic scenario stands in stark contrast to the profound differences between the modal stabilities of Newtonian pipe and channel flows,  with pipe flow being linearly stable for all $Re$ and channel flow exhibiting a linear instability at $Re = 5772$.
The Newtonian transition observed in experiments is now known to be dominated by nonlinear processes, and is similar for both the channel and pipe flow geometries. Theoretically speaking, the transition is attributed to the emergence and subsequent proliferation of ECS solutions of the Navier-Stokes equations, with increasing $Re$, in the neighborhood of the laminar state, and that drive the nonlinear transitional dynamics. The close analogy between the Newtonian pipe and channel transition scenarios, despite the aforementioned contrast in the linear stability characteristics, arises from the structural and dynamical resemblance of the underlying ECS solutions in the two cases.  On the other hand, linear stability theory appears broadly consistent with observations for the viscoelastic case, both for pipe and channel flows. As discussed below, more work, however, needs to be done with regard to the non-linear dynamics of the transition.

It is worth mentioning that the two-dimensional center-mode instability predicted here and the axisymmetric  instability predicted in our earlier work \citep{Piyush_2018,chaudharyetal_2020} are also consistent with the nature of the nonlinear state observed in simulations in these geometries: see \cite{Dubief2013,Samanta2013,Sid_2018_PRF} for the channel case and \cite{lopez_choueiri_hof_2019} for the pipe geometry. In both cases, the nonlinear elastoinertial turbulent state is dominated by span-wise structures in sharp contrast to stream-wise oriented, span-wise varying ones that dominate Newtonian transition.  This contrast between the Newtonian ECS's and the EIT structures has recently found some support in a bifurcation study \citep{page2020exact}, where the authors used an arc-length method to continue the center-mode solutions subcritically, identifying a continuous pathway from the linear threshold. Although this shows the relevance of the center-mode even in the linearly stable regime, the so-called arrowhead EIT structure found does not bear a close resemblance to the center-mode eigenfunctions, presumably due to the (strong) subcriticality.
 However,   
one expects a closer connection between the DNS structures and the linear (center-mode) eigenfunctions 
in parameter regimes where the bifurcation is supercritical
\citep{piyush_weaklynonlinear}.  
The structure identified by \cite{page2020exact}, presumably along with other new elasto-inertial structures, are likely to underlie the dynamics of the EIT state, with the EIT trajectory
sampling these novel elasto-inertial coherent states, akin to how the Newtonian turbulent trajectory samples the multitude of Newtonian ECS's \citep{budanur_etal_2017}.  Identifying the nature of the nonlinear transition mechanisms in the intermediate range of $E$'s, where the (Newtonian) ECS's are suppressed and the flow is linearly stable, is likely to be an important area for future research.  
\\

\noindent{{\bf Declaration of interests}}

The authors report no conflict of interest.


\end{document}